\documentclass[11pt, letterpaper]{article}

\usepackage[letterpaper, portrait, margin=1.6in]{geometry}
\usepackage{cite}

\usepackage{multicol}
\usepackage{xcolor}

\RequirePackage{fix-cm}
\usepackage{layout}
\def\isLong{false}
\def\onecol{true}
\newcommand{\myvspace}[1]{\vspace{0.0in}}

\usepackage{xspace}
\usepackage{helvet}
\usepackage{courier}
\usepackage{fancyhdr}
\usepackage{type1cm}         
\usepackage{subeqnarray}
\usepackage{graphicx}        % standard LaTeX graphics tool
\usepackage{enumerate}
\usepackage{amsmath, latexsym}
\usepackage{array}
\usepackage{tabularx}
\usepackage{url}
\usepackage{relsize}
\usepackage{ifthen}
\usepackage{xcolor}
\usepackage[newitem,newenum]{paralist}
 
\def\isBeamer{false}

\newcommand{\mychapterskip}{\ifthenelse{\equal{\isMPK}{true}}
{\vspace{-.8in}}{}}

\newcommand{\fullversioncolor}{\color{black!50!blue}}
\newcommand{\versions}[2]{\ifthenelse{\equal{\isfullversion}{true}}{{\fullversioncolor#1}}{#2}}
\renewcommand{\vec}[1]{\mathbf{#1}}
\newcommand{\quot}[1]{``#1''}

\newcommand{\R}[1]{\ensuremath{\mathbb{R}^{#1}\xspace}}
\newcommand{\Euc}[1]{\ensuremath{\mathbf{E}^{#1}\xspace}}
\newcommand{\Eucg}[1]{\ensuremath{E(#1)\xspace}}
\newcommand{\Eucgd}[1]{\ensuremath{E^+(#1)\xspace}}

\newcommand\proj[1]{\ensuremath{\mathbf{P}(#1)}\xspace}
\newcommand\RP[1]{\ensuremath{\mathbb{R}{P^{#1}}\xspace}}

\newcommand{\e}[1]{\vec{e}_{#1}}
\newcommand{\EE}[1]{\vec{E}_{#1}}
\newcommand{\one}{\vec{1}}
\newcommand{\eye}{\vec{I}}
\newcommand{\inert}{\mathbf{A}}

\newcommand{\MN}{metric-neutral\xspace}

\newcolumntype{Y}{X}

\newcommand{\bivo}[1]{{\vec{#1}}}
\newcommand{\pip}{\bivo{\Xi}}

\newcommand{\pvelo}{\bivo{\Gamma}}
\newcommand{\velo}{\bivo{\Omega}}
\newcommand{\momo}{\bivo{\Pi}}

\newcommand{\sigo}{\bivo{\Sigma}}

\newcommand{\grade}[2]{\langle #1 \rangle_{#2}}

\newcommand{\pgshort}{\proj{G}}
\newcommand{\pdgshort}{\proj{G^*}}
\newcommand{\pgvshort}{\proj{G(\vec{V})}}
\newcommand{\pdgvshort}{\proj{G^*(\vec{V}}}

\newcommand{\pclal}[3]{\proj{\mathbb{R}_{#1,#2,#3}}\xspace}
\newcommand{\pdclal}[3]{\proj{\mathbb{R}^*_{#1,#2,#3}}\xspace}

\newcommand{\pdclplus}[3]{\proj{\mathbb{R}^{*+}_{#1,#2,#3}}\xspace}

\newcommand{\myboldhead}[1]{\vspace{0in}\hspace{-.15in}\textbf{#1.}}
\newcommand{\myboldquestion}[1]{\vspace{0in}\hspace{-.15in}\textbf{#1?}}
\newcommand{\app}[1]{\Sec{sec:J}}

\definecolor{mypurple}{rgb}{1,0,1}

\definecolor{mygreen}{rgb}{0, .5, 0}

\newcommand{\Fig}[1]{Fig.~\ref{#1}}
\newcommand{\Tab}[1]{Table~\ref{#1}}
\newcommand{\Sec}[1]{Sect.~\ref{#1}}

\ifthenelse{\equal{\isBeamer}{true}}
{

}
{

}

\newcommand{\mycorrection}[1]{}

\renewcommand{\myboldhead}[1]{\vspace{.1in}\hspace{-.14in}\textbf{#1.}}

\usepackage{layout}
\usepackage{paracol}

\graphicspath{{.}{./pictures/}}

\usepackage{setspace}
\usepackage[protrusion=true,expansion=true]{microtype}				% Better typography
\usepackage[english]{babel}	
\usepackage{amsfonts}
\usepackage{latexsym}
\newcommand{\Line}{\mathbf{\Omega}} %boldsymbol{\ell}}
\newcommand{\Point}{\mathbf P}

\newcommand{\PSS}{\mathbf{I}}

\newcommand{\NSBV}{{\mathbf{\Theta}}} % non-simple bivector
\newcommand{\MOTOR}{motor\xspace}
\newcommand{\MOTORS}{motors\xspace}
\newcommand{\mogro}[1]{\mathfrak{M}_{#1,0,1}\xspace}

\begin{document}

\title{\vspace{-1in}Course notes\\
Geometric Algebra for Computer Graphics\footnote{Permission to make digital or hard copies of part or all of this work for personal or classroom use is granted without fee provided that copies are not made or distributed for profit or commercial advantage and that copies bear this notice and the full citation on the first page. Copyrights for third-party components of this work must be honored. For all other uses,
contact the Owner/Author. Copyright is held by the owner/author(s).\newline
SIGGRAPH '19 Courses, July 28 - August 01, 2019, Los Angeles, CA, USA\newline
ACM 978-1-4503-6307-5/19/07.\newline
10.1145/3305366.3328099}\\
SIGGRAPH 2019}

\author{Charles G. Gunn, Ph. D.\footnote{Author's address: Raum+Gegenraum, Brieselanger Weg 1, 14612 Falkensee, Germany, Email: projgeom@gmail.com}
\date{}
}
% The correct dates will be entered by the editor

%\renewcommand*\contentsname{Overview}

%\headers{Projective geometric algebra}{Charles G. Gunn}
%\pagestyle{headings}

%\layout{}
\maketitle

\begin{figure}[hb]
  \centering
  \def\xyz{1.0}
    \setlength\fboxsep{0pt}\fbox{\includegraphics[width=\xyz\columnwidth]{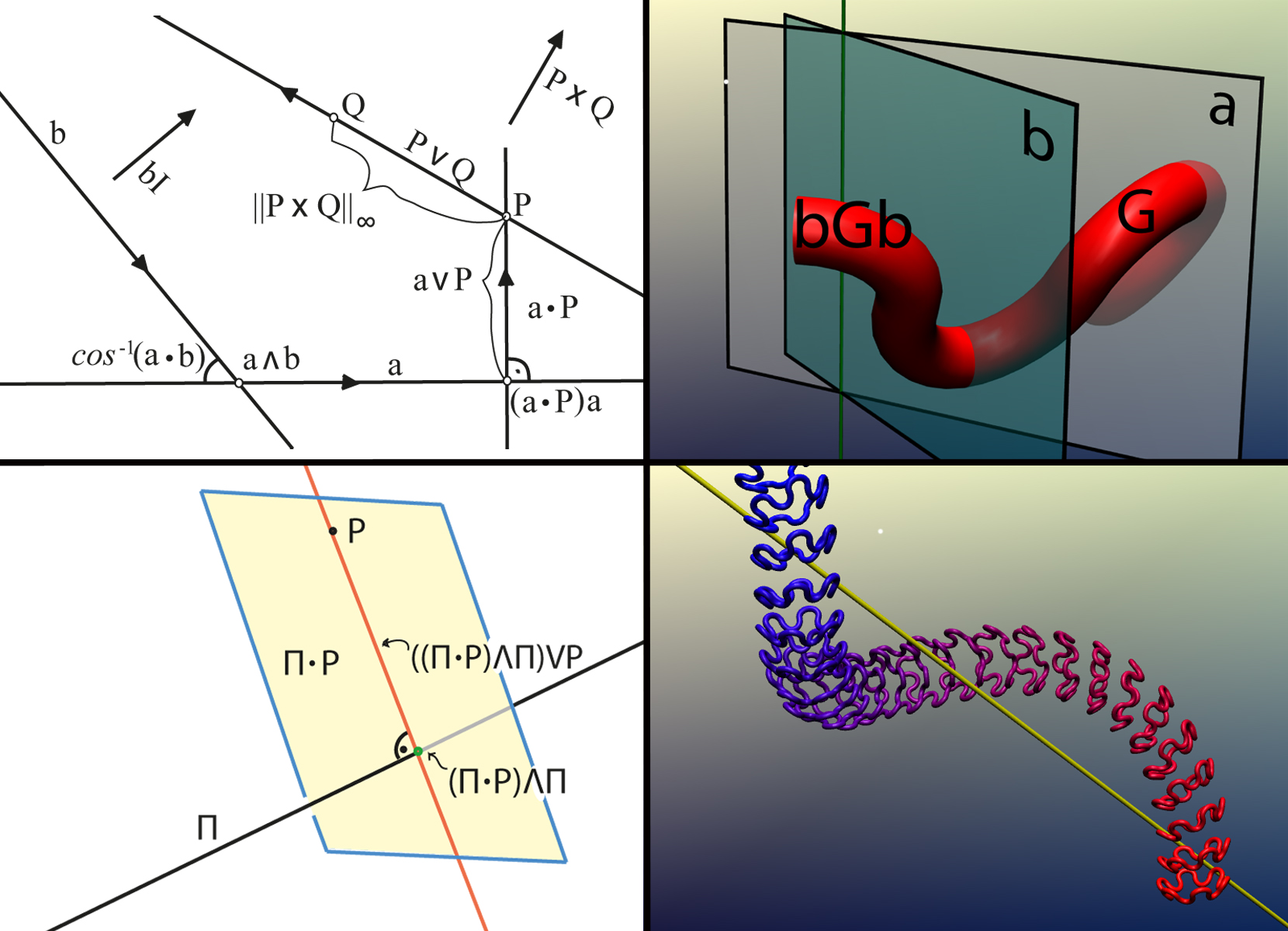}}
  %\caption{Angles of euclidean lines. }
\label{fig:sigcoll}
\end{figure} 

\newpage
\tableofcontents
%\listoffigures
%\listoftables

\section{The question}
What is the best representation  for doing euclidean geometry on computers? 
This question is a fundamental one for practitioners of computer graphics, as well as those working in  computer vision, 3D games, virtual reality, robotics, CAD, animation, geometric processing, discrete geometry, and related fields. 
While available programming languages change and develop with reassuring regularity, the underlying geometric representations   
tend to be based on \textbf{v}ector and \textbf{l}inear \textbf{a}lgebra and \textbf{a}nalytic \textbf{g}eometry  (VLAAG for short), a framework that has remained virtually unchanged for 100 years.  %More satisfactory representations such as DDG (discrete differential geometry) \cite{Crane:2013:DGP} have established themselves in niche domains (in this case surface modeling) but have little impact on the broad base of geometry practice which this article addresses. 
These notes introduce \emph{projective geometric algebra} (PGA) as a modern alternative for doing euclidean geometry and shows how it compares to VLAAG, both conceptually and practically.  % (\cite{gunnthesis}, \cite{gunn2011}, \cite{gunn2017a}, \cite{gunn2017b}) and establishes that it enjoys significant advantages over VLAAG both conceptually and practically. 
%\myboldhead{Previous work} 
In the next section we develop a basis for this comparison by drafting a  wishlist for doing euclidean geometry. % PGA in fact can be whole organism in which each of these parts first finds its true place in the scheme of things. Readers familiar with other geometric algebra approaches to euclidean geometry, for example, conformal geometric algebra (CGA) are referred to \cite{gunn2017a} for a detailed comparison.  The current article focuses on a comparison of PGA with VLAAG. 

%We begin by proposing a developers' \quot{wish-list} for doing euclidean geometry on the computer. In the rest of the article we will establish that PGA fulfills all the wishes on our list.

\myboldhead{Why fix it if it's not broken?}
The standard  approach (VLAAG) has proved itself to be a robust and resilient toolkit. Countless engineers and developers use it to do their jobs. Why should they look elsewhere for their needs?  On the other hand, long-time acquaintance and habit can blind craftsmen to limitations in their tools, and subtly restrict the solutions that they look for and find. Many programmers have had an \quot{aha} moment when learning how to use the quaternion product to represent rotations without the use of matrices, a representation in which the axis and strength of the rotation can be directly read off from the quaternion rather than laboriously extracted from the 9 entries of the matrix, and which offers better interpolation and numerical integration behavior than matrices.

\section{Wish list for doing geometry}
\label{sec:wishlist}
In the spirit of such \quot{aha!} moments we propose here a feature list for doing euclidean geometry. We believe all developers will benefit from a framework that:
\begin{compactitem}
\item  is \textbf{coordinate-free}, 
\item  has a \textbf{uniform representation for points, lines, and planes}, 
\item  can calculate \quot{parallel-safe} \textbf{meet and join} of  these geometric entities, 
\item  provides \textbf{compact  expressions} for  all classical euclidean formulas and constructions, including distances and angles, perpendiculars and parallels, orthogonal projections, and other metric operations, 
\item  has a \textbf{single, geometrically intuitive form} for euclidean motions, one with a single representation for operators \textbf{and} operands, %\footnote{which a 4x4 matrix, for example, is \textbf{not}.}?  
\item provides \textbf{automatic differentiation} of functions of one or several variables,
\item  provides a compact, efficient \textbf{model for kinematics and rigid body mechanics},  
%\item with a flip of a switch also supports \textbf{non-euclidean geometry} as well, and
\item lends itself to \textbf{efficient, practical implementation}, and
\item is \textbf{backwards-compatible} with existing representations including vector, quaternion, dual quaternion, and exterior algebras.
\end{compactitem}
%If your answer to any of these questions is 'yes',  then perhaps this article is for you!

\section{Structure of these notes}
In the rest of these notes we will introduce geometric algebra in general and PGA in particular, on the way to showing that PGA in fact fulfills the above feature list.
The treatment is devoted to dimensions $n=2$ and $n=3$, the cases of most practical interest, and focuses on examples; readers interested in theoretical foundations are referred to the bibliography.  \Sec{sec:wiga} presents an \quot{immersive} introduction to the subject in the form of three worked-out examples of PGA in action. \Sec{sec:roots} begins with a short historical account of PGA followed by a bare-bones review of the mathematical prerequisites. This culminates in \Sec{sec:gpaga} where geometric algebra and the geometric product are defined and introduced. \Sec{sec:eucplane} then delves into PGA for the euclidean plane, written $\pdclal{2}{0}{1}$, introducing most of its fundamental features in this simplified setting.
%:
%\begin{compactitem}
%\item products of pairs of elements, 
%\item formula factories from associativity, 
%\item representation of isometries using sandwiches, and
%\item automatic differentiation.
%\end{compactitem}  
\Sec{sec:eucspace} introduces PGA for euclidean 3-space,
focusing on the crucial role of lines, leading up to  the Euler equations for rigid body motion in PGA. \Sec{sec:ad} describes the native support for automatic differentiation. Sect. \ref{sec:impl} briefly discusses implementation issues. \Sec{sec:comp}  compares the results with alternative approaches, notably VLAAG, concluding that PGA is a universal solution that includes within it most if not all of the existing alternatives. Finally Sect. 12 provides an overview of available resources  for interested readers who wish to test PGA for themselves.
%Before any formal details of what geometric algebras are, we present three examples of 3D  euclidean PGA to provide a feel of what it's about. Throughout this article, points will be represented with bold Latin uppercase, lines as bold Greek uppercase, and planes as bold Latin lowercase.

\section{Immersive introduction to geometric algebra}
\label{sec:wiga}
The main idea behind geometric algebra is that \textbf{geometric primitives behave like numbers} -- for example, they can be added and multiplied, can be exponentiated and inverted,  and can  appear in algebraic equations and functions.  The resulting interplay of algebraic and geometric aspects produces a remarkable synergy that gives geometric algebra its power and charm. Each flat primitive -- point, line, and plane -- is represented by an element of the algebra. The magic lies in the geometric product defined on these elements.

 We'll define this product properly later on -- to start with we want to first give some impressions of what it's like and how it behaves.
%PGA is a relative newcomer to the applied geometric algebra scene: the idea appeared in the modern literature first in \cite{selig00} and was given its name in \cite{gunn2017a}. %and \cite{selig05} ); the present article is designed to introduce it to a wider audience.
%In comparison, the combination of algebra and geometry in analytic geometry is just a .
\subsection{Familiar components in a new setting}
To begin with it's important to note that many features of PGA are already familiar to many graphics programmers:.  
\begin{compactitem}
\item It is based on \emph{homogeneous coordinates}, widely used in computer graphics,
\item it contains within it classical \emph{vector algebra}, 
\item as well as the \emph{quaternion} and \emph{dual quaternion} algebras, increasingly popular tools for modeling kinematics and mechanics, and
\item the \emph{exterior algebra}, a powerful structure that models the flat subspaces of euclidean space. 
\end{compactitem}   
In the course of these notes we'll see that PGA in fact resembles a  organism in which each of these sub-algebras first finds its true place in the scheme of things. % 

\myboldhead{Other geometric algebra approaches}  Other geometric algebras have been proposed for doing euclidean geometry, notably \emph{conformal geometric algebra} (CGA).  Interested readers are referred to the comparison article \cite{gunn2017a}, which should shed light on the choice to base these notes on PGA.
%For readers familiar with \emph{conformal geometric algebra} (CGA), a detailed comparison of PGA and CGA for doing euclidean geometry may be found in \cite{gunn2017a} where the two algebras are assigned complementary positions in the GA ecosystem. %In particular it is shown there that for flat geometry (point, lines, planes), PGA exhibits distinct advantages.  %The focus of our comparison here is with VLAAG restricted to flat primitives (e. g., points, lines, and planes).

%\myboldhead{Geometry as numbers} T
Before turning to the formal details we present three examples of PGA at work, solving tasks in 3D euclidean geometry, to give a flavor of actual usage. Readers who prefer a more systematic introduction can  skip over to Sect. \ref{sec:roots}.

\subsection{\textbf{Example 1}: Working with lines and points in 3D}
\begin{quote}

\textbf{Task:} Given a point $\vec{P}$ and a non-incident line $\momo$ in $\Euc{3}$, find the unique line $\sigo$ passing through $\vec{P}$ which meets $\momo$ orthogonally.\footnote{In 3D PGA, lines are denoted with large Greek letters, points with large Latin letters, and planes with small Latin ones.}
\end{quote}
  \begin{figure}[hbt]
  \begin{centering}
\ifthenelse{\equal{\onecol}{true}}{ \def\xyz{.4}}{ \def\xyz{.48}}
 \def\zyx{.005}
{\setlength\fboxsep{0pt}\fbox{\includegraphics[width=\xyz\columnwidth]{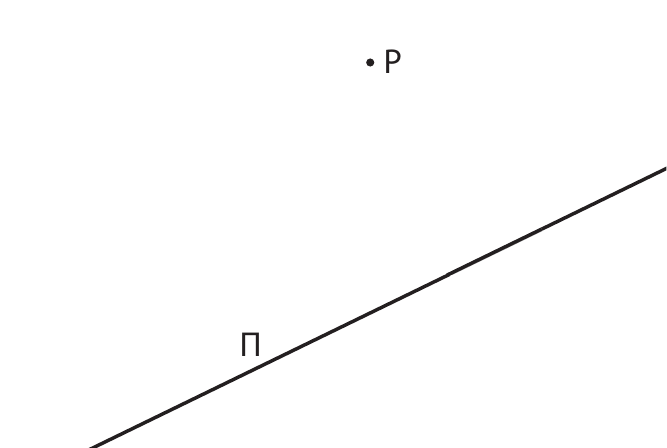}}}\hspace{\zyx\columnwidth}
{\setlength\fboxsep{0pt}\fbox{\includegraphics[width=\xyz\columnwidth]{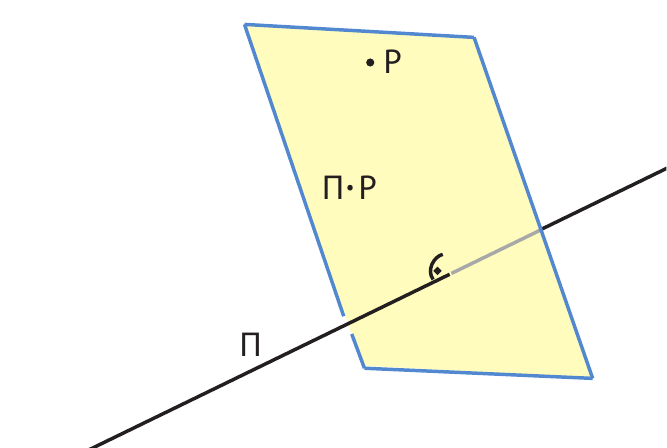}}}\hspace{\zyx\columnwidth} \\ \vspace{\zyx\columnwidth}
{\setlength\fboxsep{0pt}\fbox{\includegraphics[width=\xyz\columnwidth]{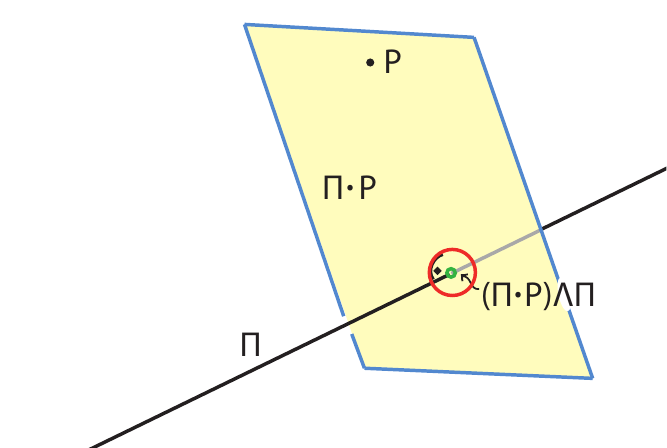}}}\hspace{\zyx\columnwidth}
{\setlength\fboxsep{0pt}\fbox{\includegraphics[width=\xyz\columnwidth]{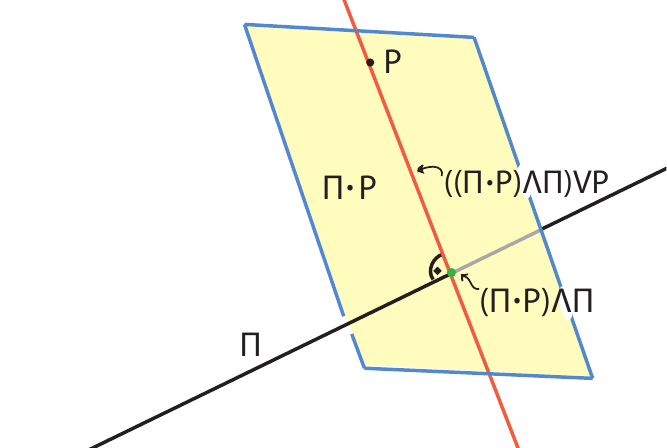}}}
\caption{Geometric construction in PGA.} 
\label{fig:perpPtLn}
\end{centering}
\end{figure}
\myvspace{-.1in}
In PGA,  geometric primitives such as points, lines, and planes, are represented by vectors of different \emph{grades},  as in an exterior algebra.  A plane is a 1-vector,  a line is a 2-vector, and a point is a 3-vector. (A scalar is a 0-vector; we'll meet 4-vectors in \Sec{sec:3dscrew}).  Hence the algebra is called a \emph{graded} algebra.

Each grade forms a vector space closed under addition and scalar multiplication. An element   of the GA is called a \emph{multivector} and is the sum of such $k$-vectors. The grade-$k$ part of a multivector $\vec{M}$ is written $\grade{\vec{M}}{k}$. The geometric relationships between primitives is expressed via  the \emph{geometric product} that we want to experience in this example.
The geometric product $\momo\vec{P} $, for example, of  a line $\momo$ (a 2-vector) and a point $\vec{P}$ (a 3-vector) consists two parts, a 1-vector and a 3-vector.\footnote{You are not expected at the point to understand \emph{why} this is so. If you know about quaternions, you've met similar behavior. Recall that the quaternion product of two imaginary quaternions $\vec{v_1}:=x_1\vec{i}+y_1\vec{j}+z_1\vec{k}$ and $\vec{v}_2:=x_2\vec{i}+y_2\vec{j}+z_2\vec{k}$ satisfies: $ \vec{v}_1\vec{v}_2 = -\vec{v}_1\cdot\vec{v}_2 + \vec{v}_1\times \vec{v}_2$. Hence, it is the sum of a scalar (the inner product) and a vector (the cross product). Something similar is going on here with the geometric product of a line and a point. We'll see why in \Sec{sec:gab} below. \Sec{sec:sphgeom} also sheds light on how the quaternions naturally occur within geometric algebra.}  We write this as:
$$ \momo \vec{P} = \grade{ \momo\vec{P}}{1} + \grade{\momo\vec{P} }{3}$$
\begin{compactenum}
\item $\grade{\momo\vec{P} }{1}$ is the plane perpendicular to $\momo$ passing through $\vec{P}$. As the lowest-grade part of the product, it is written as $\momo \cdot  \vec{P}$.  
\item $\grade{\momo\vec{P} }{3}$ is the normal direction to the plane spanned by $\momo$ and $\vec{P}$. We won't need it for this exercise.%It's written $\momo \times \vec{P} $ ($= \frac12(\momo\vec{P} -\vec{P}\momo)$, the \emph{anti-symmetric} part of the product.)
\end{compactenum}
% Rewriting with the new notation for each part: $\vec{P} \momo = \momo \cdot  \vec{P} +\vec{P} \times \momo$.
The sought-for line  $\sigo$ can then be constructed as shown in \Fig{fig:perpPtLn}: 
 \begin{compactenum}
 \item $\momo \cdot \vec{P}$ is the plane through $\vec{P}$ perpendicular to $\momo$,
 \item The point $(\momo \cdot \vec{P} )\wedge \momo)$ is the meet ($\wedge$) of $\momo \cdot \vec{P}$ with $\momo$,% and
 \item The line $\sigo := ((\momo \cdot \vec{P} )\wedge \momo)  \vee \vec{P}$ is the join ($\vee$) of this point with $\vec{P}$.
 \end{compactenum}
 The meet ($\wedge$) and joint ($\vee$) operators are part of the exterior algebra contained in the geometric algebra and are discussed in more detail below in \Sec{sec:gralg}.
%The reader is encouraged to compare this to the solution that his/her current toolkit would offer to this exercise. 

The next two examples show how euclidean motions (reflections, rotations, translations) are implemented in PGA.
 \vspace{.2in}
 
 \subsection{\textbf{Example 2}: A 3D Kaleidoscope}
\label{sec:3dkal}

   \begin{figure}[htb]
   \begin{centering}
\ifthenelse{\equal{\onecol}{true}}{ \def\xyz{.22}}{ \def\xyz{.34}}
\ifthenelse{\equal{\onecol}{true}}{ \def\xyw{.22}}{ \def\xyw{.68}}
 \def\zyx{.01}
%{\setlength\fboxsep{0pt}\fbox{\includegraphics[width=\xyz\columnwidth]{solution-01.pdf}}}\hspace{\zyx\columnwidth}
{\setlength\fboxsep{0pt}\fbox{\includegraphics[height=\xyz\columnwidth]{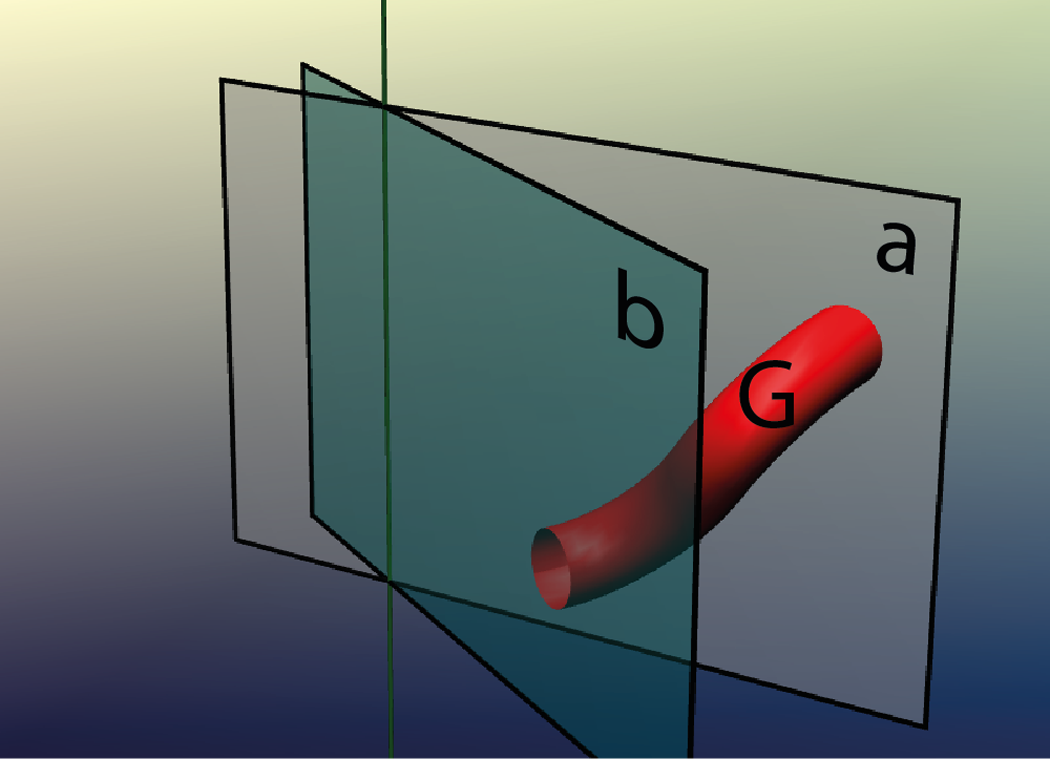}}}\hspace{\zyx\columnwidth}
{\setlength\fboxsep{0pt}\fbox{\includegraphics[height=\xyz\columnwidth]{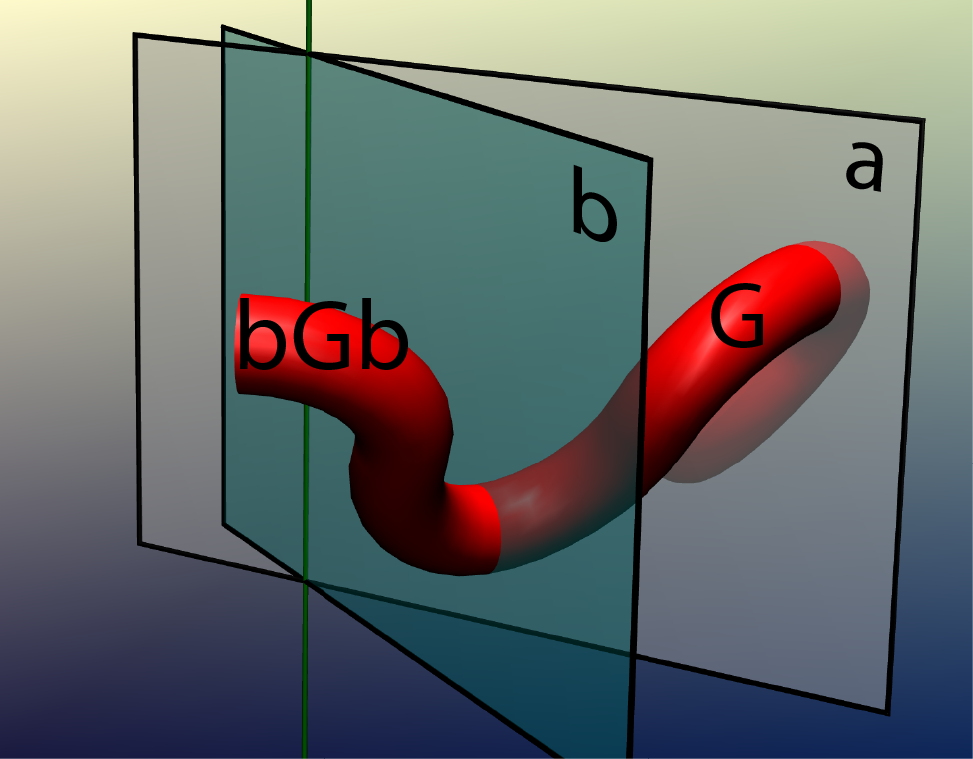}}}  \hspace{\zyx\columnwidth}%\vspace{\zyx\columnwidth}
{\setlength\fboxsep{0pt}\fbox{\includegraphics[height=\xyw\columnwidth]{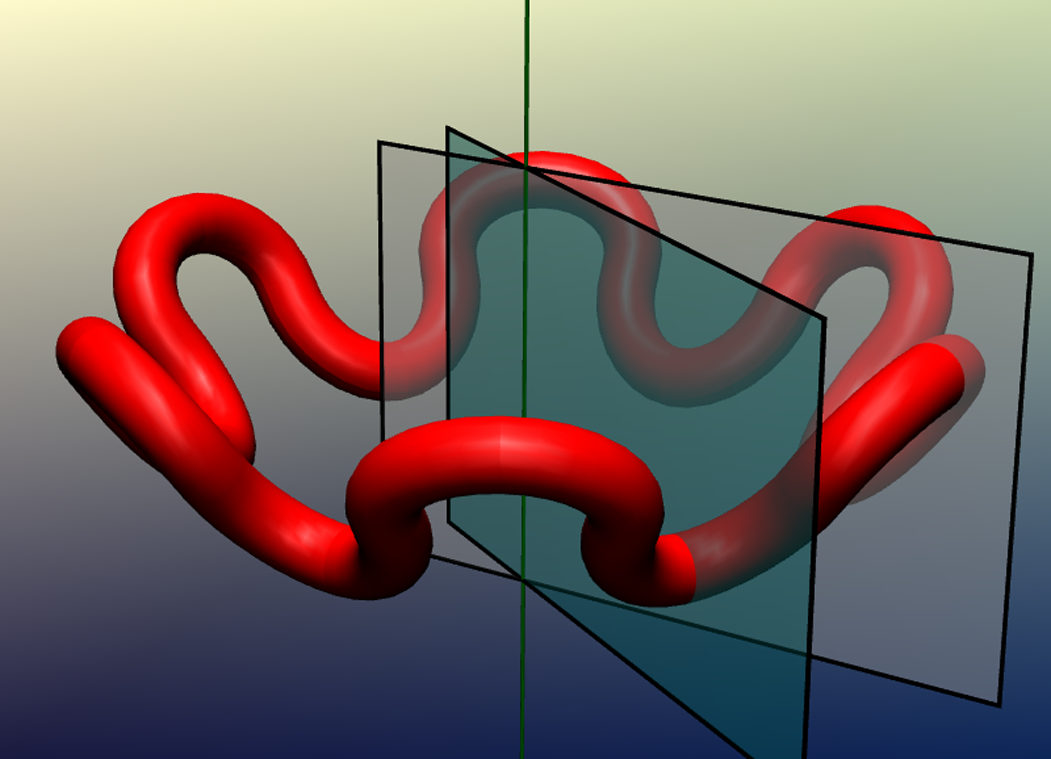}}}
\caption{Creating a 3D kaleidoscope in PGA using sandwich operators.} 
\label{fig:3dkal}
\end{centering}
\end{figure}
\myvspace{-.4in}

\begin{quote}
\textbf{Task:} A $k$-kaleidoscope is a pair of mirror planes $\vec{a}$ and $\vec{b}$  in $\Euc{3}$ that meet at an angle $\frac{\pi}{k}$. Given some geometry $\vec{G}$ generate the view of $\vec{G}$ seen in the kaleidoscope.
\end{quote}

In PGA, $\vec{a}$ is a 1-vector. We can and do normalize this 1-vector to satisfy $\vec{a}^2=1$, where $\vec{a}^2$ is  the \emph{geometric product} of $\vec{a}$ with itself. 
The geometric reflection in plane $\vec{a}$ is implemented in PGA by the \quot{sandwich} operator $\vec{a}\vec{G}\vec{a}$ (where $\vec{G}$ may be any $k$-vector -- plane, line or point). See  \Fig{fig:3dkal}. The left-most image shows the setup, where $\vec{G}$ is a red tube (modeled by some combination of 1-, 2-, and 3-vectors) stretching between the two planes.  The middle image shows the result of applying the \emph{sandwich} $\vec{b}\vec{G}\vec{b}$ to the geometry (behind plane $\vec{a}$ one can also see $\vec{a}\vec{G}\vec{a}$, unlabeled). The fact that $\vec{a}^2=1$ is consistent with the fact that repeating a reflection yields the identity.  The right image shows the result of applying all possible alternating products of the two reflections $\vec{a}$ and $\vec{b}$  to $\vec{G}$ (e. g., $\vec{b}\vec{a}\vec{G}\vec{a}\vec{b}$, etc.). Since the mirrors meet at the angle $\frac{\pi}{6}$, this process closes up in a ring consisting of 12 copies of $\vec{G}$. (To be precise, $(\vec{a}\vec{b})^6 = (\vec{b}\vec{a})^6 = 1$). 

Readers familiar with quaternions may recognize a similarity to the quaternion sandwich operators that implement 3D rotations -- but here the basic sandwiches implement reflections. %, while even products of these reflections represent rotations  (since the product of two reflections, e. g.,  $\vec{b}\vec{a}\vec{G}\vec{a}\vec{b}$, is a rotation). 
The next example derives sandwich operators for rotations without using reflections.
\myvspace{-.1in}
\subsection{\textbf{Example 3}: A continuous 3D screw motion}
\label{sec:3dscrew}
%\begin{quote}
%Generate a continuous screw motion along a line $\velo$ with pitch $\alpha$.
%\end{quote}
\begin{quote}
\textbf{Task:}  Represent a continuous screw motion in 3D.
\end{quote}
The general orientation-preserving isometry of $\Euc{3}$ is a \emph{screw motion}, that rotates around a unique fixed line (the  \emph{axis}) while translating parallel to it.  The ratio of the translation distance to the angle of rotation (in radians) is called the \emph{pitch} of the screw motion.  A rotation has pitch 0, and translation has pitch \quot{$\infty$}.

  \begin{figure}[htb]
 \def\xyz{.34}
 \def\xyw{.69}
\ifthenelse{\equal{\onecol}{true}}{ \def\xyz{.23}}{ \def\xyz{.34}}
\ifthenelse{\equal{\onecol}{true}}{ \def\xyw{.23}}{ \def\xyw{.69}}
 \def\zyx{.01}
 \begin{centering}
%{\setlength\fboxsep{0pt}\fbox{\includegraphics[width=\xyz\columnwidth]{solution-01.pdf}}}\hspace{\zyx\columnwidth}
{\setlength\fboxsep{0pt}\fbox{\includegraphics[height=\xyz\columnwidth]{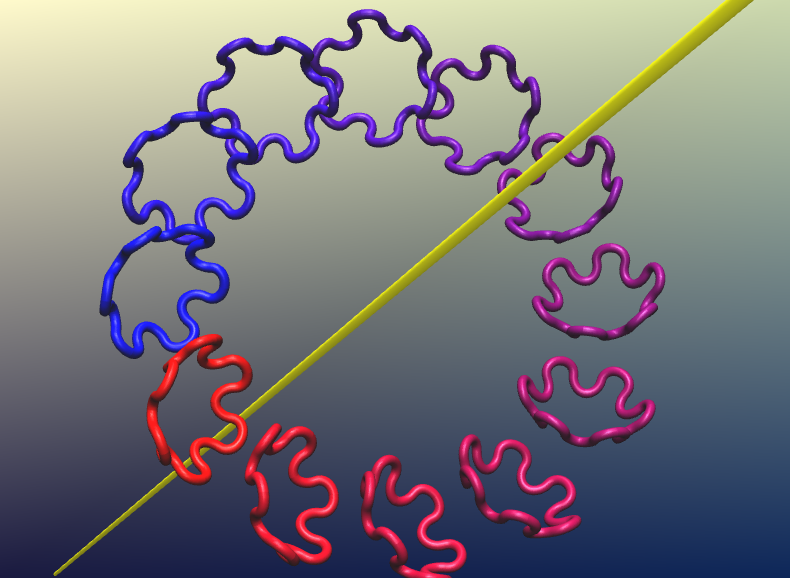}}}\hspace{\zyx\columnwidth}
{\setlength\fboxsep{0pt}\fbox{\includegraphics[height=\xyz\columnwidth]{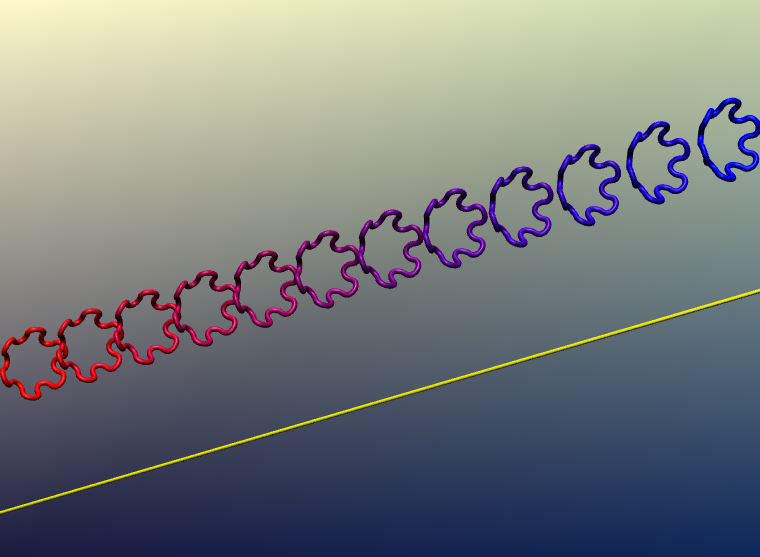}}} \hspace{\zyx\columnwidth} %\\ \vspace{\zyx\columnwidth}
{\setlength\fboxsep{0pt}\fbox{\includegraphics[height=\xyw\columnwidth]{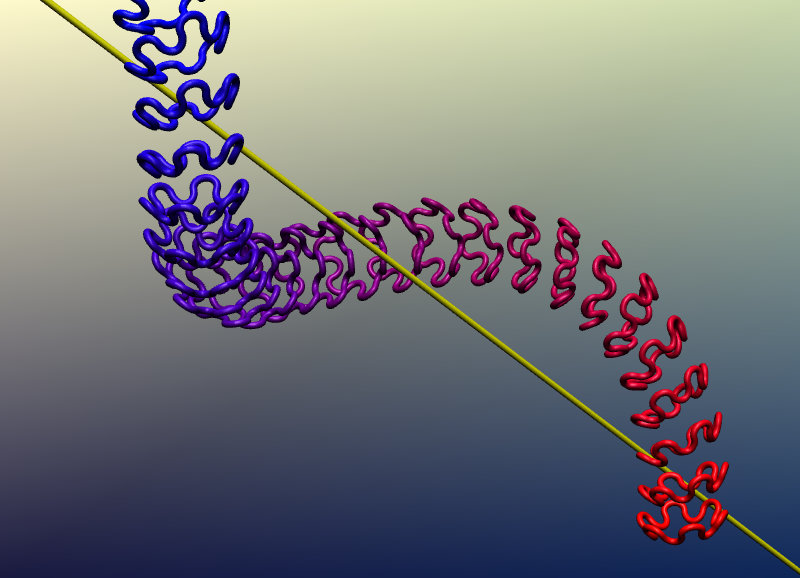}}}
\caption{Continuous rotation, translation, and screw motion in PGA by exponentiating a bivector.} 
\label{fig:screwMotion}
\end{centering}
\end{figure}

%\subsubsection{Rotation} 
The previous example already contains rotations:  a reflection in a plane $\vec{a}$ followed by a reflection in a second plane $\vec{b}$ (i. e., $\vec{b}(\vec{a}\vec{G}\vec{a})\vec{b}$) is a rotation around their common line by twice the angle between them, in this case $\frac{\pi}{3}$. Here we use a different approach to obtain a desired rotation directly from its axis of rotation.  A line in $\Euc{3}$, passing through the point $\vec{P}$ with direction vector $\vec{V}$, is given by the join operation $\velo := \vec{P} \vee \vec{V}$ (yellow line in Fig. \ref{fig:screwMotion}). We can and do normalize $\velo$  to satisfy $\velo^2=-1$. (Where $\velo^2$ means multiply $\velo$ by itself using the geometric product.) To obtain the  \emph{rotation} around $\velo$ of angle $\alpha$  define the \emph{\MOTOR} $e^{t \velo}$.  The exponential function is  evaluated using the geometric product in the formal power series of $e(x)$; it behaves like the imaginary exponential $e^{ti}$ since $\velo^2=-1$. The \emph{sandwich operator} $e^{t \velo}\vec{G} e^{-t \velo}$  implements the continuous rotation around $\velo$ applied to $\vec{G}$, parametrized by $t$.  At $t=0$ it is the identity; and at $t=\frac{\alpha}{2}$ it represents the rotation of angle $\alpha$ around $\velo$. See the left image above, which shows the result for a sequence of $t$-values between $0$ and $\pi$. Readers familiar with the quaternion representation of rotations should recognize the similarity of these formulas. This isn't accidental -- see \Sec{sec:qadq} below.

%\subsubsection{Translation}
To obtain instead a translation in the direction of $\velo$, we used a different line, obtained by applying the \emph{polarity} operator of PGA to $\velo$ to produce $\velo^\perp$. $\velo^\perp$ is the \emph{orthogonal complement} of $\velo$, an \emph{ideal} line, or so-called \quot{line at infinity}. It  consists of all directions perpendicular to $\velo$.  If $\velo$ is thought of as a vertical axis, then $\velo^\perp$ is the horizon line. 
The orthogonal complement is obtained in PGA by multiplying by a special 4-vector, the unit \emph{pseudoscalar} $\eye$: $\velo^\perp := \velo \eye$.\footnote{The pseudoscalar is one of the most powerful but mysterious features of geometric algebra.}  A continuous translation in the direction of $\velo$ is then given by a sandwich with the \emph{translator} $e^{t \velo^\perp} $.  See the middle image above. 

%\subsubsection{Screw motion}
Let the pitch of the screw motion be $p\in \mathbf{R}$.  Then the desired screw motion is given by a sandwich operator with  the \emph{\MOTOR}  $e^{t(\velo+p\velo^\perp)}$.  This motion can be factored as the product of a pure rotation and a pure translation in either order: $$e^{t(\velo+p\velo^\perp)} = e^{t\velo}e^{tp\velo^\perp} = e^{tp\velo^\perp}e^{t\velo}$$. %The \MOTOR is also obtainable as the product of the rotator and translator -- in either order.  
See image on the right above.

We hope these examples have whetted your appetite to explore further. We now turn to a quick exposition of the history of PGA followed by a modern formulation of its mathematical foundations.
\myvspace{-.1in}
%\section{Roots of PGA}
\section{Mathematical foundations}
\label{sec:roots}

\subsection{Historical overview}
Both the standard approach to doing euclidean geometry and the geometric algebra approach described here can be traced back to 16th century France. 
The analytic geometry of Ren\'{e} Descartes (1596-1650) leads to the standard toolkit used today based on Cartesian coordinates and analytic geometry.   His contemporary and friend Girard Desargues (1591-1661), an architect, confronted with the riddles of the newly-discovered perspective painting, invented \emph{projective geometry}, containing additional, so-called \emph{ideal}, points where parallel lines meet.  Projective geometry is characterized by a deep symmetry called \emph{duality}, that asserts that every statement in projective geometry has a dual partner statement, in which, for example, the roles of point and plane, and of join and intersect, are exchanged. More importantly, the truth content of a statement is preserved under duality. We will see below that duality plays an important role in PGA. 

Mathematicians in the 19th century (Cayley and Klein) showed how, using an algebraic structure called a \emph{quadratic form},  the euclidean metric could be built back into projective space. (The same technique also worked to model the newly discovered non-euclidean metrics of hyperbolic and elliptic geometry in projective space.) This \emph{Cayley-Klein model} of metric geometry forms an essential foundation  of PGA. While these developments were underway in geometry, William Hamilton  and Herman Grassmann  discovered surprising new algebraic structures for geometry. All these dramatic developments flowed together into William Clifford's invention of geometric algebra in 1878 (\cite{clifford78}). We now turn to studying from a modern perspective the ingredients of geometric algebra.   %While Cartesian geometry does introduce algebra into geometry, it is limited by being built directly on the euclidean geometry inherited from the Greeks.  PGA is geometric algebra built on top of projective geometry and then specialized for a specific metric, for example, the euclidean metric that we deal with here. 
%We focus to start with on the euclidean plane, where we can get to know all the interesting behavior before moving on to $n=3$, where the real interest lies, but also much more complex behavior.    We next turn to how to represent the subspaces of projective space in an algebraic structure.

%In the plane, euclidean points are given coordinates $(x,y,1)$ while ideal points (sometimes called \quot{points at infinity}) have the form $(x,y,0)$ and together form a special line, the ideal line (or \quot{line at infinity}).  The same principle applies in any dimension.
%We describe it briefly here.
%\subsection{Projective model of euclidean space}
%One uses homogeneous coordinates for projective space, that is, one has one more coordinate than the dimension of the underlying space; non-zero multiples of the same $(n+1)$-tuple correspond to the same point. 
%A point in euclidean space with coordinates $(x,y,z)$ is given normalized coordinates $(x,y,z,1)$.  An ideal point in the direction $(x,y,z)$ is given coordinates $(x,y,z,0)$. The set of ideal points forms a plane, the ideal plane. Together, the euclidean points and the ideal points fill out projective space. 
\begin{figure}
 \def\xyz{.3}
 \def\zyx{.02}
%{\setlength\fboxsep{0pt}\fbox{\includegraphics[width=\xyz\columnwidth]{solution-01.pdf}}}\hspace{\zyx\columnwidth}
{\setlength\fboxsep{0pt}{\includegraphics[height=\xyz\columnwidth]{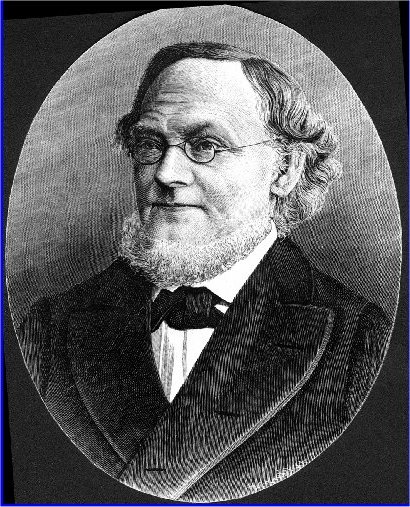}}}\hspace{\zyx\columnwidth}
{\setlength\fboxsep{0pt}{\includegraphics[height=\xyz\columnwidth]{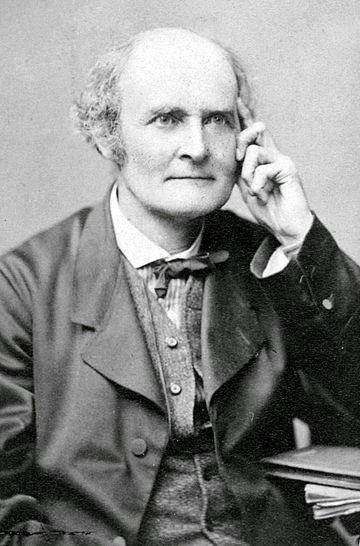}}}\hspace{\zyx\columnwidth}
{\setlength\fboxsep{0pt}{\includegraphics[height=\xyz\columnwidth]{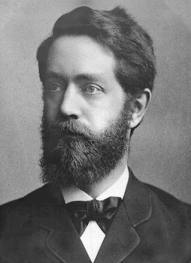}}}\hspace{\zyx\columnwidth}
{\setlength\fboxsep{0pt}{\includegraphics[height=\xyz\columnwidth]{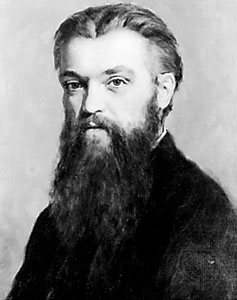}}}
\caption{Important figures in the development of PGA (l. to r.): Hermann Grassmann (1809-1877), Arthur Cayley (1821-1895), Felix Klein (1849-1925), William Clifford (1845-1879).} 
\label{fig:inventors}
\end{figure}

\label{sec:mathfound}
%Here we review the mathematical foundations involved in the developments described above.

\subsection{Vector spaces} We assume that the reader is familiar with the concept of a \emph{real vector space of dimension $n$}, where $n$ is the cardinality of a maximal linearly independent set of elements, called \emph{vectors}. Vectors are often thought of as $n$-tuples of numbers: these arise through the choice of a basis for the vector space, and represent the coordinates of that vector with respect to the basis.  A vector space is closed under addition and scalar multiplication.  For each vector space $\vec V$ there exists an isomorphic \emph{dual} vector space $\vec{V}^*$, consisting of dual vectors, or \emph{co-vectors}. A co-vector $\vec \theta$ is a \emph{linear functional} that can be \emph{evaluated} at a vector $\mathbf{v}$ to produce a real number: $\langle \vec\theta, \vec{v} \rangle \in \mathbb{R}$. This evaluation map is bilinear. It is \textbf{not} an inner product, that is defined on pairs of vectors. See the next section below.

\myboldhead{Example}. When $n=3$, $\vec{v}$ can be interpreted as a line through the origin and   $\vec\theta$, as a plane through the origin, and $\langle \vec\theta, \vec{v} \rangle \in \mathbb{R} = 0 \leftrightarrow \vec{v}$ lies in the plane $\vec\theta$.
\subsection{Normed vector spaces}
A real vector space $\vec V$ of dimension $n$ has no  
way to measure angles or distances between elements. 
For that, introduce a \emph{symmetric bilinear form} $B: \vec{V} \times \vec{V} \rightarrow \mathbb{R}$. $B$ is a map satisfying
\begin{compactenum}
\item $B(\alpha\vec{u}_1+\beta \vec{u}_2,\vec{v}) = \alpha B(\vec{u}_1,\vec{v}) + \beta B(\vec{u}_2,\vec{v})$ (bilinearity), and
\item $B(\vec{u},\vec{v})=B(\vec{v},\vec{u})$ (symmetry).
\end{compactenum}
A symmetric bilinear form $B$ can be rewritten as an \emph{inner product} on vectors: $\vec{u} \cdot \vec{v} := B(\vec{u},\vec{v})$ and used to define a \emph{norm}, or length-function, on vectors: $\| \vec{u} \| := \sqrt{|{\vec{u} \cdot \vec{u}}|}$. $\R{n}$ is a normed vector space. The next section classifies symmetric bilinear forms.

\subsection{Sylvester signature theorem}
Symmetric bilinear forms of dimension $n$ can be completely characterized by three positive integers $(p,m,z)$ satisfying $p+m+z=n$. Sylvester's Theorem asserts that for any such $B$ there is a unique choice of $(p,m,z)$ and a basis $\{\vec{e}_i\}$ for $\vec{V}$ such that 
\begin{compactenum}
\item $\vec{e}_i \cdot \vec{e}_j = 0$ for $i \neq j$ (orthogonal basis), and
\item \[\vec{e}_i \cdot \vec{e}_i = 
\begin{cases} 
1  &~\text{for}~ 1 \leq i \leq p \\
-1 &~\text{for}~ p <  i \leq p+m ~~~~~~~ \text{(normalized basis)}  \\
 0 &~\text{for}~ p+m < i \leq n  
\end{cases}\]
\end{compactenum}

\myboldhead{Example} Taking $n=3$ and $(p,m,z) = (3,0,0)$ we arrive at the familiar euclidean vector space $\R{3}$ with norm $\|(x,y,z)\| = \sqrt{x^2+y^2+z^2}$ where $(x,y,z)$ are coordinates in an orthonormal basis.

\subsection{Euclidean space $\Euc{n}$}
We can transform the vector space $\R{n}$ into the metric space $\Euc{n}$ by identifying each  vector of the former (also the zero vector $O$) with a point of the latter. Then define a distance function on the resulting points with $d(\vec{P}, \vec{Q}) := \| \vec{P} - \vec{Q} \|$.  This distance function produces a differentiable manifold $\Euc{n}$ whose tangent space at every point is $\R{n}$. 

% \emph{quadratic form} $\vec Q$ on $\vec V$  . 
\myboldhead{Terminology alert} When we say \emph{doing euclidean geometry} we are referring to the geometry of \emph{euclidean space }$\Euc{n}$, not the \emph{euclidean vector space} $\R{n}$. The elements of $\Euc{n}$ are points, those of $\R{n}$ are vectors; the motions of $\Euc{n}$ include translations \textbf{and} rotations, those of $\R{n}$ are rotations preserving the origin $O$. $\Euc{n}$ is intrinsically more complex than $\R{n}$: the tangent space at each point is $\R{n}$.  See \cite{gunn2017a}, \S 4, for a deeper analysis of this issue. 
We will see that euclidean PGA includes both $\Euc{n}$ and $\R{n}$ in an organic whole.

\subsection{The tensor algebra of a vector space} %of subspaces of $\RP{n}$}
\label{sec:tenalg}
Vector spaces have linear subspaces. The subspace structure is mirrored in the algebraic structure of the 
\emph{exterior algebra} defined over the vector space. To define the exterior algebra cleanly, we need first to introduce the \emph{tensor algebra} $T(\vec{V})$ over $\vec{V}$. This algebra is generated by multiplying arbitrary sequences of vectors together to generate a graded algebra. This product is called the \emph{tensor product} and is written $\otimes$. It is bilinear. The tensor product of $k$ vectors is called a $k$-vector. The $k$-vectors form a vector space $T^k$. $T^0$ is the underlying field $\mathbb{R}$. $T(\vec{V})$ can be written as the direct sum of these vector spaces: $$T = \bigoplus_{i=0}^\infty T^k$$ Obviously this is a very big and somewhat unwieldy structure, but necessary for a clean definition of important algebras below.

\myboldhead{Example $n=2$}
The tensor algebra of a 2-dimensional vector space with basis $\{\vec{u}, \vec{v}\}$ has the basis:
\begin{compactitem}
\item $T^0$: $\{\vec{1}\}$
\item $T^1$: $\{\vec{u}, \vec{v}\}$
\item $T^2$:$\{\vec{u}\otimes\vec{u}, \vec{u}\otimes\vec{v}, \vec{v}\otimes\vec{u}, \vec{v}\otimes\vec{v}\}$
\item $T^3$:$\{\vec{u}\otimes\vec{u}\otimes\vec{u}, \vec{u}\otimes\vec{u}\otimes\vec{v}, \vec{u}\otimes\vec{v}\otimes\vec{u}, \vec{u}\otimes\vec{v}\otimes\vec{v}, \vec{v}\otimes\vec{u}\otimes\vec{u}, \vec{v}\otimes\vec{u}\otimes\vec{v}, \vec{v}\otimes\vec{v}\otimes\vec{u}, \vec{v}\otimes\vec{v}\otimes\vec{v}\}$
\item etc.
\end{compactitem}

\subsection{Exterior algebra of a vector space} 
\label{sec:gralg}
The exterior algebra is obtained from the tensor algebra by declaring elements of the form $\vec{v}\otimes\vec{v}$ (where $\vec{u}$ and $\vec{v}$ are 1-vectors), to be equivalent to 0, that is, squares of 1-vectors vanish. By bilinearity, this implies 
$$(\vec{u+v})\otimes(\vec{u+v}) = \vec{u}\otimes\vec{u}+\vec{v}\otimes\vec{v} + \vec{u}\otimes\vec{v} +\vec{v}\otimes\vec{u} \cong 0$$ 
implying $\vec{u}\otimes\vec{v} \cong -\vec{v}\otimes\vec{u}$ since $\vec{u}\otimes\vec{u}\cong 0$ and $\vec{v}\otimes\vec{v} \cong 0$. Thus the quotient product is anti-symmetric on 1-vectors.  The resulting quotient algebra is called the \emph{exterior} algebra and its product is the \emph{exterior} or \emph{wedge} product, written as $\vec{X}\wedge \vec{Y}$. The product is associative, anti-symmetric on 1-vectors and distributes over addition. In general, the wedge of a $k$-vector $\vec X$ and an $m$-vector  $\vec Y$ will vanish $\iff \vec X$ and $\vec Y$ are linearly-dependent subspaces, otherwise it is the $(k+m)$-vector representing the subspace span of $\vec{X}$ and $\vec{Y}$.

The exterior algebra $G(\vec{V})$ mirrors the subspace structure of $\vec{V}$. Two $k$-vectors $\vec{v}$ and $\alpha\vec{v}$ that are non-zero multiples of each other represent the same subspace but have different \emph{weights}, or intensities.  $G(\vec{V})$ is finite-dimensional since any $m$-vector in the tensor algebra with $m>n$ vanishes in the exterior algebra since any product of $n+1$ basis 1-vectors will have a repeated factor, and this is equivalent to 0. It can be written as a direct sum of its non-vanishing grades: $$G(\vec{V}) = \bigoplus_{i=0}^n {{\bigwedge}^i}$$ The dimension of each grade is given by $dim\left(\bigwedge^k\right) = \binom{n}{k}$, so the total dimension of the algebra is $\Sigma_{i=0}^n \binom{n}{k} = 2^n$. 

\myboldhead{Example $n=2$}
The exterior algebra of a 2-dimensional vector space with basis $\{\vec{u}, \vec{v}\}$ is a 4-dimensional graded algebra:
\begin{compactitem}
\item $\bigwedge^0$: $\{\vec{1}\}$
\item $\bigwedge^1$: $\{\vec{u}, \vec{v}\}$
\item $\bigwedge^2$: $\{ \vec{u}\wedge\vec{v}\}$
\end{compactitem}

 Exterior algebras were, like so many other results in this field, discovered by Hermann Grassmann (\cite{grassmann44}) and are sometimes called \emph{Grassmann} algebras. %An exterior algebra (as used here) mirrors the subspace structure of projective space $\RP{n}$. 
 
 \subsection{The dual exterior algebra}
 Important for PGA: the dual vector space $\vec{V}^*$ generates its own exterior algebra $G(\mathbf{V^*}) = G^*(\vec{V})$. The standard exterior algebra represents the subspace structure based on subspace \textbf{join}, where the 1-vectors are vectors (or lines through the origin). The dual exterior algebra represents  the subspace structure \quot{turned on its head}: the 1-vectors represent hyperplanes through the origin and the wedge operation is subspace \textbf{meet}. 
 %You can build up the subspaces of  projective $n$-space  by \emph{joining points}; or  by \emph{intersecting hyperplanes}.   
 The principle of duality ensures that these two approaches are completely equivalent and neither \emph{a priori} is to be preferred.  Each construction produces a separate exterior algebra. The dual exterior algebra is important for PGA.
 
 The next step on our way to PGA is \emph{projective geometry}.
 
 \subsection{Projective space of a vector space}
 An $n$-dimensional real vector space $\vec{V}$ can be projectivized to produce $(n-1)$ dimensional real projective space $\RP{n-1}$. This is a quotient space constuction as in the case of the exterior algebra. Here the equivalence relation on vectors of  $\vec{V}$ is $$\vec{u} \cong \vec{v} \leftrightarrow \exists \lambda \neq 0 \in \mathbb{R} ~~\text{such that}~~ \vec{u} = \lambda \vec{v}$$ One sometimes says, the \textbf{points} of $\RP{n}$ are the \textbf{lines} through the origin of $\vec{V}$. 
 
 \myboldhead{Example} $\RP{2}$ is called the projective plane. We consider it as arising from projectivizing $\R{3}$ (although the norm on $\R{3}$ plays \textbf{no} role in the construction).  Take $\R{3}$ with standard basis $\{\e{0},\e{1},\e{2}$ vectors pointing in the $x$-, $y$-, and $z$-directions, resp.) Each point $\vec{P}$ of the $z=1$ plane represents the line through the origin obtained by joining $\vec{P}$ to the origin. Hence $\vec{P}$ corresponds to a point of $\RP{2}$. The only  points of $\RP{2}$ not accounted for in this way arise from lines through the origin lying in the $z=0$ plane, since such lines don't intersect the $z=1$ plane. However in projective geometry they correspond to points; it is useful to speak of \emph{ideal} points of $\RP{2}$ where these lines intersect the plane $z=1$. The intersection of parallel planes yields in the same way an \emph{ideal} line. The interplay of euclidean and ideal elements in PGA is essential to its effectiveness. 
 
 \newcommand{\dc}{\textcolor{red}}
 
 \begin{figure}[!htb]
   \centering
 \def\xyz{.48}
 \def\zyx{.03}
%{\setlength\fboxsep{0pt}\fbox{\includegraphics[width=\xyz\columnwidth]{solution-01.pdf}}}\hspace{\zyx\columnwidth}
{\setlength\fboxsep{1pt}{\includegraphics[height=\xyz\columnwidth]{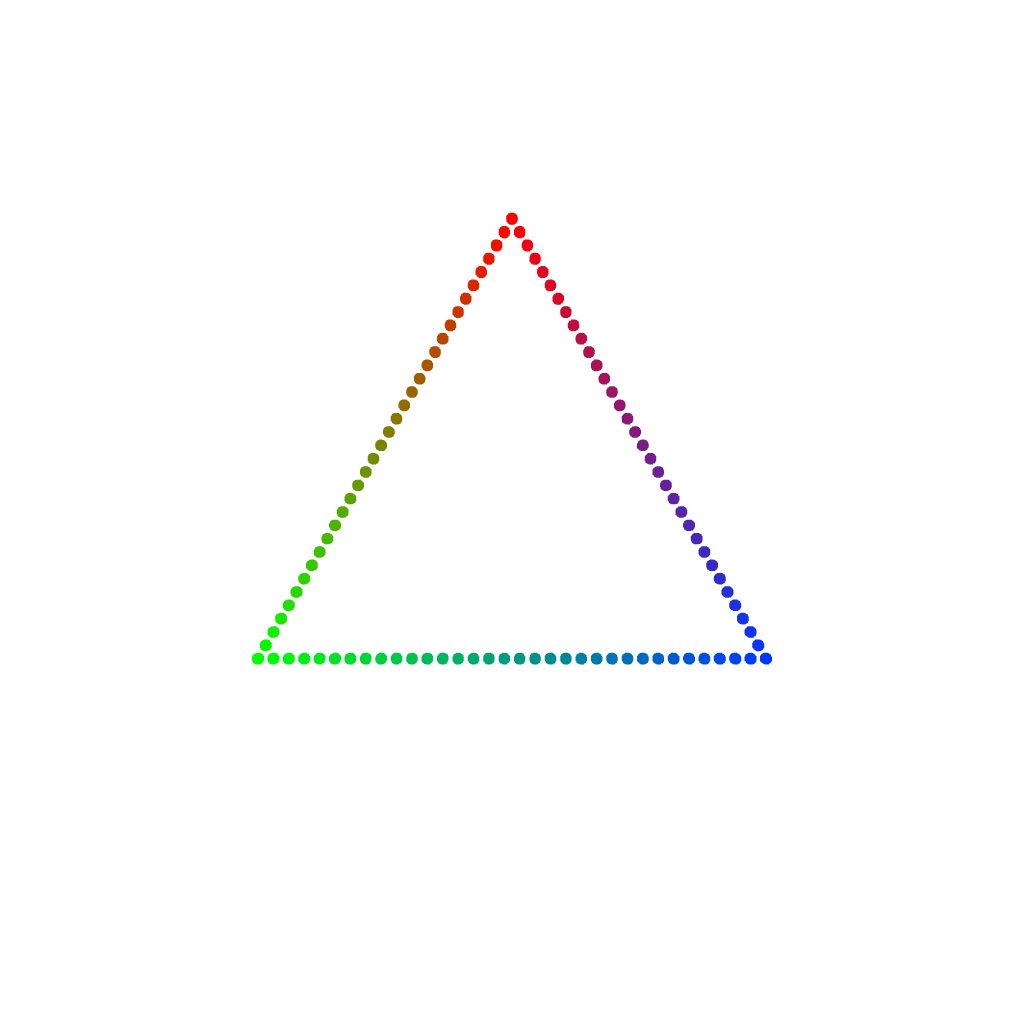}}}\hspace{\zyx\columnwidth}
{\setlength\fboxsep{1pt}{\includegraphics[height=\xyz\columnwidth]{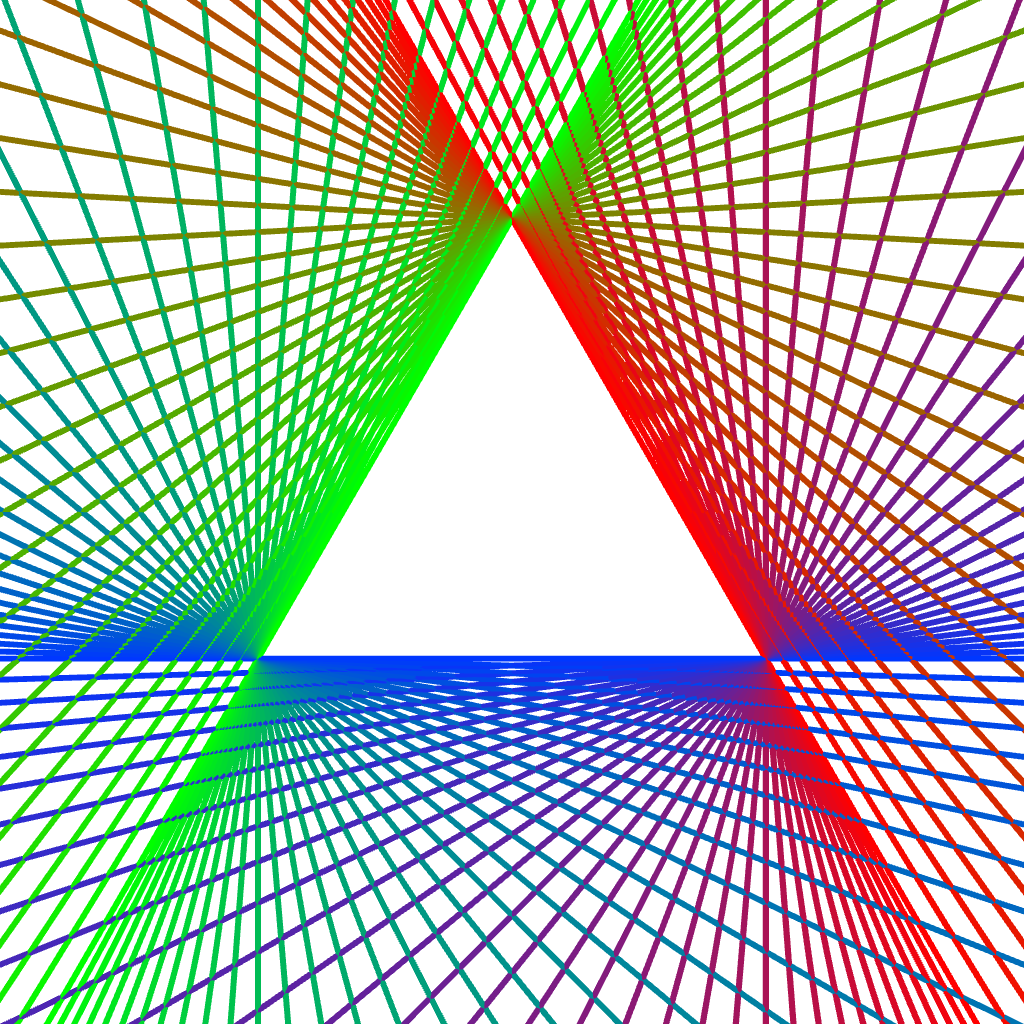}}}
\caption{ Traversing the boundary of a triangle (\emph{left}) and a trilateral (\emph{right}).}
\label{fig:duality}
\end{figure}

\myboldhead{Example of duality in $\RP{2}$}
 Because duality plays an essential role in PGA, we include an example here to show how it works. Following the pattern established in the $19^{th}$ century literature we use a two-column format to present, on the left, a geometric configuration in the projective plane and, on the right, its dual configuration. Dualized terms have been highlighted in color. \Fig{fig:duality} illustrates this example. 

\begin{multicols}{2}
A \dc{triangle} is determined by three \dc{points}, called its \dc{vertices}. The pair-wise \dc{joining} \dc{lines} of the \dc{vertices} are the \dc{sides} of the \dc{triangle}. To traverse the boundary of the \dc{triangle}, \dc{move} a \dc{point} from one \dc{vertex} to the next \dc{vertex} along their common \dc{side}, then \dc{take a turn} and continue \dc{moving along} the next \dc{side}.  Continue until arriving back at the original \dc{vertex}. 

A \dc{trilateral} is determined by three \dc{lines}, called its \dc{sides}. The pair-wise \dc{intersection} \dc{points} of the \dc{sides} are the \dc{vertices} of the \dc{trilateral}. To traverse the boundary of the \dc{trilateral}, \dc{rotate} a \dc{line} from one \dc{side} to the next around their common \dc{vertex}, then shift over and continue \dc{rotating round} the next \dc{vertex}. Continue until arriving back at the original \dc{side}. 
%\end{minipage}
\end{multicols}

Perhaps you can experience that the left-hand example is somehow more familiar than the right-hand side. After all, we learn about triangles in school, not trilaterals. This seems to be related to the fact that we think of points as being the basic elements of geometry (and reality) out of which other elements (lines, planes) are built. We'll see below in \Sec{sec:ipeg} however that PGA in important respects challenges us to think in the right-hand mode.

\myboldquestion{Why projectivize} Working in projective space guarantees that the meet of parallel lines and planes, as well as the join of euclidean and ideal elements, are handled seamlessly, without \quot{special casing} -- one of the features on our initial wish-list. Furthermore we'll see that only in projective space can we represent translations.

 \subsection{Projective exterior algebras}
 \label{sec:pea}
 The same construction applied to create $\RP{n}$ from $\vec{V}$ can be applied to the Grassmann algebras $G(\vec{V})$ and $G^*(\vec{V})$ to obtain projective exterior algebras. We denote these projectivized versions as $\pgvshort$ and $\pdgvshort$.  Here we use an $(n+1)$-dimensional $\vec{V}$ so that we obtain $\RP{n}$ by projectivizing. The resulting exterior algebras mirror the subspace structure of $\RP{n}$: 1-vectors in $G$ represent points in $\pgshort$, 2-vectors represent lines, etc., and $\wedge$ is projective join. In the dual algebra $G^*$, 1-vectors are hyperplanes ($(n-1)$-dimensional subspaces), and $n$-vectors represent points, while $\wedge$ is the meet operator.  More generally: in a {standard} projective exterior algebra $\pgshort$, the elements of grade $k$ for $k = 1, 2, ... n$,  represent the subspaces of dimension $k-1$.  For example, for $n=2$, the 1-vectors are points, and the 2-vectors are lines.    The graded algebra also has  elements of grade 0, the  scalars (the real numbers $\mathbb{R}$); and elements of grade $(n+1)$ (the highest non-zero grade), the \emph{pseudoscalars}.
 
  %which %represent the space itself and 
% will be described in more detail below in \Sec{sec:eucplane}.
 %All elements of the exterior algebra have projective coordinates; so each has a non-zero weight which can be freely chosen and, as we will see, often expresses important geometric information. The space of $k$-vectors is a vector space written $\bigwedge^{k}$.
   %The basis pseudoscalar is written $\eye$; any pseudoscalar can then be written as $\alpha \eye$ where $\alpha \in \mathbf{R}$.

  \begin{figure}[!hbt]
   \centering
 \def\xyz{.45}
 \def\zyx{.05}
%{\setlength\fboxsep{0pt}\fbox{\includegraphics[width=\xyz\columnwidth]{solution-01.pdf}}}\hspace{\zyx\columnwidth}
{\setlength\fboxsep{0pt}{\includegraphics[height=\xyz\columnwidth]{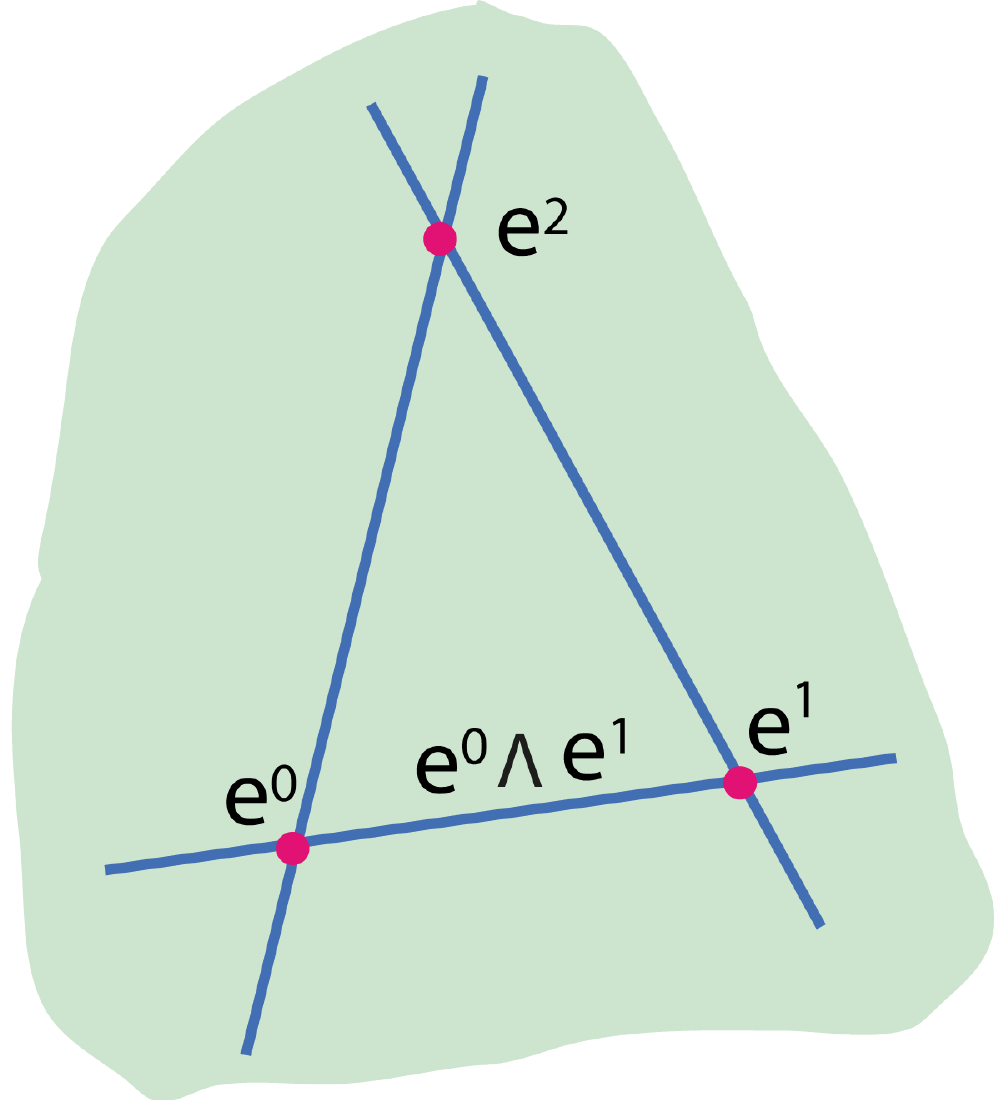}}}\hspace{\zyx\columnwidth}
{\setlength\fboxsep{0pt}{\includegraphics[height=\xyz\columnwidth]{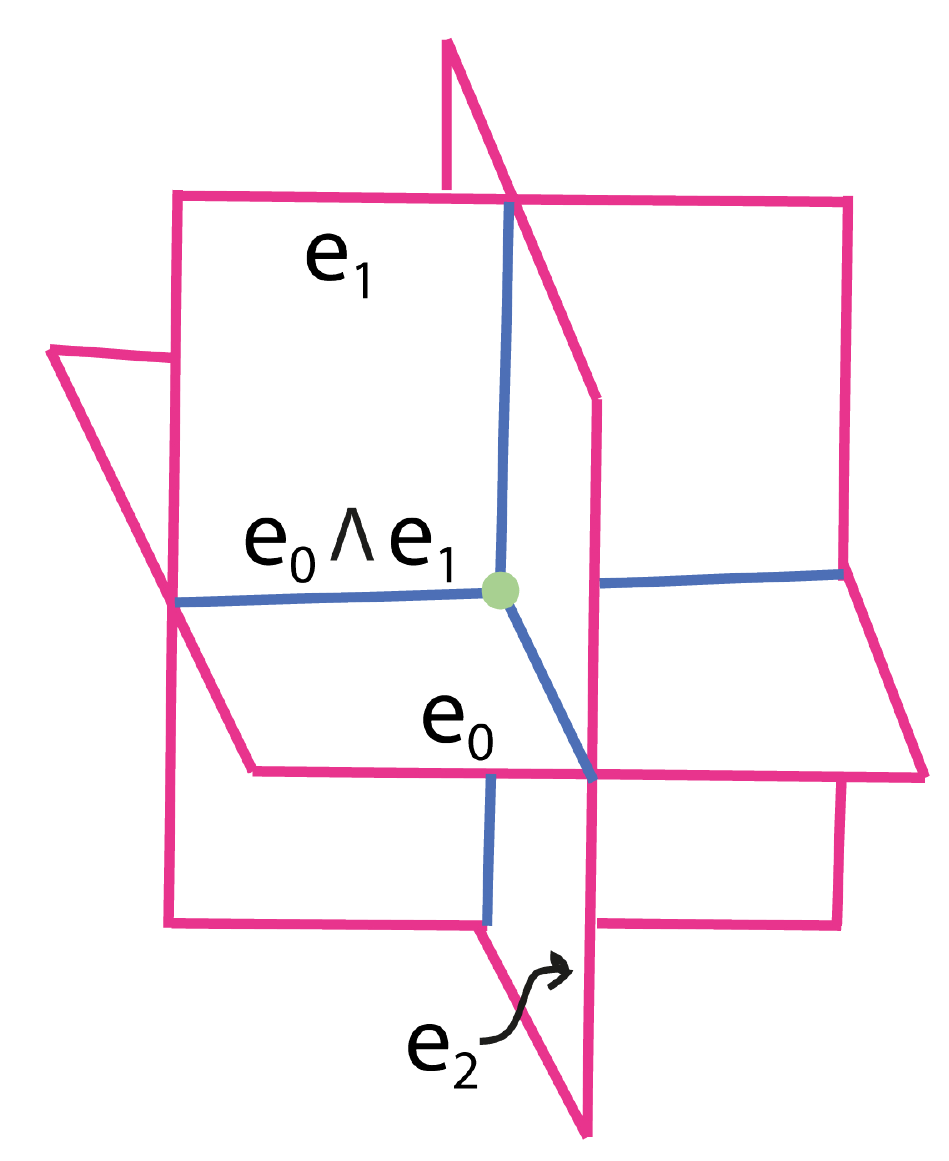}}}
\caption{\emph{Left:} The plane $\vec{e}^0\wedge\vec{e}^1\wedge\vec{e}^2$ (green) created by joining 3 points in $\pgshort$, the standard exterior algebra (written with raised indices).  \emph{Right:} The meeting point $\vec{e}_0\wedge\vec{e}_1\wedge\vec{e}_2$ (green) of three planes in $\pdgshort$, the dual exterior algebra (written with lowered indices).} 
\label{fig:dualAufbau}
\end{figure}
\myvspace{-.1in}
%  The \emph{outer}, or \emph{wedge}, {product}, in an exterior algebra is anti-symmetric, associative, and bilinear.  The outer product of a linearly independent $k$- and $m$-vector is the $(k+m-1)$-dimensional subspace that they span, otherwise it is 0.   In the \emph{dual} exterior algebra G*, on the other hand,  elements of grade $k$ represent the subspaces of dimension $n-k$, and the outer product is the \emph{meet} {operator}.  For example, for $n=3$, the 1-vectors are planes, the 2-vectors, lines, and the 3-vectors, points.  
\myboldhead{Example}\Fig{fig:dualAufbau} shows how the wedge product of three points in $\pgshort$ is a plane, while the wedge product of three  planes in $\pdgshort$ is a point. Notice the use of subscripts and superscripts to distinguish between the two algebras. 
 
\subsubsection{Dimensions of projective subspaces} 

It's important for what follows to clarify the notion of the dimension of a subspace. We are accustomed to say that a point in $\RP{n}$ is a 0-dimensional subspace.  This is indeed the case in the context of the standard exterior algebra where points are represented by 1-vectors. Then all other linear subspaces are built up out of the 1-vectors by wedging (joining) points together. The dimension counts how many 1-vectors are needed to generate a subspace. For example, a line (2-vector) can be represented as the join of two points $\ell = \vec{A} \wedge \vec{B}$. $\ell$ is 1-dimensional since there is a one-parameter set of points incident with the line, given by $\alpha \vec{A} + \beta \vec{B}$ where only the ratio $\alpha : \beta$ matters. A line considered as a set of incident points is called a \emph{point range}. In general, if you wedge together $k$ linearly independent points you obtain a $k-1$-dimensional subspace.  For let $\vec{X} = \vec{P}_1 \wedge ... \wedge \vec{P}_k$. Then $\vec{X} \wedge \vec{P} = 0 \iff \vec{P} \cong \alpha_1 \vec{P}_1 + ... + \alpha_k \vec{P}_k$ for real constants $\{\alpha_i\}$. Since we are working in projective space, this is a $(k-1)$-dimensional set of points ($\{\alpha_i\} \equiv \beta\{\alpha_i\}$ for non-zero $\beta$). 

When we apply this reasoning to the dual exterior algebra, we are led to the surprising conlusion that a plane (a 1-vector) is 0-dimensional, since all the other linear subspaces are built up from planes by the meet operation. That is, in the dual algebra planes are simple and indivisible, just as a point in the standard algebra is. A line (2-vector) is the meet of two planes  $\ell = \vec{a} \wedge \vec{b}$. $\ell$ is 1-dimensional since there is a one-parameter set of planes incident with the line, given by $\alpha \vec{a} + \beta \vec{b}$ where only the ratio $\alpha : \beta$ matters. A line considered as a set of incident planes is called a \emph{plane pencil}. It's the form you get if you spin a plane around one of its lines. The meet of three planes is a point. The set of all planes incident with the point is 2-dimensional, called a plane bundle, etc. To think in this way you have to overcome certain habits that associate dimension with extensive \quot{size}. %The dimensionality of a subspace is essentially one less than the cardinality of a minimal spanning set (of points!). By spanning set $\mathfrak{S}$ we mean every point in the subspace can be written as a linear combination of the elements of $\mathfrak{S}$. For example, a line (2-vector) can be represented as the join of two points $\ell = \vec{A} \wedge \vec{B}$. $\ell$ is 1-dimensional since there is a one-parameter set of points incident with the line, given by $\alpha \vec{A} + \beta \vec{B}$ where only the ratio $\alpha : \beta$ matters. A plane (3-vector) is 2-dimensional while there is in the same way a 2-parameter set of points incident with it. 

\myboldhead{Take-away} The dimension of a geometric primitive depends on whether it is viewed in the standard exterior algebra or the dual exterior algebra. The 1-vectors serve as the \quot{building block} in both cases. For example, in the standard algebra a point is 0-dimensional, simple, and indivisible. In the dual algebra, however, it is two-dimensional, since it is created by wedging together three planes, or, what is the same, there is a two-parameter family of planes incident with it.

\subsubsection{Poincar\'{e} duality} Every geometric entity $x$ (e.g., point, line, plane) occurs once in each exterior algebra, say as $\vec{x} \in \pgshort$ and as $\vec{x}^* \in \pdgshort$. The \emph{Poincar\'{e} duality} map $J: \pgshort \rightarrow \pdgshort$ is defined by $\vec{x} \rightarrow \vec{x}^*$. It is essentially an identity map, sometimes called the \quot{dual coordinates} map.  In particular it is invertible. When often use $J$ for both maps when there is no danger of confusion.
$J$ is a grade-reversing map, that is a vector space isomorphism $\bigwedge^k \leftrightarrow \bigwedge^{n+1-k}$ for all $k$. See  \cite{gunnthesis} \S 2.3.1 for details.

\subsubsection{The regressive product}
 Using $J$,  it's possible to \quot{import} the outer product from one algebra into the the other. This imported product is sometimes called the \emph{regressive} product to distinguish it from the native wedge product.   For example, it possible to define a join operator $\vee$ in $\pdgshort$ by $$\vec{X}\vee\vec{Y} := J^{-1}(J(\vec{X}) \wedge J(\vec{Y}))$$
where the $\wedge$ on the right-hand side is that of the algebra $\pgshort$.  In this way, join and meet are available within a single algebra.  
We'll see below in Sect. \ref{sec:ipeg} why this is important for PGA . 
 %In the following we focus on $\pdgshort$, since as we'll see below in \Sec{sec:ipeg}, euclidean PGA is built using $\pdgshort$, not $\pgshort$. %\footnote{Why this is G* instead of G is a matter for theology, not mathematics. Using G instead leads to dual euclidean geometry, which is qualitatively different from euclidean geometry.}  
We write the  outer product of $\pdgshort$, the meet operator, as $\wedge$, and the  join operator, imported from $\pgshort$, as $\vee$.  That's easy to remember due to their similarity to the set operations $\cap$ and $\cup$.  

%There is an alternative way to obtain the regressive product using the canonical dual basis.
\myboldhead{References} The above mathematical prerequisites can be well-studied on Wikipedia in the articles on: vector space, bilinear form, quadratic form, tensor algebra,  exterior algebra, and projective space. %, and geometric algebra. 
%Now that we have covered the mathematical prerequisites 
We turn now to the geometric product and associated geometric product.

\section{Geometric product and geometric algebra }
\label{sec:gpaga}
%\item 

The exterior algebra of $\RP{n}$ answers questions regarding incidence (meet and join) of projective subspaces. That's an important step and yields uniform representation of points, lines, and planes as well as a \quot{parallel-safe} meet and join operators, both features from our wish-list. 

However the exterior algebra knows nothing about measurement, such as angle and distance, crucial to euclidean geometry. To overcome this we %need a larger algebra, one that is sensitive to these quantities. We obtain this by 
refine the equivalence relation that we used to produce the exterior algebra from the tensor algebra T. Instead of requiring that $\vec{v}\otimes\vec{v} \cong 0$ we require that  $$\vec{v}\otimes\vec{v} - B(\vec{v},\vec{v}) \cong 0$$
where $B$ is a symmetric bilinear form, that is $\vec{v}\otimes\vec{v}$ is equivalent to a scalar but not necessarily to 0 as in an exterior algebra.
We define the \emph{geometric algebra}\footnote{Sometimes called a Clifford algebra in honor of its discoverer \cite{clifford78}. Clifford however called it a geometric algebra, and we follow him.} with inner product $B$ to be the quotient of the tensor algebra by this new equivalence relation. Since this relation encodes an inner product on vectors, the geometric product contains more information than the exterior product.  We write the geometric product using simple juxtaposition: $\vec{X}\vec{Y}$.

Since the square of every 1-vector reduces to a scalar (0-vector), we obtain the same finite-dimensional graded algebra structure for the geometric algebra as for the exterior algebra, described in Sect. \ref{sec:gralg}. In fact, as we now show, one can also construct the geometric algebra by extending the exterior algebra.

\myboldhead{Alternative formulation}. Define the geometric product of two 1-vectors $\vec{u}$ and $\vec{v}$ to be%\footnote{Readers familiar with quaternions may recognize a similarity to the quaternion product of two imaginary quaternions. This is not accidental; the quaternion algebra often appears as a sub-algebra of geometric algebras.} 
$$\vec{u}\vec{v} := \vec{u}\cdot\vec{v} + \vec{u}\wedge\vec{v}$$
where $\cdot$ is the inner product associated to $B$ and $\wedge$ is the wedge product in the associated exterior algebra. (I. e., skip the tensor algebra formulation entirely.) Then it's possible to show that this geometric product has a unique extension to the whole graded algebra that agrees with the geometric product obtained above using the more abstract tensor product construction. 
%You can show for example that 
%\begin{align*}
 %   \vec{u}\cdot\vec{v} &= \frac12(\vec{u}\vec{v}+\vec{v}\vec{u}) \\
%    \vec{u}\wedge\vec{v} &= \frac12(\vec{u}\vec{v}-\vec{v}\vec{u}) 
%\end{align*}

\myboldhead{Connection to exterior algebra} The geometric algebra reduces to the exterior algebra when $B$ is trivial: $B(\vec{u}, \vec{v})=0$, equivalent to a signature of $(0,0,n)$.

%This distinguishes it from VGA (vector geometric algebra), that is build on $n$-dimensional vector space coordinates; and CGA (conformal geometric algebra) which uses $(n+2)$-dimensional coordinates to model $n$-dimensional euclidean space (introduced in \cite{hlr01}, see also \cite{dfm07}).  
%There are also non-euclidean versions of PGA for hyperbolic and elliptic space; interested readers can consult \cite{gunnthesis}.

\subsection{Projective geometric algebra} 
In order to apply the Cayley-Klein construction for modeling metric spaces such as euclidean space, we work in projective space. That is, we interpret the geometric product in  a projective setting just as we did with the wedge product in the projectivized exterior algebra. We call the result a \emph{projective} geometric algebra or PGA for short.  It uses $(n+1)$-dimensional coordinates to model $n-$ dimensional euclidean geometry. The standard geometric algebra based on $\pgshort$ with signature $(p,m,z)$ is denoted $\pclal{p}{m}{z}$. The dual version of the same (based on $\pdgshort$) is written $\pdclal{p}{m}{z}$. 

\myboldhead{Remark} PGA is actually a whole family of geometric algebras, one for each signature; the rest of these notes concern finding and exploring the member of this family that models euclidean geometry. We often write \quot{PGA} for this one algebra -- we sometimes use the more precise \quot{EPGA} for "euclidean" PGA to avoid confusion.

\subsection{Geometric algebra basics}\label{sec:gab} In general, the geometric product of a $k$-vector and an $m$-vector is  a sum of components of different grades, each expressing a different geometric aspect of the product, as in the geometric product of  two 1-vectors above.  A general element containing different grades is  called a \emph{multivector}. A multivector $\vec{M}$ can be written then as a sum of different grades: $\vec{M} = \sum_{i=0}^n\grade{\vec{M}}{i}$.  For example, we can write the above geometric product of two 1-vectors as:  $\vec{a} \vec{b} := \grade{\vec{a} \vec{b}}{0} +\grade{ \vec{a}  \vec{b}}{2}$.  The product of two multivectors can be reduced to a sum of products of single-grade vectors, so we concentrate our discussions on the latter.

The highest grade part  of the product is the $(k+m)$-grade part, and coincides with the $\wedge$ product in the exterior algebra. All the other parts of the product involve some \quot{contraction} due to the square of a 1-vector reducing to a scalar (0-vector), which drops the dimension of the product down by two for each such square. We define the lowest-grade part of the geometric product of a $k$-vector and an $m$-vector to be the \emph{inner product} and write $ \vec{a} \cdot \vec{b}$ (it does not have to be a scalar!). It has grade $|k - m|$.  

We will occasionally also need the commutator product $\vec{X}\times\vec{Y} := \frac12(\vec{X}\vec{Y}-\vec{Y}\vec{X})$, the so-called \emph{anti-symmetric} part of the geometric product. %In what follows, we introduce notation for other grade parts of the geometric product as the situation requires. 
A $k$-vector which can be written as the product of 1-vectors is called a \emph{simple} $k$-vector.  Note that then all the 1-vectors are orthogonal to each other and the product is equal to the wedge product of the 1-vectors. Any multi-vector can be written as a sum of simple $k$-vectors. We sometimes call 2-vectors \emph{bivectors}, and 3-vectors, \emph{trivectors}. 

We'll also need the \emph{reverse} operator $\widetilde{\vec{X}}$, that reverses the order of the products of 1-vectors in a simple $k$-vector. If the simple $k$-vector is $\vec{X}$, then the reverse $\widetilde{\vec{X}} = (-1)^{k\choose 2}\vec{X}$. The exponent counts how many \quot{neighbor flips} are required to reverse a string with $k$ characters (since for orthogonal 1-vectors $\vec{a}$ and $\vec{b}$, $\vec{b}\vec{a} = -\vec{a}\vec{b}$).

We first explore the algebra $\pclal{3}{0}{0}$ in order to warm up in a familiar setting.

\subsection{Example: Spherical geometry via $\pclal{3}{0}{0}$}
\label{sec:sphgeom}

This is the projectivized  geometric algebra of $\R{3}$, the familiar 3D euclidean vector space. Take an orthonormal basis $\{\e{0}, \e{1},\e{2})$.  Then a general 1-vector is given by $\vec{u} = x\e{0}+y\e{1}+z\e{2}$. It satisfies $\vec{u}^2=x^2+y^2+z^2 = \| \vec{u} \|^2$. The set of 1-vectors satisfying $\| \vec{u} \| = 1$ forms the unit sphere (whereby $\vec{u}$ and $-\vec{u}$ represent the same projective point in the algebra). We saw above, the product of two normalized 1-vectors is given by $\vec{u}\vec{v} := \vec{u}\cdot\vec{v} + \vec{u}\wedge\vec{v}$. Here $\vec{u}\cdot\vec{v} = \cos{\alpha}$ where $\alpha$ is the angle between the spherical points $\vec{u}$ and $\vec{v}$, and $ \vec{u}\wedge\vec{v}$ is the line (2-vector) spanned by the points (represented by a great circle joining the points.)

An orthonormal basis for the 2-vectors is given by $$\{\EE{0}:=\e{1}\e{2},~ \EE{1}:=\e{2}\e{0},~ \EE{2}:=\e{0}\e{1}\}$$ These are three mutually perpendicular great circles. The unit pseudo-scalar is $\eye  := \e{012}:= \e{0}\e{1}\e{2}$. Multiplication of either a $1$- or $2$-vector with $\eye$ produces the orthogonal complement $\vec{X}^\perp$ of the argument $\vec{X}$. That is, $\vec{u}^\perp = \vec{u}\eye$ is the great circle that forms the \quot{equator} to the \quot{pole} point represented by $\vec{u}$; $\vec{U}\eye$ for a 2-vector $\vec{U}$ produces the polar point of the \quot{equator} represented by $\vec{U}$. 
The complete $8x8$ multiplication table is shown in 
Table \ref{tab:cl300}.

\myboldhead{Exercise.} Check in the multiplication table that the products $\e{i}\eye = \EE{i}$ and $\EE{i}\eye=-\e{i}$ for $i\in\{1,2,3\}$ and verify that these results confirm that multiplication by $\eye$ is the \quot{orthogonal complement} operator.

\begin{table}[t]

\centering
\renewcommand{\arraystretch}{1.1}
\begin{tabularx}{\columnwidth} {| Y  || Y | Y  |  Y  | Y | Y | Y | Y | Y  |} \hline 
          & $\one$ & $\e{0}$ & $\e{1}$ & $\e{2}$ & $\EE{0}$ & $\EE{1}$ & $\EE{2}$ & $\eye$  \\ \hline \hline
$\one$        & $\one$ & $\e{0}$ & $\e{1}$ & $\e{2}$ & $\EE{0}$ & $\EE{1}$ & $\EE{2}$ & $\eye$  \\ \hline
$\e{0}$  & $\e{0}$ & $1$     & $\EE{2}$  & $-\EE{1}$ & $\eye$    & $-\e{2}$ & $\e{1}$  & $\EE{0}$  \\\hline
$\e{1}$  & $\e{1}$ & $-\EE{2}$ & $\one$& $\EE{0}$ & $\e{2}$ & $\eye$ & $-\e{0}$ & $\EE{1}$ \\ \hline
$\e{2}$  & $\e{2} $  & $\EE{1}$ & $-\EE{0}$ & $\one$ & $-\e{1}$ & $\e{0}$ & $\eye$ & $\EE{2}$ \\ \hline
$\EE{0}$  & $\EE{0}$ & $\eye$ & $-\e{2}$  & $\e{1}$ & $-\one$ & $-\EE{2}$ & $\EE{1}$ & $-\e{0}$ \\ \hline 
$\EE{1}$  & $\EE{1}$ & $\e{2}$ & $\eye$   & $-\e{0}$  & $\EE{2}$ & $-\one$ & $-\EE{0}$ & $-\e{1}$  \\\hline
$\EE{2}$  & $\EE{2}$ & $-\e{1}$ & $\e{0}$   & $\eye$ & $-\EE{1}$ & $\EE{0}$ & $-\one$ & $-\e{2}$  \\\hline
$\eye$    & $\eye$     & $\EE{0}$     & $\EE{1}$ & $\EE{2}$ & $-\e{0}$ & $-\e{1}$ & $-\e{2}$ & $-\one$ \\ \hline
\end{tabularx}
\myvspace{.1in}
\caption{Multiplication table for  $\pclal{3}{0}{0}$, the geometric algebra of the sphere.}
\label{tab:cl300}
\end{table}

\myboldhead{Exercise} Show that the angle $\alpha$ between two normalized 2-vectors (great circles) in $\pclal{3}{0}{0}$ is given by $\alpha=\cos^{-1}{(\vec{U}\cdot\vec{V})}$.

\myboldhead{Exercise} Verify that the elements $\{1,\e{12},\e{20},\e{01}\}$ generates a sub-algebra of $\pclal{3}{0}{0}$ that is isomorphic to Hamilton's quaternion algebra $\mathbb{H}$ generated by $\{1,i,j,k\}$. %Deduce that a rotation around the normalized axis $\vec{u} = a\e{0}+b\e{1}+c\e{2}$ through angle $2\alpha$ is given by the sandwich operator $(\cos{\alpha}+\sin{\al$

\myboldhead{Exercise} Find as many formulas of spherical geometry/trigonometry as you can within $\pclal{3}{0}{0}$.

\myboldhead{Exercise} $\pdclal{3}{0}{0}$ is the same algebra as above but uses the dual construction where the 1-vectors are lines (great circles). Show that it also provides a model for spherical geometry, one in which the $\vec{U}\cdot\vec{V} = \cos{\alpha}$ for normalized 1-vectors $\vec{U}$ and $\vec{V}$ meeting at angle $\alpha$.

The above discussion gives a rudimentary demonstration of how the signature $(3,0,0)$ leads to a model of   spherical geometry in both the standard and dual constructions%\footnote{which will be fleshed out in future versions of these notes! cg, May 28, 2019}.  

We now turn to the question of which member of the PGA family models the euclidean plane. That is, we need to determine a signature and, possibly, choose between the standard and dual construction. The existence of parallel lines in euclidean geometry plays an essential role in this search.

\subsection{Determining the signature for euclidean geometry}
\label{sec:ipeg}
We saw that the inner product of 1-vectors in $\pdclal{3}{0}{0}$ can be used to compute the angle between two lines in spherical geometry. What does the analogous question in the euclidean plane yield?
% How can one calculate the angle between two lines in the euclidean plane?  
Let \[
a_{0}x + b_{0}y + c_{0} = 0,~~~~~~~
a_{1}x + b_{1}y + c_{1} = 0
\]  be two oriented lines which intersect at an angle $\alpha$.  We can assume without loss of generality that the coefficients satisfy ${a_{i}^{2} + b_{i}^{2} = 1}$. Then it is not difficult to show that $a_{0} a_{1} + b_{0} b_{1} = \cos{\alpha}$. One can observe for example that the direction of line $i$ is $(-b_i, a_i)$ and calculate the angle of these direction vectors.
%\begin{figwindow}[0,r, \fbox{\includegraphics[width=2.5in]{eucMetricViaLines.png}},{Angles of euclidean lines}]

\begin{figure}
  \centering
  \def\xyz{0.8}
    \setlength\fboxsep{0pt}\fbox{\includegraphics[width=\xyz\columnwidth]{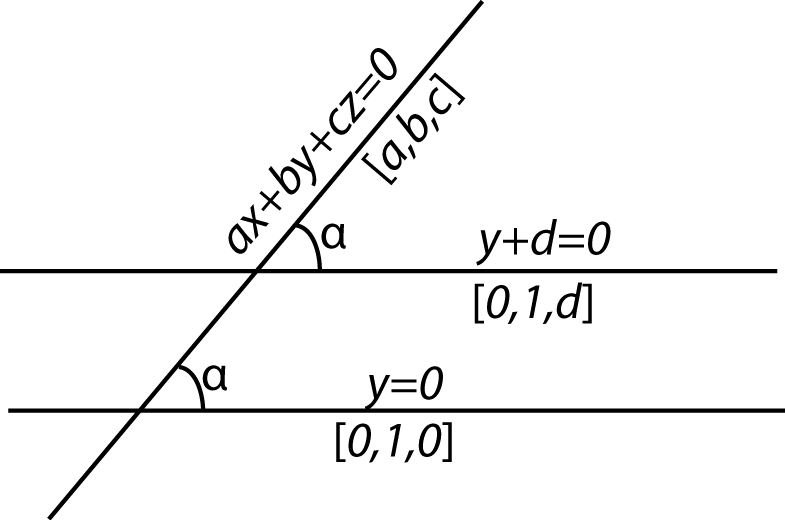}}
  \caption{Angles of euclidean lines. }
\label{fig:anglelines}
\end{figure} 

\myboldhead{The superfluous coordinate} The third coordinate of the lines makes no difference in the angle calculation! Indeed,  translating a line changes only its third coordinate,  leaving  the angle between the lines unchanged.  Refer to \Fig{fig:anglelines} which shows an example involving a general line and a pair of horizontal lines. Choose a basis for the (dual) projective plane so that $\e{1}$ corresponds to the line $x=0$, $\e{2}$ to $y=0$, and $\e{0}$ to $z=0$.\footnote{The unusual ordering is chosen since it is more convenient if in every dimension the \quot{superfluous} coordinate always has the same index.} Then the line given by $ax + by + c = 0$ corresponds to the 1-vector $c\e{0}+a\e{1}+b\e{2}$. If the geometric product of two such 1-vectors is to produce $a_1 a_2+b_1 b_2$ then the signature has to be $(2,0,1)$.  Hence the proper PGA for $\Euc{2}$ is $\pdclal{2}{0}{1}$.  %The final '1' reflects the fact that the $c$-coordinate can be ignored in the inner product.  
 Such a signature, or metric, is called \emph{degenerate} since $z\neq 0$. 

\hspace{-.15in}\textbf{Reminder:} The $*$ in the name says that the algebra is built on $\pdgshort$, the dual exterior algebra, since the inner product is defined on lines instead of points. %Unlike with non-degenerate metrics, We'll see how to measure the distance between points later.
A similar argument applies in dimension $n$, yielding  the  signature  $(n,0,1)$ for $\Euc{n}$. 
$\pclal{n}{0}{1}$ models a qualitatively different metric space called \emph{dual euclidean space}.

\hspace{-.15in}\textbf{Degenerate metric: asset or liability?}
PGA's development reflects the fact that much of the existing literature on geometric algebras deals only with non-degenerate metrics, reflecting widespread prejudices regarding degenerate metrics. (See \cite{gunn2017a} for a thorough analysis and refutation of these misconceptions.)  After long experience we are convinced that the degenerate metric, far from being a liability, is an important part of PGA's success -- exactly the degenerate metric models the metric relationships of euclidean geometry faithfully (see \cite{gunn2017a}, \S 5.3). %We hope to  convince the reader in what follows that degenerate metrics have a positive contribution to make to euclidean geometry.

\section{PGA for the euclidean plane: $\pdclal{2}{0}{1}$}
\label{sec:eucplane}

We give now a brief introduction to PGA by looking more closely at euclidean plane geometry.  Readers can find more details in \cite{gunn2017b}. The approach presented here can be carried out in a coordinate-free way (\cite{gunn2017b}, Appendix).  But for an introduction it's easier and also helpful to refer occasionally to coordinates.  The coordinates we'll use are sometimes called \emph{affine coordinates} for euclidean geometry. We add an extra coordinate to standard $n$-dimensional coordinates. For $n=2$:
\begin{compactitem}
\item \textbf{Point:} $(x,y)\rightarrow (x,y,1)$
\item \textbf{Direction:} $(x,y)\rightarrow (x,y,0)$
\end{compactitem} 

\begin{figure}
  \centering
  \def\xyz{0.65}
    \setlength\fboxsep{0pt}\fbox{\includegraphics[width=\xyz\columnwidth]{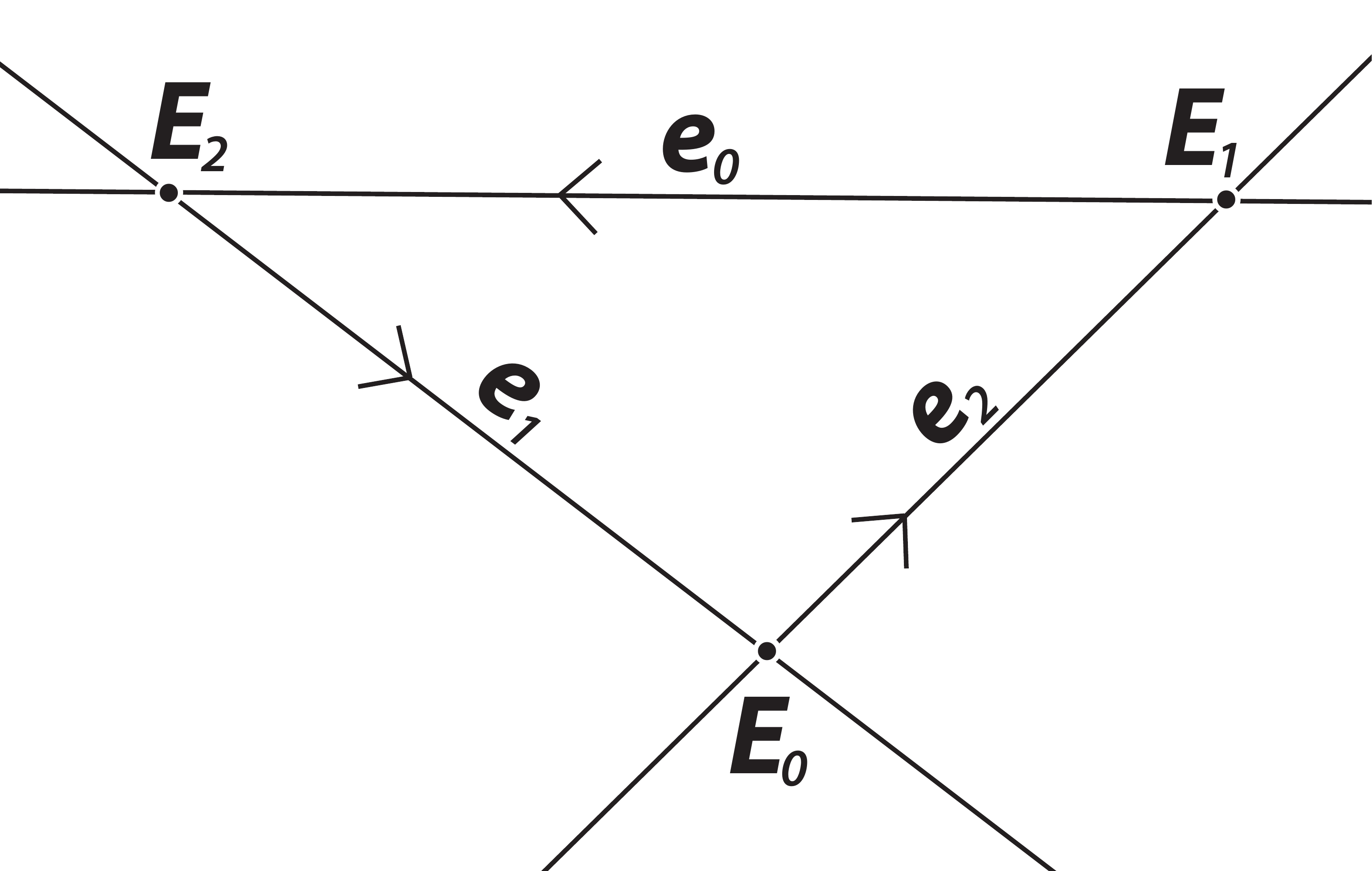}}
  \caption{Fundamental triangle of coordinate system. }
\label{fig:fundtri}
\end{figure} 

A perspective figure of the basis elements is shown in \Fig{fig:fundtri}.  The basis 1-vector $\e{0}$ represents the ideal line, sometimes called the \quot{line at infinity} and written $\omega$ to remind us that it is defined in a coordinate-free way. 
$\e{1}$ and $\e{2}$ represent the coordinate lines $x=0$ and $y=0$, resp.  
These basis vectors satisfy $\e{0}^2 = 0$ and $\e{1}^2=\e{2}^2 = 1$, consistent with the signature $(2,0,1)$. Note that by orthogonality, $\e{i}\e{j} = \e{i} \wedge \e{j}$ when $i\neq j$. A basis for the 2-vectors is given by the products (i. e., intersection points) of these orthogonal basis lines: \[\EE{0} := \e{1}\e{2},~~~ \EE{1}:=\e{2}\e{0}, ~~~\EE{2}:=\e{0}\e{1}\] whereby $\EE{0}$ is the origin, $\EE{1}$ and $\EE{2}$ are the $x$- and $y-$ directions (ideal points), resp.  They satisfy $\EE{0}^2=-1$ while $\EE{1}^2=\EE{2}^2=0$. That is, the signature on the 2-vectors is more degenerate: $(1,0,2)$. Finally, the unit pseudoscalar $\eye := \e{0}\e{1}\e{2}$ represents the whole plane and satisfies $\eye^2=0$.  The full 8x8 multiplication table of these basis elements can be found in \Tab{tab:cl201}. 

 \myboldhead{Exercise} 
 1) For a 1-vector $\vec{m} = a\e{1}+b\e{2}+c\e{0}$,  $\vec{m}^2=a^2+b^2$. 2) For a 2-vector $\vec{P} = x\EE{1}+y\EE{2}+z\EE{0}$, $\vec{P}^2=-z^2$.
 %These exercises establish that every $k$-vector in $\pdclal{2}{0}{1}$ is simple and hence satisfies $\vec{v}^2 \in \mathbb{R}$.
% \begin{compactenum}
% \item This is trivial for $k\in \{0,1,3\}$. 
% \item For $k=2$, show that a linear combination of simple bivectors is simple. Since $\EE{0}, \EE{1},$ and $\EE{2}$ are simple, deduce that every bivector is simple.
% \item For a simple $k$-vector $\vec{v}$, $\vec{v}\widetilde{\vec{v}} \in \mathbb{R}$.
% \item The \emph{inverse} $\vec{M}^{-1}$ of a multivector $\vec{M}$ satisfies $\vec{M}\vec{M}^{-1} = 1$. Show that for a simple $k$-vector, $\vec{v}^{-1} = \pm \cfrac{\vec{v}}{\vec{v}\widetilde{\vec{v}}}$ hence $\vec{v}^2 \in \mathbb{R}$.
% \end{compactenum}

\begin{table}

\centering
\renewcommand{\arraystretch}{1.1}
\begin{tabularx}{\columnwidth} {| Y  || Y | Y  |  Y  | Y | Y | Y | Y | Y  |} \hline 
          & $\one$ & $\e{0}$ & $\e{1}$ & $\e{2}$ & $\EE{0}$ & $\EE{1}$ & $\EE{2}$ & $\eye$  \\ \hline \hline
$\one$        & $\one$ & $\e{0}$ & $\e{1}$ & $\e{2}$ & $\EE{0}$ & $\EE{1}$ & $\EE{2}$ & $\eye$  \\ \hline
$\e{0}$  & $\e{0}$ & $0$     & $\EE{2}$  & $-\EE{1}$ & $\eye$    & $0$      & $0$          & $0      $  \\ \hline
$\e{1}$  & $\e{1}$ & $-\EE{2}$ & $\one$& $\EE{0}$ & $\e{2}$ & $\eye$ & $-\e{0}$ & $\EE{1}$ \\ \hline
$\e{2}$  & $\e{2} $  & $\EE{1}$ & $-\EE{0}$ & $\one$ & $-\e{1}$ & $\e{0}$ & $\eye$ & $\EE{2}$ \\ \hline
$\EE{0}$  & $\EE{0}$ & $\eye$ & $-\e{2}$   & $\e{1}$   & $-\one$ & $-\EE{2}$ & $\EE{1}$ & $-\e{0}$  \\ \hline 
$\EE{1}$  & $\EE{1}$ & $0$     & $\eye$     & $-\e{0}$    & $\EE{2}$ & $0$ & $0$ & $0$  \\ \hline
$\EE{2}$  & $\EE{2}$ & $0$     & $\e{0}$   & $\eye$    & $-\EE{1}$ & $0$ & $0$ & $0$  \\ \hline
$\eye$    & $\eye$     & $0$     & $\EE{1}$ & $\EE{2}$ & $-\e{0}$ & $0$ & $0$ & $0$ \\ \hline
\end{tabularx}
\myvspace{.1in}
\caption{Multiplication table for the geometric product in $\pdclal{2}{0}{1}$}
\label{tab:cl201}
\end{table}
%\myvspace{-.1in}

\subsection{Normalizing $k$-vectors}

From the previous exercise, the square of any $k$-vector is a scalar. When it is non-zero, the element is said to be \emph{euclidean}, otherwise it is \emph{ideal}. Just as with euclidean vectors in $\R{n}$, it's possible and often preferable to normalize simple $k$-vectors.  Euclidean $k$-vectors can be normalized by the formula $$\widehat{\vec{X}} := \cfrac{\vec{X}}{\sqrt{| \vec{X}^2|}}$$ Then $\widehat{\vec{X}}$ satisfies $\vec{X}^2 = \pm 1$.

 For a euclidean line $\vec{a}$, the element $\widehat{\vec{a}} := \dfrac{\vec{a}}{\sqrt{\vec{a}^2}}$ represents the same line but is normalized so that $\widehat{\vec{a}}^2=1$. A euclidean point $\vec{P}=x\EE{1}+y\EE{2}+\EE{0}$ is normalized and satisfies $\vec{P}^2 = -1$.\footnote{The point $-\vec{P}$ is a normalized form for $\vec{P}$ also but we use positive $z$-coordinate wherever possible.} This gives rise to a standard norm on euclidean $k$-vectors $\vec{X}$ that we write $\|\vec{X} \|$.  %In terms of coordinates, the homogeneous coordinate is $\pm1$. %For such a vector $\vec{X}$, $\vec{X}^2 \in \mathbf{R} ~\text{and} \neq 0$, hence $\dfrac{\vec{X}}{\sqrt{\mid \vec{X}^2 \mid}}$ is a projectively equivalent to $\vec{X}$ and its square equals $\pm1$.  We define the norm $\| \vec{X} \| := \sqrt{\mid \vec{X}^2 \mid}$.  

\subsubsection{The ideal norm} Such a normalization is not possible for ideal elements, since these satisfy $\vec{X}^2=0$.  It turns out that there is a  \quot{natural} non-zero norm on ideal elements that arises by requiring that the inner product of any 2 $(n-1)$-dimensional ideal flats should be the same as the inner product between any two euclidean $n$-dimensional flats whose intersections with the ideal plane are these two ideal flats. For example, when $n=2$ this means that the inner product of ideal points is the same as the inner product of any pair of lines meeting the ideal line in these points. The reader can check that this is well-defined by recalling that moving a line parallel to itself does not change its angle to other lines. 

If the two lines are $a_i\e{1} + b_i\e{2} + c_i\e{0})$ their inner product is $(a_0 a_1 + b_0 b_1)$ and their ideal points are $a_i\EE{01}+b_i\EE{02}$. In order for the inner product of these two lines to be $(a_0 a_1 + b_0 b_1)$ it's clear that the signature on the ideal line has to be $(2,0,0)$, and in general, $(n,0,0)$. In this way the set of ideal elements are given the structure of an $(n-1)$-dimensional dual PGA with signature $(n,0,0)$, the standard positive definite metric of $\mathbb{R}^n$: ideal points are identical with euclidean vectors, a fact already recognized by Clifford \cite{clifford73}. In the projective setting we say that the ideal plane has an \emph{elliptic} metric.

In fact, rather than starting with the euclidean planes and deducing the “induced” inner product on ideal lines as sketched above, it is also possible to start with this inner product on the ideal elements and “extend” it onto the euclidean elements (i. e., the inner product of two euclidean lines is defined to be the inner product of their two ideal points).
This approach to the ideal norm is sketched in the appendix of \cite{gunn2017b}. %This appendix in general shows how to develop PGA without using coordinates and assumes some familiarity with modern abstract algebra.

In the case of $n-2$, this yields an \emph{ideal} norm with the following properties. %These two norms, euclidean and ideal, harmonize to a surprising degree; the sequel presents many examples.
\begin{compactitem}
\item\textbf{Point}
In terms of the coordinates introduced above, for an ideal point $\vec{V} = x\EE{1}+y\EE{2}$, $\|\vec{V}\|_\infty := \sqrt{x^2+y^2}$. A coordinate-free definition of the ideal norm of an ideal point $\vec{V}$ is given by $\|\vec{V}\|_\infty := \| \vec{V} \vee \vec{P}\|$ for any normalized euclidean point $\vec{P}$. %This agrees with the standard Euclidean vector space norm restricted to the  subspace satisfying $z=0$. In fact, an ideal point is essentially a free vector, a fact already recognized by Clifford \cite{clifford73}.
\item \textbf{Line} The ideal norm for an ideal line $\vec{m} = c\e{0}$ is given by $\| \vec{m} \|_\infty := c$. This can be obtained in a coordinate-free way via the formula $\|\vec{m}\|_\infty = \vec{m}\vee \vec{P}$ where $\vec{P}$ is \textbf{any} normalized euclidean point. Using the $\vee$ operator instead of $\wedge$ produces a scalar directly instead of a pseudoscalar with the same numerical value.
\item\textbf{Pseudoscalar} We can also consider the pseudoscalar as an ideal element since since $\eye^2=0$. The ideal norm for a pseudoscalar $a\eye$ is $\| a\eye \|_\infty = a$.
\end{compactitem}
Note that the ideal norms for lines and pseudoscalars  are signed magnitudes. This is due to the fact that they belong to 1-dimensional subspaces that allow such a coordinate-free signed magnitude (based on the single generator).  To distinguish them from traditional (non-negative) norms we call them \emph{numerical values} but use the same notation $\|...\|_\infty$ for both.

\subsubsection{Ideal norm via Poincar\'{e} duality}
Another neat way to compute the ideal norm is provided by Poincar\'{e} duality. The discussion of Poincar\'{e} duality above in \Sec{sec:pea} took place at the level of the Grassmann algebra.  It's possible to consider this map to be between geometric algebras, in this case, $J: \pdclal{2}{0}{1} \rightarrow \pclal{2}{0}{1}$.  We leave it as an exercise for the reader to verify that for ideal $\vec{x} \in \pdclal{2}{0}{1}$, $\| \vec{x} \|_\infty = \| J(\vec{x})\|$ (where by sleight-of-hand the scalar on the right-hand side is interpreted as a scalar in $\pdclal{2}{0}{1}$).  That is, the ideal norm in the euclidean plane is the ordinary norm in the dual euclidean plane. Naturally the same holds for arbitrary dimension. Whether this \quot{trick} has a deeper meaning remains a subject of research.  

We will see  that the two norms -- euclidean and ideal -- harmonize remarkably with each other, producing \emph{polymorphic} formulas -- formulas that produce correct results for any combination of euclidean and ideal arguments. The sequel presents numerous examples.%We meet such an example in the product of two euclidean lines in the following section.

\myboldhead{Weight of a vector} Regardless of the type of norm, if an element satisfies $\| \vec{X} \| = d \in \mathbb{R}$, we say it has \emph{weight} $d$. The normed elements have weight 1. A typical computation requires that the arguments are normalized; the weight of the result then gives important insight into the calculation.  That means, we don't work strictly projectively, but use the weight to distinguish between elements that are projectively equivalent. We will see this below, in the section on 2-way products.  
%The geometric product generates a rich gamut of formulas for euclidean geometry: 
%measuring angles between planes and between lines, 
%measuring distances between points, points and lines, points and planes, etc., 
%construction of perpendiculars and parallels, and a wide variety of orthogonal projections.
In the discussions below, we assume that all the arguments have been  normalized with the appropriate norm since, just as in $\R{n}$, it simplifies many formulas.  

%\textbf{Notation conventions}:  bold large Latin letters ($\vec{P}$) always represent $n$-vectors (points).  Bold small Latin letters ($\vec{a}$) represent 1-vectors (lines in 2D and planes in 3D). Bold large Greek letters ($\sigo$) represent 2-vectors (lines) in 3D. 

\subsection{Examples: Products of pairs of elements in 2D}
\label{sec:prodpr}
\begin{figure}[!hbt]
 \centering
 \def\xyz{.95}
{\setlength\fboxsep{0pt}\fbox{\includegraphics[width=\xyz\columnwidth]{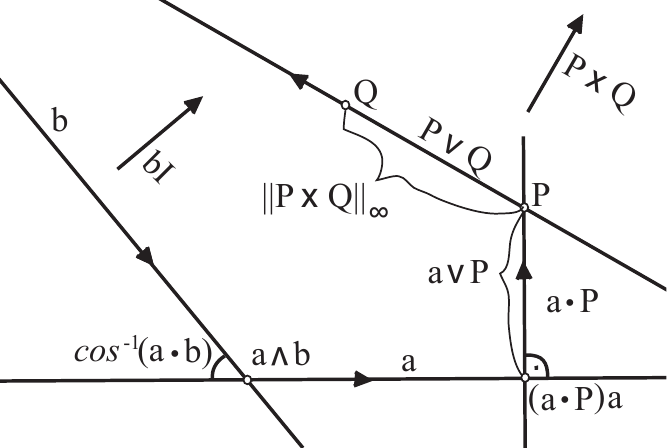}}}
\caption{Selected geometric products of pairs of simple vectors.}
\label{fig:geomprod}
\end{figure}

%To start with, we focus on the simpler case of $n=2$, the euclidean plane.  
 We get to know the geometric product better by considering basic products.  We consider first multiplication by the pseudoscalar $\eye$, then turn to products of pairs of normalized euclidean points and lines. It may be helpful to refer to the multiplication table (Table \ref{tab:cl201}) while reading this section.
 Also, consult \Fig{fig:geomprod} which illustrates many of the products discussed below. A fuller discussion can be found in \cite{gunn2017b}.

%\begin{compactitem}
\myboldhead{Multiplication by the pseudoscalar} Multiplication by the pseudoscalar $\eye$ maps a $k$-vector onto its orthogonal complement with respect to the euclidean metric.  For a euclidean line $\vec{a}$, $\vec{a}^\perp := \vec{a}\eye$ is an ideal point perpendicular to the direction of $\vec{a}$.  For a euclidean point $\vec{P}$, $\vec{P} ^\perp := \vec{P}\eye$ is the ideal line $\vec{e}_0$. Multiplication by $\eye$ is also called the \emph{polarity on the metric quadric}, or just the polarity operator.

\myboldhead{Product of two euclidean lines} 
We saw above that this product can be used as the starting point for the geometric product:
$$ \vec{a} \vec{b} = \vec{a} \cdot \vec{b} + \vec{a} \wedge \vec{b} $$
$\vec{a} \cdot \vec{b} = \grade{\vec{a} \vec{b}}{0} = \cos{\alpha}$, where $\alpha$ is the oriented angle between the two lines ($\pm1$ when they coincide or are parallel), while $\vec{a}\wedge \vec{b} = \grade{ \vec{a}  \vec{b}}{2}$ is their  intersection point.  If we call the normalized intersection point  $\vec{P}$  (using the appropriate norm), then $\grade{ \vec{a}  \vec{b}}{2} = (\sin{\alpha})\vec{P}$ when the lines intersect and $\grade{ \vec{a}  \vec{b}}{2} = d_{ab}\vec{P}$  when the lines are parallel and are separated by a distance $d_{ab}$. Here we see the remarkable functional polymorphism mentioned earlier, reflecting the harmonious interaction of the two norms.

\myboldhead{Product of two euclidean points}
\begin{align*}
\vec{P}\vec{Q} = \grade{\vec{P} \vec{Q}}{0} + \grade{\vec{P} \vec{Q}}{2} = -1 + d_{PQ}\vec{V}
\end{align*} 
The inner product of any two normalized euclidean points is -1. This illustrates the degeneracy of the metric on points: every other point yields the same inner product with a given point! The grade-2 part is more interesting: it is the direction (ideal point) perpendicular to the joining line $\vec{P} \vee \vec{Q}$.  It's easy to verify that $\grade{\vec{P} \vec{Q}}{2} = \vec{P} \times \vec{Q}$.  $\vec{V}$ in the formula is the normalized form of $\vec{P} \times \vec{Q}$. Then the formula shows that the distance $d_{PQ}$ between the two points satisfies $d_{PQ} = \|\vec{P} \times \vec{Q} \|_\infty$: while the inner product of two points cannot be used to obtain their distance,  $\langle \vec{P}\vec{Q}\rangle_2$  can. Here are two further formulas that yield this  distance: $d_{PQ} = \| \vec{P}\vee\vec{Q}\| = \|\vec{P} - \vec{Q}\|_\infty$.

\myboldhead{Product of euclidean point and euclidean line}
This yields a line and a pseudoscalar, both of which contain important geometric information:
\myvspace{-.05in}
\begin{align*}
\vec{a}\vec{P} = \grade{\vec{a} \vec{P}}{1} + \grade{\vec{a} \vec{P}}{3} &= \vec{a} \cdot \vec{P} + \vec{a} \wedge \vec{P} \\
&= \vec{a}_P^\perp + d_{aP}\eye
\end{align*} 
Here $\vec{a}_P^\perp:=\vec{a} \cdot \vec{P}$ is the line passing through $\vec{P}$ perpendicular to $\vec{a}$, %(it is normalized when both $\vec{a}$ and $\vec{P}$ are), 
while the pseudoscalar part has weight $d_{aP}$, the euclidean distance between the point and the line. Note that this inner product is anti-symmetric: $\vec{P} \cdot \vec{a} = - \vec{a} \cdot \vec{P}$.

\myboldhead{Practice in thinking dually: more about $\vec{a} \cdot \vec{P}$} You might be wondering, why is $\vec{a}\cdot\vec{P}$  a line through $\vec{P}$ perpendicular to $\vec{a}$? This is a good opportunity to practice thinking in the dual algebra. We are used to thinking of lines as being composed of points. That however is only valid in the standard algebra $\pgshort$.  In the dual algebra, we have to think of points as being composed of lines! The 1-vectors (lines) are the building blocks; they create points via the meet operator. A point \quot{consists} of the lines that pass through it -- called the \emph{line pencil} in $\vec{P}$. This is analogous to thinking of a line as consisting of all the points that lie on it -- called the \emph{point range} on the line. Consider $\vec{a} \cdot \vec{P}$ in this light. 

When $\vec{P}$ lies on $\vec{a}$ then we can write $\vec{P} = \vec{a}\vec{b}$ for the orthogonal line $\vec{b}$ through $\vec{P}$. Then $\vec{a}\vec{P} = \vec{a}\vec{a}\vec{b} = \vec{b}$ since $\vec{a}^2=1$. Hence the claim is proven. When $\vec{P}$ does not lie on $\vec{a}$ %: the line through $\vec{P}$ parallel to $\vec{a}$ is of the form $\vec{a}+k\omega$. We factor $\vec{P}$ orthogonally as before to obtain $\vec{P} = (\vec{a}+k\omega}\vec{b}$. This gives $$\vec{a}\cdot\vec{P} = \grade{(\vec{a}+k\omega}\vec{b}}{1}
the multiplication removes the line  through $\vec{P}$ parallel to $\vec{a}$ from the grade-1 part of the product, leaving as before the line $\vec{b}$ orthogonal to $\vec{a}$. We leave the details as an exercise for the reader. (Hint: any line parallel to $\vec{a}$ is of the form $\vec{a} + k\e{0}$.) %\footnote{Any line parallel to $\vec{a}$ is of the form $\vec{a} +k\omega$. Since $\omega$ is \quot{invisible} to the $(2,0,1)$ metric, removing $\vec{a}$ also removes all parallel lines.}  When this line is \emph{completely} removed from $\vec{P}$ it leaves only the line  perpendicular to $\vec{a}$. 
This example shows why the inner product is often called a \emph{contraction} since it reduces the dimension by removing common subspaces. 
 
\myboldhead{Remarks regarding 2-way products} In the above results, you can also allow one or both of the arguments to be ideal; one obtains in all cases meaningful, \quot{polymorphic} results. We leave this as an exercise for the interested reader. Interested readers can consult \cite{gunn2017b}.  The above formulas have been collected in \Tab{tab:pga2d}.  Note that the formulas assume \emph{normalized} arguments. 

%\subsection{3-way products}
After this brief excursion into the world 2-way products, we turn our attention to 3-way products with a repeated factor.  First, we look at products of the form $\vec{XXY}$ (where $\vec{X}$ and $\vec{Y}$ are either $1$- or $2$-vectors).  Applying the associativity of the geometric product produces \quot{formula factories}, yielding  a wide variety of important geometric identities. Secondly,  products of the form $\vec{aba}$ for 1-vectors $\vec{a}$ and $\vec{b}$ are used to develop an elegant representation of euclidean motions in PGA based on so-called \emph{sandwich} operators.  \cite{gunn2017b} contains more about general 3-way products in $\pdclal{2}{0}{1}$.

\myvspace{-.2in}
\subsection{Formula factories through associativity} %{Orthogonal projections in the plane}
  
First recall that for a normalized euclidean point or line, $\vec{X}^2=\pm1$. Use this and associativity to write 
\[ \vec{Y} = \pm(\vec{X}\vec{X})\vec{Y} = \pm\vec{X}(\vec{X}\vec{Y})\] where $\vec{Y}$ is also a normalized euclidean $1$- or $2$-vector.
 The right-hand side yields an \emph{orthogonal decomposition} of $\vec{Y}$ in terms of $\vec{X}$.   Associativity of the geometric product shows itself here to be a powerful tool.  These decompositions are not only useful in their own right, they provide the basis for a family of other constructions, for example,  \quot{the point on a given line closest to a given point}, or \quot{the line through a given point parallel to a given line} (see also \Tab{tab:pga2d}).
 
 Note that the grade of the two vectors can differ.   We work out below three orthogonal projections. % (details can be found in \cite{gunn2017b}). 
 As in the above discussions, we  assume the given points and lines  have been normalized, so their coefficients carry unambiguous metric information.  
 
 \begin{figure}[t]
  \centering
\def\xyq{.38}
\ifthenelse{\equal{\onecol}{true}}{ \def\xyq{.32}}{ \def\xyq{.38}}
\ifthenelse{\equal{\onecol}{true}}{ \def\xyr{.32}}{ \def\xyqr{.68}}
{\setlength\fboxsep{0pt}\fbox{\includegraphics[width=\xyr\columnwidth]{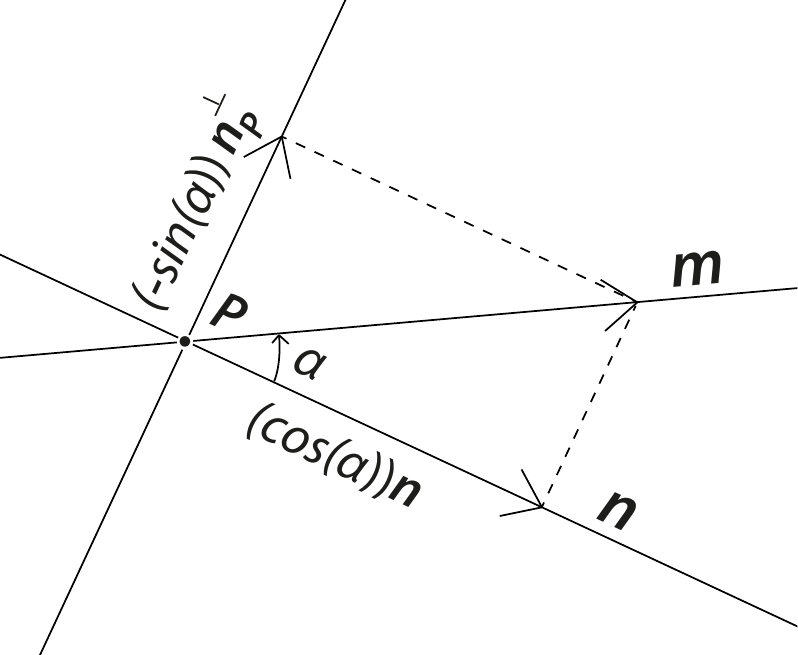}}} \hspace{.005in} 
{\setlength\fboxsep{0pt}\fbox{\includegraphics[width=\xyq\columnwidth]{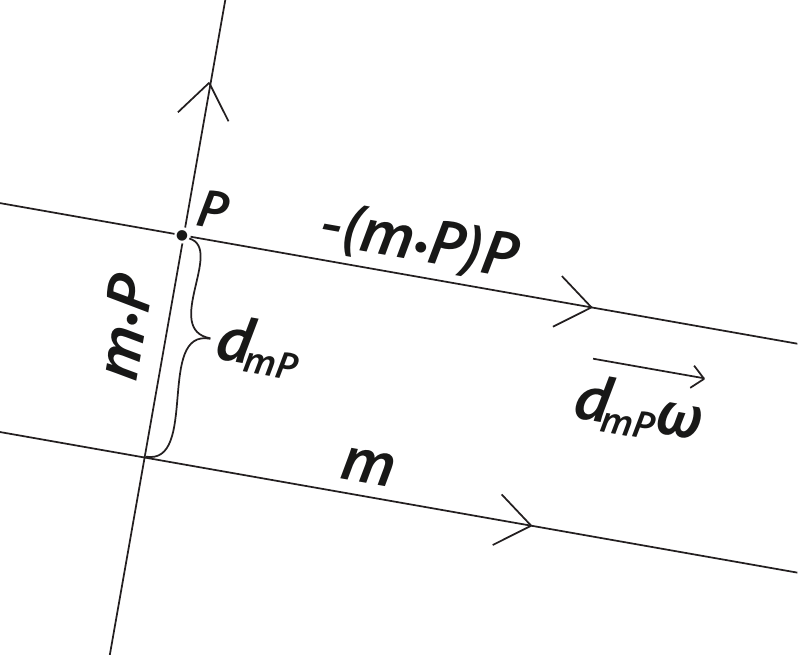}}}\hspace{.005in}
{\setlength\fboxsep{0pt}\fbox{\includegraphics[width=\xyq\columnwidth]{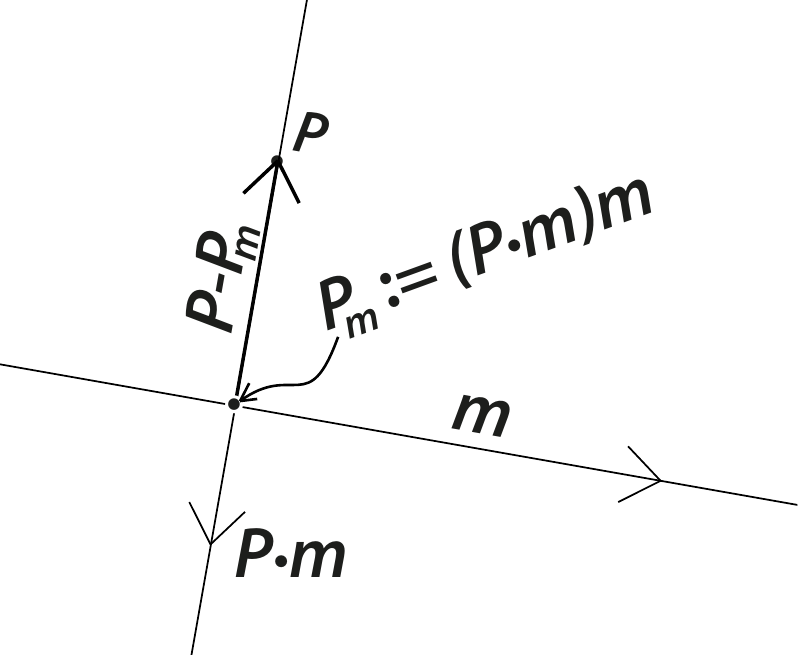}}} %\\ \vspace{.007in}
\caption{Orthogonal projections (l. to r.):, line $\vec{m}$ onto line $\vec{n}$, line $\vec{m}$ onto point $\vec{P}$, point  $\vec{P}$  onto line $\vec{m}$.}
\label{fig:orthProj}
\end{figure}

%\begin{compactenum}
%\item Projecting a line onto a line
%\item Projecting a point onto a line
%\item Projecting a line onto a point
%\item Projecting a point onto a point
%\end{compactenum}
%\myvspace{-.1in}

\myboldhead{Project line onto line}
%Assume both lines are euclidean. 
Assume both lines are euclidean and they they intersect in a euclidean point. 
Multiply \[\vec{m}\vec{n} = \vec{m}\cdot \vec{n} + \vec{m} \wedge \vec{n}\] with $\vec{n}$ on the right and use $\vec{n}^{2}=1$ to obtain 
%\myvspace{-.1in}
\begin{align*}
\vec{m} &= (\vec{m} \cdot \vec{n})\vec{n} + (\vec{m} \wedge \vec{n}) \vec{n} \\
&= (\cos{\alpha}) \vec{n} + (\sin{\alpha} )\vec{P}\vec{n} \\
&= (\cos{\alpha}) \vec{n} - (\sin{\alpha}) \vec{n}^{\perp}_{\vec{P}}
\end{align*}
\myvspace{-.2in}
%Note that $\vec{P}\vec{n} =-\vec{n}^{\perp}_{\vec{P}}$ since $\vec{P}\wedge\vec{n} = 0$. 
In the second line, $\vec{P}$ is the normalized intersection point of the two lines. Thus one obtains a decomposition of $\vec{m}$ as the linear combination of $\vec{n}$ and the perpendicular line $\vec{n}^{\perp}_{\vec{P}}$ through $\vec{P}$.  See \Fig{fig:orthProj}, left.

\myboldhead{Exercise}  If the lines are parallel one obtains $\vec{m} = \vec{n} + d_{\vec{m}\vec{n}}\omega$.  %\mydogblue

%\vspace{-.3in}
\myboldhead{Project line onto point}
%Assume both point and line are euclidean.
Multiply $\vec{m}\vec{P} = \vec{m}\cdot \vec{P} + \vec{m} \wedge \vec{P}$ with $\vec{P}$ on the right and use $\vec{P}^{2}=-1$ to obtain 
\begin{align*}
\vec{m} &= -(\vec{m} \cdot \vec{P})\vec{P} - (\vec{m} \wedge \vec{P}) \vec{P} \\
&= -\vec{m}^{\perp}_{\vec{P}}\vec{P} - (d_{\vec{m}\vec{P}}\eye) \vec{P} \\
%&= \vec{m}^{||}_{\vec{P}} - d_{\vec{m}\vec{P}}\vec{P}^{\perp} \\
&= \vec{m}^{||}_{\vec{P}} - d_{\vec{m}\vec{P}} \vec{\omega}
\end{align*}
%\myvspace{-.1in}
In the third equation, $\vec{m}^{||}_{\vec{P}}$ is the line through $\vec{P}$ parallel to $\vec{m}$, with the same orientation.  Thus one obtains a decomposition of $\vec{m}$ as the sum of a line through $\vec{P}$ parallel to $\vec{m}$ and a multiple of the ideal line.  Note that just as adding an ideal point (\quot{vector}) to a point translates the point, adding an ideal line to a line translates the line.  %%(adding which, as noted above in \Sec{sec:sumdiff}, translates euclidean lines parallel to themselves). 
See \Fig{fig:orthProj},  middle.

%\myvspace{-.05in}

\myboldhead{Project point onto line}
\label{sec:ppol}
Finally one can project a point $\vec{P}$ onto a line $\vec{m}$.  One obtains thereby a decomposition of $\vec{P}$ as $\vec{P}_m$, the point on $\vec{m}$ closest to $\vec{P}$,   plus a vector perpendicular to $\vec{m}$. See \Fig{fig:orthProj}, right. %Details are in \cite{gunn2017b}.

\myvspace{-.05in}

\subsection{Representing isometries as sandwiches}
\label{sec:isom}

  \begin{figure}
   \centering
   \def\xyz{.5}
{\setlength\fboxsep{0pt}\fbox{\includegraphics[width=\xyz\columnwidth]{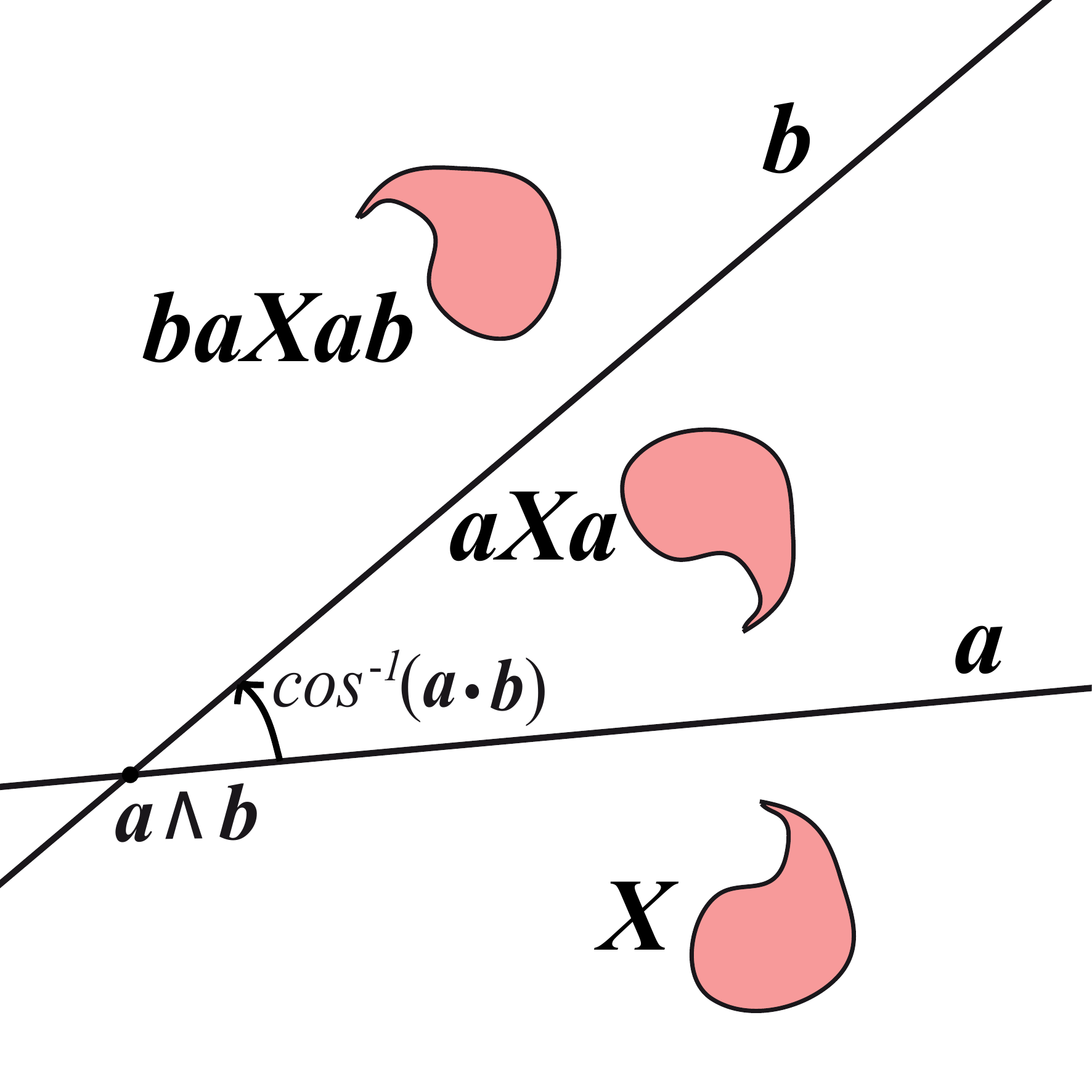}}}
\caption{The reflection in the line $\vec{a}$ followed by reflection in line $\vec{b}$.}
\label{fig:rotationab}
\end{figure}

Three-way products of the form $\vec{a}\vec{b}\vec{a}$ for euclidean 1-vectors $\vec{a}$ and $\vec{b}$ turn out to represent the reflection of the line $\vec{b}$ in the line $\vec{a}$, and form the basis for an elegant realization of euclidean motions as sandwich operators.  We sketch this here. %Besides having a uniform representation for geometric primitives with a rich set of formulas, the geometric product allows us to represent euclidean motions, such as rotations and translations, in a simple, universal form.
%\item

Let $\vec{a}$ and $\vec{b}$ be normalized 1-vectors representing different lines. Then 
\begin{align*}
\vec{a}\vec{b}\vec{a} &= \vec{a}(\vec{b}\vec{a}) = \vec{a}(\vec{b} \cdot \vec{a} + \vec{b} \wedge \vec{a}) \\
&= \cos(\alpha)  \vec{a} + \vec{a}(\vec{b} \wedge \vec{a})\\
&= \cos(\alpha)  \vec{a} + \sin(\alpha)\vec{a}\vec{P}  \\
&= \cos(\alpha)  \vec{a} + \sin(\alpha)\vec{a}\cdot \vec{P}  \\
&= \cos(\alpha)  \vec{a} +  \sin(\alpha) \vec{a}^\perp_\vec{P}
\end{align*}
We use the symmetry of the inner product in line 2. In line 3 we replace $\vec{a}\wedge\vec{b}$ with the normalized point $\vec{P}$ and weight $\sin{\alpha}$. Line 4 is justified by the fact that $\vec{a}\wedge\vec{P} = 0$, and line 5 uses the definition of  $\vec{a}^\perp_\vec{P}$. 
Compare this with the orthogonal decomposition for $\vec{b}$ obtained above in \Sec{sec:ppol}: $$\vec{b}=\cos(\alpha) \vec{a} - \sin(\alpha) \vec{a}^\perp_\vec{P}$$

Using the fact that $\vec{a}^\perp_\vec{P}$ is a line perpendicular to $\vec{a}$ leads to the conclusion that $\vec{a}\vec{b}\vec{a}$ must be the reflection of $\vec{b}$ in $\vec{a}$, since the reflection in $\vec{a}$ is the unique linear map fixing $\vec{a}$ and $\omega$ and mapping $\vec{a}^\perp_\vec{P}$ to $-\vec{a}^\perp_\vec{P}$. (\textbf{Exercise} Prove that $\vec{a}\omega\vec{a} = -\omega$.) We call this the \emph{sandwich operator} corresponding to $\vec{a}$ since $\vec{a}$ appears on both sides of the expression. It's not hard to show that for a euclidean point $\vec{P}$, $\vec{a}\vec{P}\vec{a}$ is the reflection of $\vec{P}$ in the line $\vec{a}$.\footnote{Hint: write $\vec{P} = \vec{p}_1\vec{p}_2$ where $\vec{p}_1\cdot\vec{p}_2=0$}
%= 0$. Then 
%\begin{align*}
%\vec{a}\vec{P}\vec{a}\ &= vec{a}\vec{p}_1(1)\vec{p}_2\vec{a} \\
%&= vec{a}\vec{p}_1\vec{a}\vec{a}\vec{p}_2\vec{a}
%\end{align*} 
Similar results apply in higher dimensions: the same sandwich form for a reflection works regardless of the grade of the \quot{meat} of the sandwich.
\begin{table}
\begin{center}
\normalsize %scriptsize %\larger
\renewcommand{\arraystretch}{1.25}
\begin{tabular}[h]{||  l  |    c ||}  \hline
Operation  & PGA \\ \hline  \hline
Intersection point of two lines &   $ \vec{a} \wedge \vec{b}$  \\  \hline
%Angle of two intersecting lines &  $ \cos^{-1}(\vec{a} \cdot \vec{b} ), \sin^{-1}(\|\vec{a}\wedge\vec{b}\|) $ \\ \hline
Angle of two intersecting lines &  $ \cos^{-1}(\vec{a} \cdot \vec{b} ) $ \\ \cline{2-2}
&  $\sin^{-1}(\|\vec{a}\wedge\vec{b}\|) $ \\ \hline
Distance of two $||$ lines &  $\| \vec{a} \wedge \vec{b}\|_\infty $ \\ \hline
Joining line of two points  &     $\vec{P} \vee \vec{Q}$ \\ \hline
$\perp$ direction to join of two points  &     $\vec{P} \times \vec{Q}$ \\ \hline
%Distance between two euclidean points &  $\|\vec{P}\vee \vec{Q}\|$,$~~\| \vec{P} \times \vec{Q} \|_\infty $ \\ \hline
Distance between two  points &  $\|\vec{P}\vee \vec{Q}\|$,$~~\| \vec{P} \times \vec{Q} \|_\infty $ \\ \hline
Oriented distance  point to line  &  $\| \vec{a} \wedge \vec{P} \| $  \\ \hline
Angle of ideal point to line  &  $\sin^{-1}{(\| \vec{a} \wedge \vec{P} \|_\infty)} $  \\ \hline
Line through point $\perp$ to line &   $\vec{P} \cdot \vec{a} $ \\ \hline
Nearest point on line to point &   $(\vec{P} \cdot \vec{a}) \vec{a}$ \\ \hline
Line through point $||$ to line &   $(\vec{P} \cdot \vec{a})  \vec{P}$ \\ \hline
Oriented area of triangle $ABC$ & $\frac12(\vec{A} \vee \vec{B} \vee \vec{C} )$ \\ \hline
Length of closed loop $\vec{P}_1\vec{P}_2...\vec{P}_n$ & $ \Sigma_{i=1}^n \| \vec{P}_i \vee \vec{P}_{i+1}\|$  \\ \hline
Oriented area of closed loop $\vec{P}_1\vec{P}_2...\vec{P}_n$ & $\| \Sigma_{i=1}^n (\vec{P}_i \vee \vec{P}_{i+1})\|$  \\ \hline
Reflection in line ($\vec{X} =$ point or line) & $\vec{a} \vec{X} \vec{a}$ \\ \hline \renewcommand{\arraystretch}{1.3} 
Rotation around point of angle $2 \alpha$ & $\vec{R} \vec{X} \widetilde{\vec{R}} ~~ (\vec{R} :=  e^{\alpha \vec{P}}$) \\ \hline
Translation by $2d$ in direction $\vec{V}^\perp$& $\vec{T} \vec{X} \widetilde{\vec{T}} ~~ (\vec{T} := 1 + d{\vec{V}})$  \\ \hline
Motor moving line $\vec{a}_1$ to $\vec{a}_2$ & $1 + \widehat{\vec{a}_2\vec{a}_1}$ \\ \hline
Logarithm of \MOTOR $\vec{g}$ & $\cos^{-1}(\langle \vec{g} \rangle_0) \widehat{\langle \vec{g} \rangle_2}$ \\ \hline
\end{tabular}
\end{center}
\myvspace{.1in}
\caption{A sample of geometric constructions and formulas in the euclidean plane using PGA (assuming normalized arguments, all arguments euclidean unless otherwise stated). }
\label{tab:pga2d}
\end{table}

\myboldhead{Rotations and translations} It is well-known that all isometries of euclidean space are generated by reflections.   The sandwich $\vec{b}(\vec{a}\vec{X}\vec{a})\vec{b}$ represents the composition of reflection in line $\vec{a}$ followed by reflection in line $\vec{b}$.  See \Fig{fig:rotationab}. When the lines meet at angle $\frac{\alpha}2$, this is well-known to be  a rotation around the point $\vec{P}$ through of angle $\alpha$. $\vec{R} := \vec{a}\vec{b} = \cos{\frac{\alpha}{2}}+\sin{\frac{\alpha}{2}}\vec{P}$ by the above formula. The  rotation can be expressed as $\vec{R} \vec{X} \widetilde{\vec{R}}$. (Here, $\widetilde{\vec{R}}$ is the \emph{reversal} of $\vec{R}$, obtained by writing all products in the reverse order. When $\vec{R}$ is normalized, it's also the inverse of $\vec{R}$.) 

  When $\vec{a}$ and $\vec{b}$ are parallel, $\vec{R}$ generates the \emph{translation} in the direction perpendicular to the two lines, of twice the distance between them -- once again, PGA polymorphism in action.
A product of $k$ euclidean 1-vectors is called a $k$-\emph{versor}; hence the sandwich operator is sometimes called a \emph{versor} form for the isometry.  
When $\vec{R}$ is normalized so that $\vec{R}\widetilde{\vec{R}} = 1$, it's called a \emph{\MOTOR}.  A \MOTOR is either a rotator (when its fixed point is euclidean) or a translator (when it's ideal).

\myboldhead{Exponential form for \MOTORS} Motors can be generated directly from the normalized center point $\vec{P}$ and angle of rotation $\alpha$ using the exponential form $$\vec{R} = e^{\frac{\alpha}{2}\vec{P}}=\cos{\frac{\alpha}{2}}+\sin{\frac{\alpha}{2}}\vec{P}$$.  This is another standard technique in geometric algebra: The exponential behaves like the exponential of a complex number since, as we noted above, a normalized euclidean point satisfies $\vec{P}^2=-1$.  When $\vec{P}$ is ideal ($\vec{P}^2=0$), the same process yields a translation through distance $d$ perpendicular to the direction of $\vec{P}$, by means of the formula $\vec{T} = e^{\frac{d}{2}\vec{P}} = 1 + \frac{d}{2}\vec{P}$. 

\myboldhead{Motor moving one line to another} Given two lines $\vec{l}_1$ and $\vec{l}_2$, there is a unique direct isometry that moves $\vec{l}_1$ to $\vec{l}_2$ and fixes their intersection point $\vec{P} := \vec{l}_1 \wedge \vec{l}_2$. Indeed, we know that when $\vec{P}$ is euclidean and the angle of intersection is $\alpha$, the product $\vec{g} := \vec{l}_2 \vec{l}_1$ is a \MOTOR that rotates by $2 \alpha$ around the intersection point $\vec{l}_1 \wedge \vec{l}_2$. Hence the desired \MOTOR can be written $\sqrt{\vec{g}}$. (\textbf{Exercise.} When $\vec{g}$ has been normalized to satisfy $\| \vec{g} \| = 1$, then $\sqrt{\vec{g}} = \widehat{1+\vec{g}}$. [Hint: The proof is similar to that of the statement: Given $P =(\cos{t},\sin{t})$ and $Q = (1,0)$ on the unit circle, $\dfrac{P+Q}{2}$ lies on the angle bisector of central angle $POQ$.]) This result is true also when $\vec{P}$ is ideal.

 This concludes our treatment of the euclidean plane.\Tab{tab:pga2d} contains an overview of formulas available in $\pdclal{2}{0}{1}$, most of which have been introduced in the above discussions. We are not aware of any other frameworks offering comparably concise and polymorphic formulas for plane geometry.

%\subsection{Reflections on ideal points}
%\label{sec:idealisfree}
%Before leaving the 2D case, we want to reflect on the relationship of euclidean and ideal points.  What are ideal points after all?  An ideal point behaves just like a free vector or direction, a fact already recognized by Clifford \cite{clifford73}.  Hence, the ideal line can be treated as the euclidean vector space $\R{2}$ that coexists in PGA with  $\Euc{n}$, a relationship mediated by the two norms.  This tight  integration of points and vectors yields many diverse formulas (as shown in Table \ref{tab:pga2d}), in comparison to which the integration in standard vector algebra is but a pale shadow. The strength of the PGA approach is, in brief: in PGA one can add \textbf{and} multiply points and vectors, while in VA one can only add them.  

%\subsection{Examples of euclidean isometries in 2D}
%[Insert worked-out examples]
\myvspace{-.2in}
\section{PGA for euclidean space: $\pdclal{3}{0}{1}$}
\label{sec:eucspace}
If you have followed the treatment of plane geometry using PGA, then you are well-prepared to tackle the 3D version $\pdclal{3}{0}{1}$.  
Naturally in 3D one has points, lines, \textbf{and} planes, with the planes taking over the role of lines in 2D (as dual to points); the lines represent a new, middle element not present in 2D. 
With a little work one can derive similar results to the ones given above for 2-way products, for orthogonal decompositions, and for isometries.  For example, $\vec{a}\cdot\vec{b}$ is the angled between two planes $\vec{a}$ and $\vec{b}$. 
A look at the tables of formulas for 3D (\Tab{tab:pga3d}, \Tab{tab:pga3d2}) confirms that many of the 2D formulas reappear, with planes substituting for lines.
If you re-read Examples \ref{sec:3dkal} and  \ref{sec:3dscrew} now you should understand much better how 3D isometries are represented in PGA, based on what you've learned about 2D sandwiches.

\myboldhead{Notation and foundations}
We continue to use large roman letters for points. Dual to points, planes are now written with small roman letters. Lines (and in general 2-vectors) are written with large Greek letters.
Now $\e{0}$ is an ideal plane instead of ideal line, and there are three ideal points $\EE{1}$, $\EE{2}$ and $\EE{3}$ representing the $x$-, $y$-, and $z$-directions instead of just three.  Bivectors have 6 coordinates corresponding to the six intersection lines of the four basis planes. The lines $\e{01},\e{02},\e{03}$ are ideal lines, and represent the intersections of the 3 euclidean basis planes with the ideal plane. The lines $\e{23},\e{31},\e{12}$ are lines through the origin in the $(x,y,z)$-directions, resp.  Hence, every bivector can be trivially written as the sum of an ideal line and a line through the origin.
%[This begins the second half of the tutorial.  It has not been worked out in as much detail as the first half. -cg]

In the interests of space, we leave it to the reader to confirm the similarities of the 3D case to the 2D case. We focus our energy for the remainder of this section on one important difference to 2D:   bivectors of $\pdclal{3}{0}{1}$, which, as we mentioned above, have no direct analogy in $\pdclal{2}{0}{1}$.  %Then, building on this we give a brief account of how to map kinematics and rigid body mechanics into this framework.

\begin{table}
\begin{center}
\normalsize %\scriptsize %\larger
\renewcommand{\arraystretch}{1.25}
\begin{tabular}[h]{||  l  |    c ||}  \hline
\textbf{Operation}  & \textbf{formula} \\ \hline  \hline
Intersection line of two planes &   $ \vec{a} \wedge \vec{b}$  \\  \hline
Angle of two intersecting planes &  $ \cos^{-1}(\vec{a} \cdot \vec{b} ) $ \\ \cline{2-2}
 &  $\sin^{-1}(\|\vec{a}\wedge\vec{b}\|) $ \\ \hline
Distance of two $||$ planes &  $\| \vec{a} \wedge \vec{b} \|_\infty$ \\ \hline
Joining line of two points  &     $\vec{P} \vee \vec{Q}$\\ \hline
Intersection point of three planes &    $\vec{a} \wedge \vec{b} \wedge \vec{c}$ \\ \hline
Joining plane of three points &    $\vec{P} \vee \vec{Q} \vee \vec{R}$ \\ \hline
Intersection of line and plane & $\velo \wedge \vec{a}$ \\ \hline
Joining plane of point and line &   $\vec{P} \vee \velo $ \\ \hline
Distance from  point to plane  &  $\| \vec{a} \wedge \vec{P} \|$  \\ \hline
Angle of ideal point to plane  &  $\sin^{-1}{(\| \vec{a} \wedge \vec{P} \|_\infty)}$  \\ \hline
%$\perp$ line to joining line of two euc. points &  $ \vec{P} \times \vec{Q}$ \\ \hline
$\perp$ line to join  of two  points &  $ \vec{P} \times \vec{Q}$ \\ \hline
Distance of two points &  $\|\vec{P}\vee \vec{Q}\|$,$~~\| \vec{P} \times \vec{Q} \|_\infty $\\ \hline
Line through point $\perp$ to plane  &   $\vec{P} \cdot \vec{a} $ \\ \hline
Project point onto plane &   $(\vec{P} \cdot \vec{a}) \vec{a}$ \\ \hline
Project plane onto point &   $(\vec{P} \cdot \vec{a})  \vec{P}$ \\ \hline
Plane through line $\perp$ to plane  &   $\velo \cdot \vec{a} $ \\ \hline
Project line onto plane &   $(\velo \cdot \vec{a})\vec{a} $ \\ \hline
Project plane onto line &   $(\velo \cdot \vec{a})\velo $ \\ \hline
Plane through point $\perp$ to line  & $ \vec{P} \cdot  \velo $ \\ \hline
Project point onto line & $ ( \vec{P} \cdot \velo )  \velo $ \\ \hline
Project line onto point&   $(\vec{P} \cdot \velo)  \vec{P}$ \\ \hline
Line through point $\perp$ to line &   $((\vec{P} \cdot \velo)  \velo)\vee \vec{P}$ \\ \hline
Oriented volume of tetrahedron $ABCD$ & $\frac13 \|\vec{A} \vee \vec{B} \vee \vec{C}\vee\vec{D} \| $ \\ \hline
Area of triangle mesh $M$ & $\frac 1 2  \sum\limits_{\Delta_i \in M}\| \hat \Point_{i1}\vee \hat\Point_{i2}\vee \hat\Point_{i3}\|$ \\ \hline
Volume of closed triangle mesh $M$ & $\frac 1 3 \| (\sum\limits_{\Delta_i \in M} \hat \Point_{i1}\vee \hat\Point_{i2}\vee \hat\Point_{i3}) \|_\infty$ \\ \hline
%Reflection in plane ($\vec{X} =$ point, line, or plane)  & $\vec{a} \vec{X} \vec{a}$ \\ \hline \renewcommand{\arraystretch}{1.3}
\end{tabular}
\end{center}
\myvspace{.1in}
\caption{A sample of geometric constructions and formulas in 3D using PGA (assuming normalized arguments).}
\label{tab:pga3d}
\end{table}

\begin{table}
\begin{center}
\normalsize %\scriptsize %\larger
\renewcommand{\arraystretch}{1.25}
\begin{tabular}[h]{||  l  |    c ||}  \hline
\textbf{Operation}  & \textbf{formula} \\ \hline  \hline
Common normal line to $\Line_1, \Line_2$ & $\widehat{\Line_1 \times \Line_2}$ \\ \hline
Angle $\alpha$ between $\Line_1, \Line_2$ & $\alpha = \cos^{-1}{(\hat \Line_1 \cdot \hat \Line_2)}$ \\ \hline
Distance between $\Line_1, \Line_2$ & $ d_{\Line_1 \Line_2} = {\csc{\alpha}\,{ (\hat \Line_1 \vee \hat \Line_2)}}$ \\ \hline
Refl. in plane ($\vec{X} =$ pt, ln, or pl)  & $\vec{a} \vec{X} \vec{a}$ \\ \hline %\renewcommand{\arraystretch}{1.4}
Rotation with axis $\velo$ by angle $2 \alpha$ & $\vec{R} \vec{X} \widetilde{\vec{R}} ~~ (\vec{R} :=  e^{\alpha \velo}$) \\ \hline
Translation by $2d$ in direction $\vec{V}$ & $\vec{T} \vec{X} \widetilde{\vec{T}} ~~ (\vec{T} := {(\EE{0} \vee d\vec{V})\eye})$  \\ \hline
Screw with axis $\velo$ and pitch $p$ &$ \vec{S} \vec{X} \widetilde{\vec{S}} ~~ (\vec{S} := e^{t(1+p\eye)\velo})$  \\ \hline
Logarithm of \MOTOR $\vec{m}$  & $\mathbf b = \langle \mathbf m \rangle_2, s = \sqrt{-\mathbf b \cdot \mathbf b}, p=\frac{-\mathbf b \wedge \mathbf b}{2s} $\\
& $\widehat{\vec{b}} = \frac{s-p}{s^2}\mathbf b $\\
& $\log \mathbf m = \big ( \tan^{-1}(\frac{s}{\langle \mathbf m \rangle_0}) + \frac{p}{\langle \mathbf m \rangle_0}\big ) \widehat{\vec{b}}$\\ \hline
\end{tabular}
\end{center}
\myvspace{.1in}
\caption{More formulas in 3D using PGA focused on \MOTORS and bivectors.}
\label{tab:pga3d2}
\end{table}

%\myvspace{-.2in}

\subsection{Simple and non-simple bivectors in 3D}
\label{sec:l2v2d}

%\subsubsection{Introduction to non-simple bivectors}
%\subsection{Characterizing non-simple bivectors}
In $\pdclal{2}{0}{1}$, all $k$-vectors are \emph{simple}, that is, they can be written as the product of $k$ 1-vectors.  This is no longer the case in $\pdclal{3}{0}{1}$.  A simple bivector $\sigo$ in 3D is the geometric product of two perpendicular planes $\sigo = \vec{p}_1 \wedge \vec{p}_2$ and represents their intersection line. Then clearly $\sigo\wedge \sigo = 0$.   
Let $\sigo_1$ and $\sigo_2$  be two simple bivectors that represent \emph{skew} lines\footnote{Skew lines are lines that do not intersect. Remember: parallel lines meet at ideal points and so are not skew.}.   We claim that the bivector sum $\sigo := \sigo_1 + \sigo_2$ is a non-simple.   First note that since $\sigo_1$ and $\sigo_2$ are skew, they are linearly independent, implying $\sigo_1\wedge\sigo_2 \neq 0$. Then, using bilinearity and symmetry of the wedge product (on bivectors!), one obtains directly $\sigo \wedge \sigo = 2 \sigo_1 \wedge \sigo_2 \neq 0$. We saw above however that a simple 2-vector $\sigo$ satisfies $\sigo \wedge \sigo = 0$.   Hence $\sigo$ must be non-simple.
In fact, as the next section shows, \emph{most} bivectors are non-simple.

\myboldhead{Exponentials of simple bivectors}
In the sequel we will need to know the exponential of a simple bivector. The situation is exactly analogous to the 2D case handled above and yields:
For a simple euclidean bivector $\velo$,  $e^{\alpha}\velo = \cos{\alpha} + \sin{\alpha}\velo$. For a simple ideal bivector $\velo_\infty$,  $e^{d\velo_\infty} = 1 + d\velo_\infty$.

 \subsubsection{The space of bivectors and Pl\"{u}cker's line quadric} 
 As noted above, the space of bivectors $\bigwedge^2$ is  spanned by the 6 basis elements $\e{ij} := \e{i} \e{j}$  %The $\e{ij}$ can be thought of as the lines of intersection of the 4 basis planes. The three elements $\e{0i}$ are ideal lines %at right angles to each other, 
 %while $(\e{23}, \e{31}, \e{12})$ are lines through the origin in the $(x, y, z)$-directions, resp. 
 and forms a 5-dimensional projective space $\proj{\bigwedge^2}$. %The condition that a bivector is simple (represents a line) can be translated into a quadratic constraint on the coordinates of the bivector a result due to  Pl\"{u}cker (1801-1868). 
 From the above discussion we can see the condition that a bivector $\velo$ is a line can be written as $\velo \wedge \velo = 0$. (In terms of coordinates, the bivector $\Sigma a_{ij}\e{ij}$ is a line $\iff$  $a_{01}a_{23}+a_{02}a_{31}+a_{03}a_{12} = 0$.)  This defines the \emph{Pl\"{u}cker quadric} $\mathcal{L}$, a 4D quadric surface (with signature $(3,3,0)$) sitting inside $\proj{\bigwedge^2}$, and giving rise to the well-known \emph{Pl\"{u}cker coordinates} for lines. Points not on the quadric are non-simple bivectors, also known as \emph{linear line complexes}. Consult Figure \ref{fig:complexspace}. Linear line complexes were first introduced by \cite{moebius37} in his early studies of statics under the name \emph{null systems}.
 
 \subsubsection{Product of two euclidean lines}
\label{sec:prtwoli}
\begin{figure}[t]
   \centering
   \def\xyy{.8}
   \def\xyz{.52} %.38}  %
   \def\xyw{.41} %.294} %
{\setlength\fboxsep{0pt}\fbox{\includegraphics[width=\xyz\columnwidth,trim={5mm 0 9mm 0}, clip]{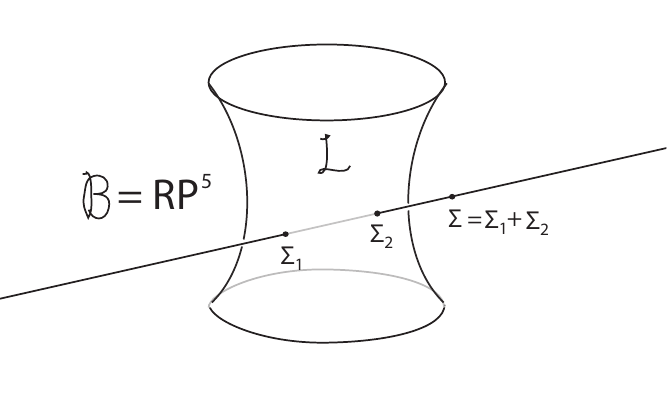}}} \hspace{.04in}  %.02in
{\setlength\fboxsep{0pt}\fbox{\includegraphics[width=\xyw\columnwidth,trim={6cm 0 2cm 0},clip]{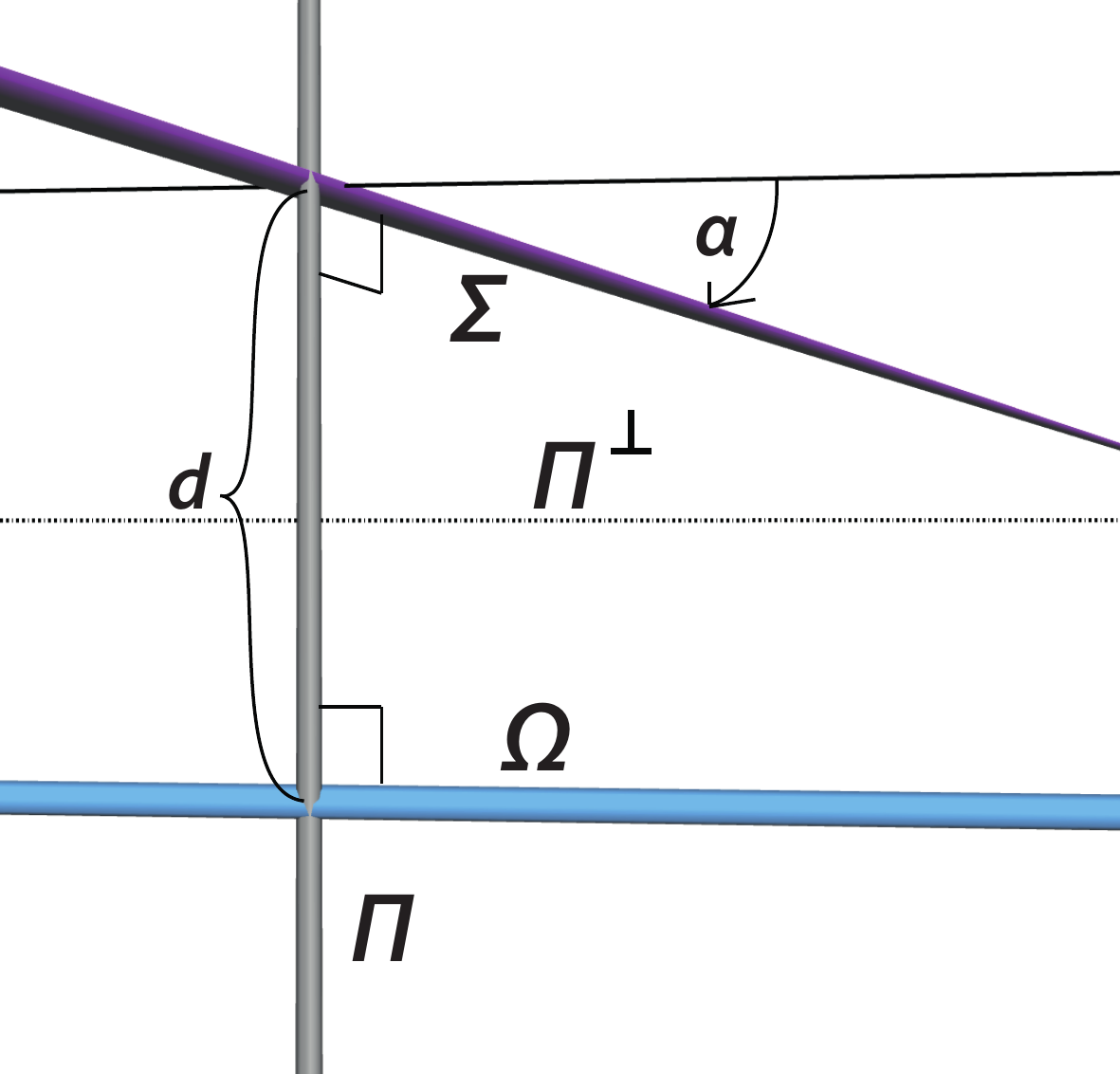}}}
\caption{\emph{Left:}The space of lines sits inside the space of 2-vectors as a quadric surface $\mathcal{L}$. \emph{Right}: Product of two skew lines $\velo$ and $\sigo$ involving the common normals $\momo$ (euclidean) and $\momo^\perp$ (ideal).}
\label{fig:complexspace}
\end{figure}

%We apply what we've learned above to calculating the geometric product of two lines.
Here  we present an account of the geometric product of two euclidean lines. 
Justifications for the claims made can be found in the subsequent sections. 
Let the two lines be $\velo$ and $\sigo$.  Assume they are euclidean and normalized, i.e., $\velo \wedge \sigo \neq 0$ and $\velo^2 = \sigo^2 = -1$. Two euclidean lines determine in general a unique third euclidean line that is perpendicular to both, call it $\momo$.   Consult \Fig{fig:complexspace}, right. %It plays an important role in the geometric product $ \velo\sigo$. % = \grade{\velo\sigo}{0}+\grade{\velo\sigo}{2}+\grade{\velo\sigo}{4} \] consists of a scalar, a 2-vector, and a pseudoscalar. 
$\velo\sigo$ consists of 3 parts, of grades 0, 2, and 4:
\begin{align*}
\velo\sigo &= \grade{\velo\sigo}{0} + \grade{\velo\sigo}{2} + \grade{\velo\sigo}{4} \\
&= \velo \cdot \sigo + \velo \times \sigo + \velo \wedge \sigo \\
&= \cos{\alpha} + (\sin{\alpha}\momo + d \cos{\alpha}\momo^\perp) + d \sin{\alpha} \eye 
\end{align*}
 %Here the reader should recognize $\momo := \widehat{\left(\velo \times \sigo\right)}$ as the axis of $\velo \times \sigo$. 
 $\alpha$ is the angle between $\velo$ and $\sigo$, viewed along the common normal $\momo$; $d$ is the distance between the two lines measured along $\momo$ (0 when the lines intersect). $d \sin{\alpha}$ is the volume of a tetrahedron determined by unit length segments on $\velo$ and $\sigo$. Finally, $ \velo \times \sigo$ is a weighted sum of  $\momo$ and $\momo^\perp$. The appearance of $\momo^\perp$ is not so surprising,  as it is also a \quot{common normal} to $\velo$ and $\sigo$, but as an ideal line, is easily overlooked. 
 
Does $\velo \sigo$ have a geometric meaning? %\footnote{Experienced GA users know the answer is \quot{Of course!}.} 
Consider sandwich operators with bivectors, that is, products of the form $\velo\vec{X}\widetilde{\velo} $ %\footnote{We use the fact that for bivectors $\widetilde{\velo} = -\velo$.} 
for simple euclidean $\velo$. Such a product is called a \emph{turn} since it is a half-turn around the axis $\velo$ (see below, Sect. \ref{sec:simbivrandt}).  And, in turn, the turns generate the full group $\Eucgd{3}$ of direct euclidean isometries (\cite{study91}. A little reflection shows that the  composition of the two turns $\velo\sigo$ will be a screw motion that rotates around the common normal $\momo$ by $2\alpha$ while translating by $2d$ in the direction from $\sigo$ to $\velo$ (the translation is a \quot{rotation} around $\momo^\perp$).   This is analogous to the product of two reflections meeting at angle $\alpha$ discussed above in \Sec{sec:isom}.

Analogous to the 2D case, we can easily calculate the \MOTOR that carries $\sigo$ exactly onto $\velo$. This is given by $\sqrt{\velo\sigo} = \widehat{\left(\cfrac{1+\velo \sigo}{2}\right)}$. 

\myboldhead{The case of two intersecting lines}
If the two lines are not skew, they have a common point and a common plane and are linearly dependent: $\velo \wedge \sigo = 0$.   The common plane is given by $ (\velo\wedge \e{0}) \vee \sigo$;  the common point $\vec{P}$ by $ \vec{P} = ((\momo \wedge \e{0}) \vee \velo)\wedge \sigo$ where $\momo = \grade{\velo\sigo}{2}$ is the common normal.  

 We turn now to a rather detailed discussion of the structure and behavior of non-simple bivectors. Readers with limited time and interest in such a treatment are encouraged to skip ahead to Sect. \ref{sec:ad}.
 
\subsubsection{The axis of a bivector}
\label{sec:normbiv}
In 2D we used the following formula to normalize a 1- or 2-vector: $$\widehat{\vec{X}} = \cfrac{\vec{X}}{\sqrt{|\vec{X}^2|}}$$ This was made easy since $\vec{X}^2$ in all cases was a real number.
A non-simple euclidean bivector satisfies $\NSBV^2 = \NSBV\cdot\NSBV + \NSBV\wedge\NSBV = s + p\eye$ with $s, p \neq 0$. Since it's euclidean, $s<0$. We saw above in \Sec{sec:l2v2d} that $p \neq 0 \iff \NSBV$ is non-simple. A number of the form $s + p\eye$ for $s, t \in \mathbb{R}$ is called a \emph{dual} number. If we want to normalize a bivector using a formula like the one above, then we have to be able to find  the square root of dual numbers. % and also to invert the result. 

For a dual number $d = s+p\eye, s > 0, p\neq0$, define 
%the inverse:
%({s+p\eye})^{-1}=\cfrac1a - \cfrac{b}{a^2}\PSS$$
%and 
the square root $$\sqrt{s+p\eye}=\sqrt{s} +\cfrac{p}{2\sqrt{s}}\eye$$
and verify that it deserves the name.
Define $$ \|\NSBV\| = u + v\eye := \sqrt{-(\NSBV\cdot\NSBV + \NSBV\wedge\NSBV})$$
Then $\widehat{\NSBV} :=  \|\NSBV\|^{-1}\NSBV$ and $\widehat{\NSBV}^2 = -1$.
We  write $\NSBV$ in terms of $\widehat{\NSBV}$: 
\begin{align*}
\NSBV &= \|\NSBV\|\widehat{\NSBV} = (u+v\eye)\widehat{\NSBV} \\
&= u \widehat{\NSBV} + v \widehat{\NSBV}^\perp
\end{align*}

That is, we have decomposed the non-simple bivector as the sum of a euclidean line $\widehat{\NSBV}$ and its orthogonal line $\widehat{\NSBV}^\perp$.\footnote{${\NSBV}^\perp$ can be thought of as the ideal line consisting of all the directions perpendicular to $\NSBV$. For example, if $\NSBV$ is vertical, then $\NSBV^\perp$ is the horizon line.} It is easy to verify that $\widehat{\NSBV} \times \widehat{\NSBV}^\perp = 0$ so that the two bivectors commute.  We now apply this to computing the exponential of a bivector that we need below in Sect. \ref{sec:logmot}.

\myboldhead{Remarks on the axis pair} Note that $\widehat{\NSBV}$  is not a normalized vector in the traditional  sense since it is no longer projectively equivalent to the original bivector. Indeed, it arises by multiplying the latter by a dual number, not a real number. The euclidean axis has a special geometric significance that will prove to be very useful in the analysis of \MOTORS that follows. 

\myboldhead{Terminology} We call $\widehat{\NSBV}$ the \emph{euclidean axis} and $\widehat{\NSBV}^\perp$ the \emph{ideal axis}  of the non-simple bivector $\NSBV$. Together they form the \emph{axis pair} of the bivector.  %Define $\vec{m} := e^\NSBV$.   %Each deserves the name since they are the only two lines invariant under the screw motion. 
The euclidean axis however is primary since the ideal axis can be obtained from it by polarizing: $\widehat{\NSBV}\eye$, but not vice-versa.

\subsubsection{The exponential of a non-simple bivector}
The existence of an axis pair for an non-simple bivector is the key to understanding its exponential.
Applying the decomposition of $\NSBV = u \widehat{\NSBV} + v \widehat{\NSBV}^\perp$ as an axis pair we can write $$
e^{\NSBV}= e^{v \widehat{
\NSBV} + v \widehat{\NSBV}^\perp}$$
Since $\NSBV$ and $\NSBV^\perp$ commute (see above), the exponent of the sum is the product of the exponents: $$e^{\NSBV} = e^{u \widehat{
\NSBV}+v \widehat{\NSBV}^\perp} = e^{u \widehat{
\NSBV}}e^{v \widehat{\NSBV}^\perp} = e^{v \widehat{\NSBV}^\perp}e^{u \widehat{
\NSBV}}$$
where the third equality also follows from commutivity.  We can then apply what we know about the exponential of  simple bivectors from Sect. \ref{sec:l2v2d} to obtain:
\begin{align}
e^{\NSBV} &= (\cos{u} + \sin{u}\widehat{\NSBV})(1+v\widehat{\NSBV}\eye) \\
    &= \cos{u}+\sin{u}\widehat{\NSBV} +v\cos{u}\widehat{\NSBV}\eye - v\sin{u}\eye\\
    &= (\cos{u}-v\sin{u}\eye) + (\sin{u}+v\cos{u}\eye)\widehat{\NSBV}\label{eq:motordecomp} %\\
 %   &=: (s_1+p_1\eye) + (s+p\eye)\widehat{\NSBV}
\end{align}
We will apply this formula below when we compute the logarithm of a motor $\vec{m}$.

\subsubsection{Bivectors and motions}
\label{sec:simbivrandt}
\myboldhead{Simple bivectors, simple motions}
We saw in the discussion of 2D PGA that bivectors (points) play an important role in implementing euclidean motions: every rotation (translation) can be implemented by exponentiating a euclidean (ideal) point to obtain a \MOTOR. This was a consequence of the fact that sandwiches with 1-vectors (lines) implement reflections and even compositions of reflections generate all direct isometries. The same stays true in 3D: a sandwich with a plane (1-vector) implements the reflection in that plane. Composing two such reflections generates a direct motion (rotation/translation around the intersection line) that is represented by a 3D \MOTOR, completely analogous to the 2D case. 
Using the formula for the exponential of a simple bivector from Sect. \ref{sec:l2v2d}, we derive the formulas for the rotator around a simple euclidean bivector $\velo$ by angle $\alpha$: $e^{\frac{\alpha}{2}\velo} = \cos{\frac{\alpha}2} + \sin{\frac{\alpha}2}\velo$. The translator $e^{\frac{d}{2}\velo_\infty} = 1 + \frac{d}{2}\velo_\infty$ produces a translation of length $d$ perpendicular to the ideal line $\velo_\infty$. 

\myboldhead{Non-simple bivectors, screw motions} 
But there are other possibilities in 3D for direct motions than just rotations and translations. The generic motion is a \emph{screw motion} that composes a rotation around a 3D line, called its \emph{axis}, with a translation in the direction of the line. To be precise, the axis of a screw motion is the unique euclidean line fixed by the screw motion. The motion is further characterized by its \emph{pitch}, which is the ratio of the angle turned (in radians) to the distance translated. % We state without proving: the \MOTOR for a screw motion arises from exponentiating a non-simple bivector, and vice-versa. We look in more detail at this relation in the next section.

\subsubsection{The motor group} 
\begin{figure}[t]
   \centering
   \def\xyy{.8}
{\setlength\fboxsep{0pt}\fbox{\includegraphics[width=\xyy\columnwidth]{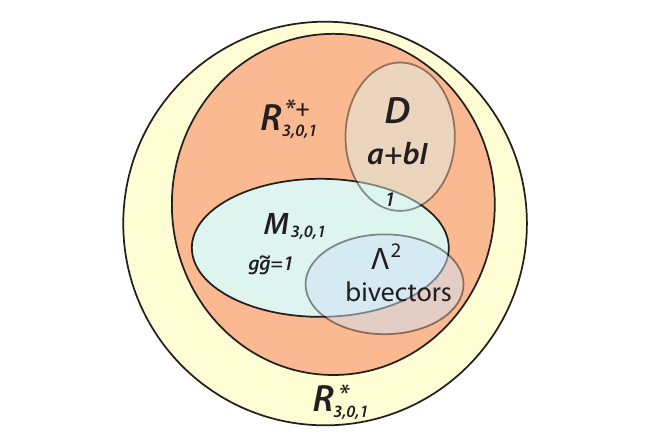}}}
\caption{Venn diagram showing the inclusion relationships of the algebra $\pdclal{3}{0}{1}$, its even subalgebra $\pdclplus{3}{0}{1}$, the dual numbers $\mathbb{D}$, the motor group $\mogro{3}$, and the bivectors $\bigwedge^2$.}
\label{fig:venn}
\end{figure}

Every direct isometry is the result of composing an even number of reflections: hence such a \MOTOR lies in the even sub-subalgebra, consisting of elements of even grade and written $\pdclplus{3}{0}{1}$.  An element  $\vec{m}$ of  the even sub-algebra  is a \MOTOR if it satisfies $\vec{m}\widetilde{\vec{m}} = 1$; such elements form a group, written  $\mogro{3} \subset \pdclplus{3}{0}{1}$ called the \emph{motor group}. The motor group is more generally called the \emph{Spin} group of the geometric algebra. It's a 2:1 cover of the direct Euclidean group $\Eucgd{3}$ since $\vec{m}$ and $-\vec{m}$ yield the same isometry. Elements of the form $e^{\velo}, \velo \in \bigwedge^2$ are in $\mogro{3}$ since $\widetilde{e^\velo} = e^{\widetilde{\velo}} = e^{-\velo}$ and $e^\velo e^{-\velo} = 1$.  \Fig{fig:venn} illustrates the various inclusions involved among the algebra, the even algebra, the motor group, the dual numbers, and the bivectors. For example, a normalized simple bivector is also a motor: used as a sandwich, it produces a rotation of $\pi$ radians around the line it represents.

\myboldhead{Calculating the logarithm of a \MOTOR}
\label{sec:logmot}
The \emph{logarithm} $\NSBV$ of a motor $\vec{m}$ is an algebra element that satisfies $\vec{m} = e^{\NSBV}$. We now show how to find such a logarithm. 
We know the normalized \MOTOR $\vec{m}$ contains only even-grade parts:
\begin{align}
    \vec{m} &= \langle \vec{m} \rangle_0 + \langle \vec{m} \rangle_2 + \langle \vec{m} \rangle_4 \\
    &= s_1 + \NSBV + p_1\eye \\
    &= (s_1 + p_1\eye) + (s_2 + p_2\eye)\widehat{\NSBV}\label{eq:motorwaxis}
\end{align}
 In the last line we have substituted  ${\NSBV} = \|\NSBV\|\widehat{\NSBV}$ (see above Sect. \ref{sec:normbiv}).
 Comparing coefficients in Eq. \ref{eq:motordecomp} and Eq. \ref{eq:motorwaxis} we see that we have an overdetermined system: from the four quantities $\{s_1, p_1, s_2, p_2\}$ we have to deduce the two parameters $\{u, v\}$. This leads to the following values for $u$ and $v$:

\vspace{-.2in}
\begin{align}
u:= \tan^{-1}(s_2,s_1), ~v:=\cfrac{p_2}{s_1}  &~~~~\text{for}~ s_1\neq 0 \\
u:= \tan^{-1}(-p_1,p_2), ~v:=\cfrac{-p_1}{s_2} &~~~~\text{otherwise} \label{eq:turnmotor}
\end{align}
Note that either $s_1 \neq 0$ or $s_2 \neq 0$ since otherwise $\vec{m}^2 = 0$.
  Then $$ e^{(u+v\eye)\widehat{\NSBV}} = \vec{m}$$
  $(u+v\PSS)\widehat{\NSBV}$ is the \emph{logarithm} of $\bf{m}$. It is unique except for adding multiples of $2\pi$ to $u$.  The pitch of the screw motion is given by the proportion $v:u$. The logarithm shows that $\vec{m}$ can be decomposed as
  $$ e^{u\widehat{\NSBV}} e^{v\widehat{\NSBV}^\perp}$$
  that is, the composition (in either order) of a rotation through angle $2u$ around the axis $\widehat{\NSBV}$ and a translation of distance $2v$ around the polar axis $\widehat{\NSBV}^\perp$. 
  
 \hspace{-.15in}\textbf{Axis of bivector or axis of screw motion?} These formulas make clear that the two uses of \emph{axis} that we have encountered are actually the same. The axis pair of a screw motion (considered as the unique pair of invariant lines) \textbf{is} the axis pair of its bivector part.
  
\myboldhead{Connection to Lie groups and Lie algebras} By establishing the logarithm function (unique up to multiples of $2\pi$), we have established that the exponential map $\exp: \bigwedge^2 \rightarrow \mogro{3}$ is invertible. Hence we are justified in identifying $\mogro{3}$ as a Lie group and $\bigwedge^2$ as its Lie algebra, and  can apply the well-developed Lie theory to this aspect of PGA. %The numerical advantages over traditional matrix approaches to the Lie group $\Eucg{3}$ become clear: in PGA $\mogro{3}$ is a 6D submanifold of the 8D vector space $\pdclplus{3}{0}{1}$ while the matrix group is a 6D submanifold of  the 12D affine group Aff$(3,\mathbb{R})$. Normalizing an element $\mogro{3}$ to satisfy $\vec{m}\widetilde{\vec{m}}=1$ brings it back onto the group submanifold; keeping a matrix on the euclidean submanifold of Aff$(3,\mathbb{R})$ is a much more challenging task. 

This concludes our introductory treatment of the geometric product in \pdclal{3}{0}{1}. We turn now to its formulation of rigid body mechanics, whose essential features were already known to Pl\"{u}cker and Klein in terms of 3D line geometry (\cite{ziegler}).

\subsection{Kinematics and Mechanics in $\pdclal{3}{0}{1}$}

%Hopefully the preceding remarks make clear the central role that bivectors play in 3D PGA. %Essentially they form the \emph{Lie algebra} of $\Eucgd{3}$, the oriented Lie group of euclidean space $\Euc{3}$.   
Here we give a very abbreviated overview of the treatment of kinematics and rigid body mechanics in PGA in the form of a bullet list.
%Symbolically,
%\begin{compactitem}
% \item $\pdclplus{3}{0}{1} := \{x \in \pdclal{3}{0}{1}~~|~~\grade{x}{1}=\grade{x}{3}=0\}$
% \item $\mogro{3} := \{ \vec{g}\in\pdclplus{3}{0}{1}~~ |~~ \vec{g}\widehat{\vec{g}} = 1\}$
% \item $\exp: \bigwedge^2 \rightarrow \mogro{3} $
% \item $\log:  \mogro{3} \rightarrow \bigwedge^2 $
%\end{compactitem}

\newcommand{\InTen}{\text{\fontfamily{qhv}\selectfont{A}}}
 
\begin{compactenum}
\item Kinematics deal with continuous motions in $\Euc{3}$, that is, paths in $\mogro{3}$. Let $\vec{g}(t)$ be such a path describing the motion of a rigid body. 
\item There are two coordinate systems for a body moving with $\vec{g}$: body and space. An entity $\vec{x}$ can be represented in either: $\vec{X}_c$/$\vec{X}_s$ represents body/space frame.
\item The \emph{velocity in the body}  $\velo_c := \widetilde{\vec{g}}\dot{\vec{g}}$; in space, $\velo_s := \dot{\vec{g}}\widetilde{\vec{g}}$. $\velo_c$ and $\velo_s$ are bivectors.
\item  $\InTen$, the inertia tensor  of the body, is a 6D symmetric bilinear form  $$J^{-1}(\InTen(\velo_c)) = \momo_c~~~\text{and}~~~\velo_c = \InTen^{-1}(J(\momo_c))$$ where $\momo_c$ is the momentum in the body and $J$ is Poincar\'{e} duality map.\footnote{$\InTen$ actually maps to the dual exterior algebra: $\InTen: \bigwedge^{2} \rightarrow \bigwedge^{2*}$, we compose it with the duality map $J$ to bring the result back to $\pdclal{3}{0}{1}$.}
\item The kinetic energy $E$ satisfies $E=\velo_c\wedge\momo_c$.
\item Let $\vec{\Phi}_c$ represent the external forces in body frame. Then $\dot{E} = -2\vec{\Phi}_c \vee \velo_c$.
\item The work done can be computed as $$w(t) = E(t)-E(0) =\int_{0}^t{\dot{E}ds} = -2\int_0^t{\vec{\Phi}_c \vee \velo_c ds}$$
\item The Euler equations of motion for the of the motion free top one obtains the following Euler equations of motion:
\begin{align}\label{eqn:eqnmot}
\dot{\vec{g}} &= \vec{g}\velo_c\\
\dot{\velo}_c &= \inert^{-1}(\vec{\Phi}_c + 2\inert(\velo_c) \times \velo_c )
\end{align}
\end{compactenum}  

\myboldhead{Theoretical discussion} The traditional separation of both velocities and momenta into linear and angular parts disappears completely in PGA, further evidence of its polymorphicity. %Velocities and accelerations correspond to plane-wise lines, or \emph{axes}, while momenta and forces correspond to point-wise ones, or \emph{spears}. 
The special, awkward role assigned to the coordinate origin in the calculation of angular quantities (moment of a force, etc.) along with many mysterious cross-products likewise disappear. 

What remains are unified velocity and momentum bivectors that represent geometric entities with intuitive significance. We have already above seen how the velocity can be decomposed into an axis pair that completely describes the instantaneous motion at time $t$. Similar remarks are valid also for momentum and force bivectors. We focus on forces now but everything we say also applies to momenta. The simple bivector representing a simple force \textbf{is} the line carrying the force; the weight of the bivector is the intensity of the force.  A force couple is a simple force carried by an ideal line (like a translation is a \quot{rotation} around an ideal line).  Systems of forces that do not reduce to a simple bivector can be decomposed into an axis pair exactly as the velocity bivector above, combining a simple force with an orthogonal force couple. This axis pair has to be interpreted however in a dynamical, not kinematical, setting. Further discussion lies outside the scope of these notes. %We leave that as an exercise in dual thinking for the reader.% But they have a reversed significance. 
%[Hint: A force-couple always gives rise to rotational movement around the center of gravity, while a simple force directed through the center of gravity always gives rise to translational motion. All forces between these two extremes give rise to motions between these two extremes, mediated by the inertia tensor.]

\myboldhead{Practical discussion} The above Euler equations equations behave particularly well numerically: the solution space has 12 dimensions (the isometry group is 6D and the momentum space (bivectors) also) while the integration space has 14 dimensions ($\pdclplus{3}{0}{1}$ has dimension 8 and the space of bivectors has 6).  Normalizing the computed \MOTOR $\vec{g}$ brings one directly back to the solution space.  In traditional matrix approaches as well as the CGA approach  (\cite{lasenby2011}), the co-dimension of the solution space within the integration space is much higher and leads typically to the use Lagrange multipliers or similar methods to maintain accuracy.  This advantage over VLAAG and CGA is typical of the PGA approach for many related computing challenges.

See \cite{gunn2011} or \cite{gunnthesis}, Ch. 9, for details on rigid body mechanics in $\pdclal{3}{0}{1}$.  For a compact, playable PGA implementation see \cite{ganjacs}.  

%To obtain the motion of the rigid body as a path $\vec{g}$ in the Lie group $SO(3)$, the rotations of $\R{3}$, one can then integrate $\dot{\vec{g}}$ using the relation $\dot{\vec{g}} = \vec{g} \vec{V}_{c}$. 
%Space limitations prohibit a deeper study of this fascinating topic.  
 % Proceeding as in  Example \ref{sec:3dscrew} above,  and setting $\vec{g} := \velo\sigo$, one can show that $\vec{g}$ is a \MOTOR with axis $\momo$  whose sandwich operator rotates by $-2\alpha$ and translates by $-2p$, that is, is \quot{twice} the screw motion that moves $\sigo$ to $\velo$. 
 \myvspace{-.15in}
 
 \section{Automatic differentiation}
\label{sec:ad}
\cite{hessob87} introduces the term \quot{geometric calculus} for the application of calculus to geometric algebras, and shows that it offers an attractive unifying framework in which many diverse results of calculus and differential geometry can be integrated.  While a treatment of geometric calculus lies outside the scope of these notes, we want to present a related result to give a flavor of what is possible in this direction. %by showing how to do automatic differentiation of real functions in $\pdclal{n}{0}{1}$.  

We have already met above, in \Sec{sec:normbiv}, the 2-dimensional sub-algebra of $\pdclal{n}{0}{1}$ consisting of scalars and pseudoscalars known as the dual numbers.  It can be abstractly characterized by the fact that $1^2=1$ while $\eye^2=0$.  Already Eduard Study, the inventor of dual numbers, realized that they can be used to do automatic differentiation  (\cite{study03}, Part II, \S 23).
A modern reference describes how \cite{wikiAD}:
\begin{quote}
Forward mode automatic differentiation is accomplished by augmenting the algebra of real numbers and obtaining a new arithmetic. An additional component is added to every number which will represent the derivative of a function at the number, and all arithmetic operators are extended for the augmented algebra. The augmented algebra is the algebra of dual numbers.
\end{quote}
This extension can be obtained by beginning with the monomials. Given $p_k(x)=x^k$, define \[p_k(x+y\eye) := (x+y\eye)^k = x^k+nx^{n-1}y\eye\] All higher terms disappear since $\eye^2=0$.  Setting $y=1$ we obtain \[{p}_k(x+\eye) = p_k(x)+\dot{p}_k(x)\eye\] That is, the scalar part is the original polynomial and the pseudoscalar, or dual, part is its derivative. In general if $u$ is a function $u(x)$ with derivative $\dot{u}$, then \[p_k(u+\dot{u}\eye) = p_k(u) + \dot{p}_k(u)\eye\] Thus, the coefficient of $\eye$ tracks the derivative of $p_k$.  Extend these definitions to polynomials by additivity in the obvious way.  Since the polynomials are dense in the analytic functions, the same \quot{dualization} can be extended to them and one obtains in this way robust, exact automatic differentiation.  One can also handle multivariable functions of $n$ variables, using the $(n)$ ideal $n-$vectors $E_i$ for $i>0$ (representing the ideal directions of euclidean $n$-space) as the nilpotent elements instead of $\eye$. %, since $E_i^2=0$ when $i\neq0$. 
For a live JavaScript demo see \cite{ganjacs}.

 \section{Implementation issues}
 \label{sec:impl}
%This concludes our introductory tour of the PGA toolkit from the user's perspective. 
 Our description would be incomplete without discussion of the practical issues of implementation.  This has  been the focus of much work and there exists a well-developed theory and practice for general geometric algebra implementations to maintain performance parity with traditional approaches.  See \cite{hildebrand13}.  PGA presents no  special challenges in this regard; in fact, it demonstrates clear advantages over other geometric algebra approaches to euclidean geometry in this regard (\cite{gunn2017a}). %Although most general-purpose software GA packages do not support degenerate metrics, such metrics do not present intrinsic numerical difficulties.  
For a full implementation of PGA in JavaScript ES6 see Steven De Keninck's ganja.js project on GitHub \cite{ganja} and the interactive example set at \cite{ganjacs}.
%The author has also implemented (non-optimized) euclidean (and non-euclidean) PGA in a variety of software frameworks, including Mathematica, Java, and Clojure -- without encountering noteworthy obstacles.  A 2D sample demo running in the browser can be found at \cite{pga2ddemo16}.
 \begin{table}
\begin{center}
\small%\scriptsize
\setlength{\extrarowheight}{4pt}
\def\xyz{3pt}
\begin{tabular}{| p{.45\columnwidth} | p{.45\columnwidth} |}
\hline
%\textbf{Projective geometric algebra} & \textbf{Vector and linear alg + analytic geom.} \\
\textbf{PGA} & \textbf{VLAAG} \\
\hline
\hline
Unified representation for points, lines, and planes based on a graded exterior algebra; all are \quot{equal citizens} in the algebra.
&
The basic primitives are  points and vectors and all other primitives are built up from these. For example, lines in 3D sometimes  parametric, sometimes w/ Pl\"{u}cker coordinates. \\[\xyz]
\hline
Projective exterior algebra provides robust meet and join operators that deal correctly with parallel entities.
&
Meet and join operators only possible when homogeneous coordinates are used, even then tend to be \emph{ad hoc} since points have distinguished role and ideal elements  rarely integrated. \\[\xyz]
\hline
Unified, high-level treatment of euclidean (\quot{finite}) and ideal (\quot{infinite}) elements of all dimensions. Unifies e.g. rotations and translations, simple forces and force couples.
&
Points (euclidean) and vectors (ideal) have their own rules, user must keep track of which is which; no higher-dimensional analogues for lines and planes.\\[\xyz]
%\hline
%Universal expressions: formulas handle varying types of arguments seamlessly.
%&
%Formulas tend to be \emph{ad hoc} with many special cases for parallel, non-parallel; separate ones for points, lines, planes, etc.\\
\hline
Unified representation of isometries based on sandwich operators which act uniformly on points, lines, and planes.
&
Matrix representation for isometries has different forms for points, lines, and planes.\\[\xyz]
\hline
Same representation for operator and operand: $\vec{m}$ is the  plane as well as the reflection in the plane.
&
Matrix representation for reflection in $\vec{m}$ is different from the vector representing the plane. \\[\xyz]
\hline
Compact, universal expressive formulas and constructions based on geometric product (see Tables \ref{tab:pga2d}, \ref{tab:pga3d}, and \ref{tab:pga3d2}) valid for wide range of argument types and dimensions.
&
Formulas and constructions are \emph{ad hoc}, complicated, many special cases, separate formulas for points/lines/planes, for example, compare   \cite{gg90}. \\[\xyz]
\hline
Well-developed theory of implementation optimizations to maintain performance parity.
&
Highly-optimized libraries, direct mapping to current GPU design. \\[\xyz]
\hline
Automatic differentiation of real-valued functions using dual numbers. %You can add \textbf{and} multiply points and vectors, lines, and planes.
&
Numerical differentiation \\%You can only add points and vectors. \\[\xyz]
%\hline
%Faithful representation of Lie algebra and Lie group of $\Euc{n}$ is geometrically intuitive and numerically optimal.
%&
%Representation of Lie algebra and Lie group of $\Euc{n}$ as subgroups of matrix group $GL(n, \mathbb{R})$ is not geometrically intuitive and is numerically problematical. \\
\hline
\end{tabular}
\myvspace{.1in}
\caption{A comparison of PGA with the standard VLAAG approach. }
\label{tab:comp}
\end{center}
\end{table}

%\myvspace{-.15in}  
\section{Comparison}
\label{sec:comp}
Table \ref{tab:comp} encapsulates the foregoing results in a feature-by-feature comparison with the standard (VLAAG) approach.   It establishes that PGA fulfills all the features on our wish-list in Sec. \ref{sec:wishlist}, while the standard approach offers almost none of them.  (For a proof that PGA is coordinate-free, see the Appendix in \cite{gunn2017b}.) 

\subsection{Conceptual differences}
How can we characterize conceptually the difference of the two approaches leading to such divergent results? \begin{compactitem}
\item First and foremost:  VLAAG is \emph{point-centric}: other geometric primitives of VLAAG such as lines and planes are built up out of points and vectors. PGA on the other hand is \emph{primitive-neutral}: the exterior algebra(s) at its base provide \emph{native} support for the subspace lattice of points, lines and planes (with respect to both join and meet operators). 
\item Secondly, the projective basis of PGA allows it to deal with points and vectors in a unified way: vectors are just ideal points, and in general, the ideal elements play a crucial role in PGA to integrate parallelism, which typically has to be treated separately in VLAAG.  The existence of the ideal norm in PGA goes beyond the purely projective treatment of incidence, producing polymorphic metric formulas that, for example, correctly handle two intersecting lines whether they intersect or are parallel (see above \Sec{sec:prodpr}).  
\item The representation of isometries using sandwich operators generated by reflections in planes (or lines in 2D) can be understood as a special case of this \quot{compact polymorphicity}: the sandwich operator $\vec{g}\vec{X}\tilde{\vec{g}}$ works no matter what $X$ is, the same representation works whether it appears as operator or as operand, and rotations and translations are handled in the same way. 
\end{compactitem}

\subsection{The expressiveness of PGA}
All these conceptual differences contribute to the astounding richness of the PGA syntax in comparison to VLAAG, a richness exemplified in the formulas of tables \ref{tab:pga2d}, \ref{tab:pga3d}, and \ref{tab:pga3d2}. Each of the conceptual differences in the above list can be thought of as a set of distinctions that are embedded in a unified form within the PGA syntax: points/lines/planes, euclidean/ideal, operator/operand, etc. This leads to having  many more basic expressions for modeling geometry than in VLAAG, and they all combine meaningfully with each other.  To the best of our knowledge these formula collections establish PGA as the \quot{world champion} among all existing frameworks for euclidean geometry with respect to compactness, completeness, and polymorphicity.  Compare  \cite{gg90} for selected VLAAG analogs. Readers who know of a competitive formula collection are urged to drop the author an email with a pointer to it.  We also expect that there are more formulas waiting to be discovered (after all, here we've only considered the 2-way products and a small subset of the 3-way products).

\myboldhead{Non-euclidean metrics} The expressiveness of PGA takes on a wider dimension when we recall that PGA is actually a family of geometric algebras. We have focused attention here on euclidean PGA. The other members of the family model non-euclidean spaces, notably, spherical and hyperbolic space. Simply by specifying a different value for $\e{0}^2$, these other PGA's can also be accessed. Many of the formulas and constructions included in these notes can be applied unchanged to these other metric spaces (for example, the treatment of rigid body mechanics included above is metric-neutral in this sense, as are many of the constructions in the tables) or with minor and instructive differences with respect to EPGA. The example of spherical geometry in Sect. \ref{sec:sphgeom} illustrates the power of this approach. \cite{gunnthesis} develops all its PGA results in the setting of all three classical metrics.

 \begin{figure}[t]
   \centering
   \def\xyy{.8}
{\setlength\fboxsep{0pt}\fbox{\includegraphics[width=\xyy\columnwidth]{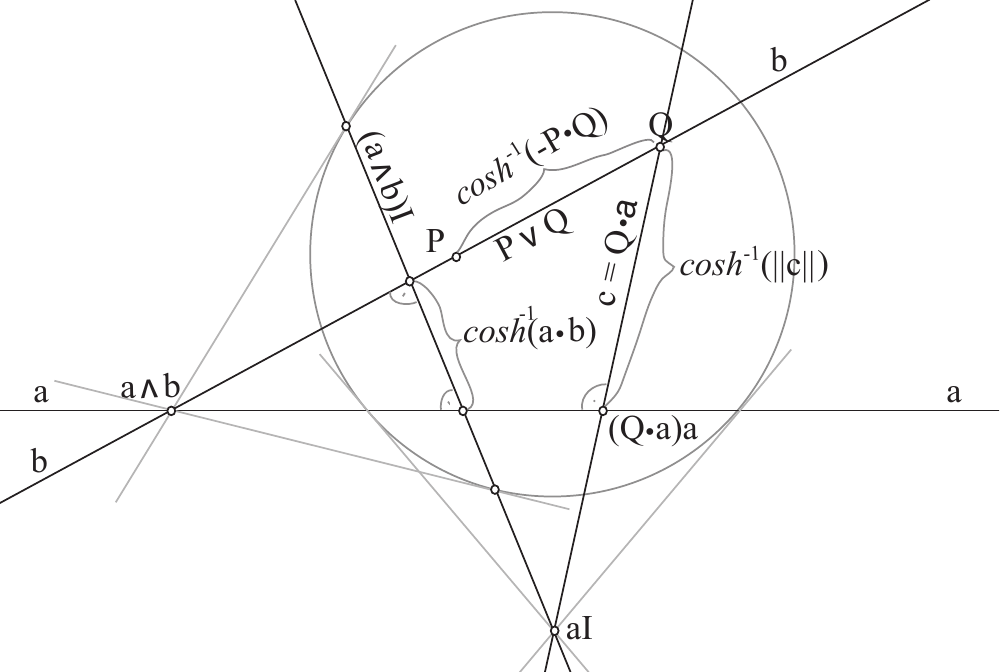}}}
\caption{Doing geometry in the hyperbolic plane using the PGA $\pdclal{2}{1}{0}$.}
\label{fig:hyperbolic}
\end{figure}

\subsection{The universality of PGA}
The previous section highlighted the structural advantages enjoyed by PGA over VLAAG.   We strengthen this argument in this section by showing that alternate approaches to euclidean geometry are largely present already in PGA as parts of the whole.

%\begin{compactitem}
\subsubsection{Vector algebra} The previous section has already suggested that VLAAG can be seen less as a direct competitor to PGA than as a restricted subset.  Indeed, restricting attention to the vector space of $n$-vectors (sometimes written $\bigwedge^n$) in PGA  essentially yields standard vector algebra. Define the \quot{points} to be euclidean $n$-vectors ($\vec{P}^2 \neq 0$) and \quot{vectors} to be ideal $n$-vectors ($\vec{P}^2 = 0$).  All the rules of vector algebra can be then derived using the  vector space structure of $\bigwedge^n$ equipped with the standard and ideal PGA norms (assuming normalized arguments as usual). The absence of the geometric product in this context makes clear why VLAAG is so much \quot{smaller} than PGA.  %Of course the geometric product can't be used  since $\bigwedge^n$ isn't closed under multiplication, only addition. That explains why the vector algebra only allows addition but not multiplication of elements. 

\myboldhead{Unified $\R{n}$ and $\Euc{n}$} This embedding of vector algebra in PGA also comes with a nice geometric intuition absent in traditional vector algebra: the vectors make up the ideal plane bounding the euclidean space of points, \emph{ i. e.},  points and vectors make up a connected, unified space (topologically equivalent to projective space $\RP{n}$).  Furthermore, intuitions developed in vector algebra such as \quot{Adding a vector to a point translates the point.} have natural extensions in PGA: adding an ideal line (plane) to a euclidean line (plane) translates the line (plane) parallel to itself\footnote{Whereby the two lines must be co-planar.}. Such patterns are legion.

\subsubsection{Linear algebra and analytic geometry} Note that PGA is fully compatible with the use of linear algebra. A linear map on the 1-vectors has  induces linear maps on all grades of the algebra that can be automatically computed and applied. The big difference to VLAAG is that linear algebra no longer is needed to implement euclidean motions -- a role for which it is not particularly well-suited. We envision the development of an analytic geometry based on the full extent of PGA, not just on the small subset present in VLAAG, and would have at its disposal the geometric calculus sketched in \Sec{sec:ad}.  Traditional analytic geometry would make up a small subset of this extended analytic geometry, like vector algebra makes up a small part of PGA proper.

\subsubsection{Exterior algebra} 
The underlying graded algebra structure of PGA can be thought of as being inherited from the exterior algebra. The wedge product is just the highest grade part of the geometric product, and implements the meet operator in PGA. The join operator is available from the dual exterior algebra via Poincar\'{e} duality (see Sect. \ref{sec:pea}).

\subsubsection{Quaternions and dual quaternions} \label{sec:qadq}
Many  aspects of PGA are present in embryonic form in quaternions and dual quaternions, but they only find their full expression and utility when embedded in the full algebra PGA. Indeed, the quaternion and dual quaternion algebras are isomorphically embedded in the even sub-algebra $\pdclplus{n}{0}{1}$ for $n \ge 3$. %(Whereby the mysterious nilpotent unit $\epsilon$  of the dual quaternions turns out to be the pseudoscalar $\eye$ of PGA.) 

\myboldhead{Integrated with points and planes} The advantage of the embedding in PGA are considerable. The full algebraic structure of PGA provides a much richer environment than these quaternion algebras alone. At the most basic level, quaternion and dual quaternion sandwich operators only model direct isometries; the embedding in PGA reveals how they arise naturally as even compositions of the reflections provided by sandwiches with 1-vectors.  Furthermore, few of the formulas in Tables \ref{tab:pga2d}, \ref{tab:pga3d}, and \ref{tab:pga3d2} are available in the quaternion algebras alone since the latter only have natural representations for primitives of even grade (essentially bivectors for $n=2$ and $n=3$). For example, in PGA, you can apply all sandwiches to geometric primitives of any grade.  In contrast, one of the \quot{mysteries} of contemporary dual quaternion usage is that there are separate \emph{ad hoc} representations for points, lines, and planes and slightly different forms of the sandwich operator for each in order to be able to apply euclidean direct isometries.  These eccentricities disappear when, as in PGA, there are native representations for points and planes, see \cite{gunn2017a}, \S 3.8.1. %We return to this embedding below in Sec. \ref{sec:l2v2d}. 

\myboldhead{Demystifying $\epsilon$ and the legacy of William Clifford} The PGA embedding clears up other otherwise mysterious aspects of current dual quaternion practice. Consider the  dual unit $\epsilon$ satisfying $\epsilon^2=0$. In the embedding map, it maps to the pseudoscalar $\eye$ of the algebra (for details see \cite{gunnthesis}, \S 7.6), perhaps tarnishing the mystique  but replacing it with a deeper understanding of the genesis of the dual quaternions.  It is also here worth noting that  William Clifford invented both dual quaternions (or biquaternions as he called them) and geometric algebra. That he did not also discover their happy reunion in EPGA is most likely due to his early death at age 34. At the time of his death neither the dual construction of the exterior algebra nor the degenerate metric (both necessary ingredients of eucidean PGA) had been introduced to the study of geometric algebras.

%\end{compactitem}
%For a detailed comparison with CGA see \cite{gunn2017a}, where it is shown that for classical flat geometry, PGA exhibits significant advantages.

 \section{Conclusion}
 We have established that euclidean PGA fulfills the developers' wish list with which we began these notes, offering numerous advantages over the existing VLAAG approach.
 The natural next question for interested developers is, what is involved in migrating to PGA from one of the alternatives discussed above?
In fact, the use of homogeneous coordinates and the inclusion of quaternions, dual quaternions, and exterior algebra in PGA means that many practitioners already familiar with these tools can expect a  gentle learning curve. Furthermore, the availability of a JavaScript implementation on GitHub (\cite{ganja}) and the existence of platforms such as Observable notebooks \cite{observable} means that interested users, equipped with the attached \quot{cheat sheets} for 2D and 3D PGA, can quickly get to work to prototype and share their applications. Readers who would like to deepen their understanding of the underlying mathematics are referred to the literature \cite{gunnFull2010}, \cite{gunnthesis},  \cite{gunn2017a}, \cite{gunn2017b}. We intend also to establish an on-line presence for PGA that will facilitate the exchange of information among the community of users, that we will announce at the course meeting in Los Angeles in July 2019.
  
%\section{Conclusion} We close with some reflections on the intimate relationship between mathematics and its applications. Naturally there are good reasons to focus on the primacy of the application, and the use of mathematics as a tool to achieve that end. And, indeed, users who persevere in mastering PGA can expect to reap the benefits established in the foregoing discussion.  Hence, existing applications in all the practical fields mentioned at the beginning of this article will, we believe, benefit in this way from exposure to PGA.   Equally exciting in our view is the prospect that PGA, as a new  way of thinking about euclidean geometry, may lead to innovative applications that were, so to speak, hidden from view using previous approaches.  So from whatever direction you are coming to the subject -- whether you are interested in improving an existing application or in learning a new approach to euclidean geometry -- PGA has plenty to offer to all.
%\section{Conclusion}
% We have in the process hopefully convinced you that PGA offers a practical and intuitive model of euclidean geometry that deserves further study of the literature and the attached cheat sheets, and coding explorations at \cite{ganjacs}. 

\section*{Acknowledgements} Thanks to Steven De Keninck for \texttt{ganja.js}, stimulating conversations and helpful feedback during the preparation of these notes.

 %; the relationship to CGA is also an open research topic (first enunciated in \cite{gunn2017a}). 
%That means that users of PGA have a much richer environment for problem formulation and description, that at the same time is amenable to efficient, practical implementation: a natural recipe for a successful modern geometric toolkit. 
%Developers and practitioners can focus on higher-level descriptions of their tasks confident that the practical performance will not suffer as a result: a promising recipe for PGA to become an essential component of the geometric toolkit of the $21^{st}$ century. 
% \ifthenelse{\equal{\isLong}{false}}{This is confirmed by comparing Table \ref{tab:pga3d} with formula collections such as \cite{gg90}, based on the standard approach.}{}

%\section{Conclusion}

\ifthenelse{\equal{\isLong}{false}}
{}
{
\section{Further reading}
Readers interested in knowing more about PGA can consult the following publications:
\small
\begin{compactitem}
\item \cite{gunn2011}: initial publication of euclidean PGA, including a self-contained treatment of  kinematics and rigid body mechanics,
\item  \cite{gunnFull2010}: an extended version of \cite{gunn2011},
\item  \cite{gunnthesis}: dissertation featuring a \MN setting for PGA including hyperbolic and elliptic space,
\item \cite{gunn2017a}: comparison of PGA  to CGA (another version of geometric algebra for euclidean geometry).
\item \cite{gunn2017b}: tutorial-like introduction to PGA applied to euclidean plane geometry,
%\item  \href{http://dgd.service.tu-berlin.de/wordpress/vismathws12/2014/02/20/829}{3-part blog posting by the presenter \quot{Introduction to projective geometric algebra}}
\end{compactitem}
\normalsize
 \section{Application areas}
 We want now to turn to consider its potential impact on three application areas: computer graphics, game engines, and 3D manifolds (a branch of topology) before wrapping up with a comparison with the standard approach to euclidean geometry.
Progress takes various forms.  It can be theoretical or practical.  But the dividing line is often not so clear as it appears.  

 \subsection{Computer graphics}
 
 Here we take \emph{computer graphics} in its wider sense, to include the ecosystem of theory and practice that includes animation, modeling, and rendering.  If one compares the formulas in Tables \ref{tab:pga3d} and \ref{tab:pga3d2} with \cite{gg90}, a similar but smaller collection based on the standard approach, the compactness and elegance of the PGA formulas is striking.\footnote{We would in fact like to issue a challenge to the computer graphics  community: is there another framework which provides these formulas so compactly (measured by symbol count) as PGA,  with the same degree of polymorphicity as PGA?  Candidates should be sent via email to the author.} Consequently, Euclidean PGA shows great promise to provide a unified geometric language for all fields related to computer graphics. Those who see a similar role for CGA are referred to \cite{gunn2017a} discusses this issue in detail.
Having a high-level polymorphic description of euclidean geometry has obvious advantages for problem-solving, since it shortens solutions and reduces the chances for errors.  Computer graphics, however, must also satisfy the constraints of real-time computation.  How does PGA behave with respect to practical implementation?   
  
 \section{Implementation issues}
 Fortunately there has been lots of work carried out to establish that, from a practical perspective, geometric algebras remain competitive with standard linear algebra approaches to the same problems.  See for example Ch. 22 of \cite{dfm07} where a comprehensive strategy is described for avoiding pitfalls in implementing geometric algebra structures. This includes keeping track of the non-zero grades of a multi-vector, automatically generating special purpose code for sub-products of the geometric product involving pure $k$-vectors, and judicious use of linear algebra methods behind the scenes in computationally-intensive loops.  Until GPU technology is based on GA models and not on VLAAG ones, GA toolkits will need to convert from their GA representation to VLAAG API's of the GPU (such as OpenGL or DirectX).  This will of course be hidden from the PGA user/developer.

 \subsection{3D Kinematics via 2-vectors}
\label{sec:kin}
 
 M\"{o}bius, in his investigations of euclidean statics \cite{moebius37}, discovered the linear line complex (which is an older name for the 2-vectors discussed here, naturally in a much less clear form) and thereby initiated the modern study of \emph{line geometry}.  Other investigators, notably Pl\"{u}cker and Felix Klein (\cite{klein71b}), pursued this direction of research and obtained an elegant and complete theory of kinematics and rigid body mechanics based on line geometry (\cite{ziegler}).  In fact, the history of geometric algebra itself is inextricably intertwined with this stream of research:  Clifford was a friend and admirer of Klein's, and his investigations of rigid body mechanics in elliptic space \cite{clifford74} using line geometry cannot be separated from his invention of dual quaternions (\cite{clifford73}) and geometric algebra (\cite{clifford78}).  Consult \cite{ziegler} for a detailed account of these fascinating chapter of $19^{th}$ century mathematics.
 
\subsubsection{The null system}

Since it plays a central role in the following discussion of kinematics and rigid body mechanics, we present M\"{o}bius's key results in \cite{moebius37} in more detail.  He was investigating 

 The basic object of kinematics is to describe the motion of bodies preserving euclidean (or some other metric) measurement.  Such a movement is called an \emph{isometry}.  We describe such a motion as a path $g(t) \subset \Eucgd{3}$, the group of orientation-preserving Euclidean isometries, %in the euclidean group $\Eucg{3}$  
 such that $g(0) = \mathbb{1}$, the identity.  We saw in the example above \Sec{sec:3dscrew} that such a path can be expressed in $\pdclal{3}{0}{1}$ as the \MOTOR $\vec{g}(t) := e^{t(\velo+p\velo^\perp)}$, representing a screw motion with axis $\velo$ and pitch $p$. And, in light of what was said regarding non-simple 2-vectors above, we see that $\velo+p\velo^\perp$ is just a general 2-vector  (with the exception of ideal 2-vectors represented as $p \velo^\perp$). So, the exponential of any 2-vector generates a euclidean motion, and any euclidean motion can be so generated.  Thus, the 2-vectors function as the \emph{Lie algebra} for the group $\Eucg{3}$, and that group itself is neatly imbedded in the even subalgebra of $\pdclal{3}{0}{1}$ (i. e., the subalgebra consisting of even-grade vectors: scalars, bivectors, and pseudoscalars).\footnote{The exponential map actually yields a 2:1 covering map of the direct Euclidean group since the \MOTORS $\vec{R}$ and $-\vec{R}$ yield the same isometry.}
 
 Also the infinitesimal aspects of kinematics are elegantly represented here. For simplicity write $\vec{g}(t) := e^{t \velo}$ for a general bivector $\velo$. Then $\vec{g}(t)$ is a euclidean motion and $\dot{\vec{g}}(t) = \velo \vec{g}(t)$, so $\dot{\vec{g}}(0) = \velo$. That is, the derivative of the motion at time $t=0$ is $\velo$ itself. We call $\velo$ in this context the \emph{global velocity state}. Also forces and momenta are represented by 2-vectors.  For example, a simple force is represented by the line $\momo$ upon which it acts.   %Already M\"{o}bius had considered the relationship of a velocity $\velo$ and a force $\momo$ interacting, and had introduced the term \emph{null line} of $\velo$ for a line $\sigo$ such that a force along $\sigo$ achieves no virtual work in the presence of the velocity state $\velo$. He also showed that the null lines through a point $\vec{P}$ lie in a plane through $\vec{P}$ called the \emph{null plane} of $\vec{P}$.  
%Consider  the motion of the point  $\vec{P}$ under the motion:  $\vec{P}(t) = \vec{g}(t)\vec{P}\widetilde{\vec{g}}(t)$.  In what direction does the motion $\vec{g}(t)$ move $\vec{P}$ at time $t=0$? To answer this question we need a theorem which will also be important for the discussion of rigid body motion.  
The following theorem shows how $\velo$ provides the basis for calculating all kinematic quantities in PGA. Consult \cite{gunnthesis}, \S 8.3 for a proof.  The subscripts $c$ and $s$ represent the body and space coordinate systems (space limitations prohibit a fuller discussion of this essential element of kinematics). 

\begin{theorem} \label{thm:liebracket}
For  time-varying multi-vector $\vec{X}$ subject to the motion $\vec{g}$ with velocity in the body $\velo_c$,
\[
\dot{\vec{X}}_s = \vec{g}(\dot{\vec{X}}_c+2(\velo_c \times \vec{X}_c )) \tilde{\vec{g}} = \vec{g}\dot{\vec{X}}_c\tilde{\vec{g}} +2(\velo_s\times \vec{X}_s)
\]
\end{theorem}

     \begin{figure}
   \centering
{\setlength\fboxsep{0pt}{\includegraphics[width=.3\columnwidth]{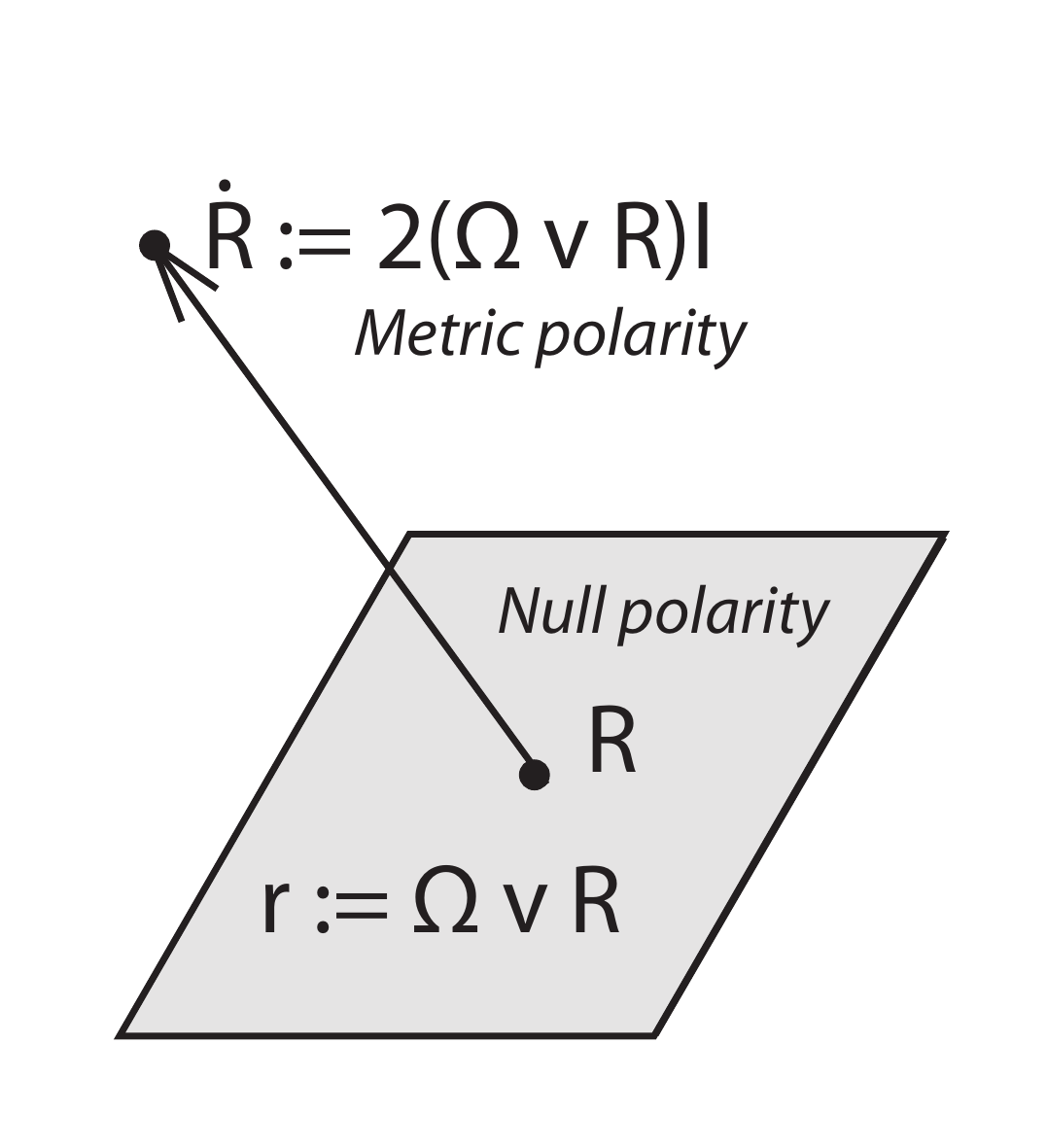}}}
{\setlength\fboxsep{0pt}{\includegraphics[width=.65\columnwidth]{threeNullPlanesWithPerpLabeled-02.pdf}}}
\caption{\emph{Left:} The infinitesimal direction of motion of a point is the composition of the null polarity of the velocity state $\velo$ and the metric polarity.\emph{Right}: A similar picture arises at each point of a null line.}
\label{fig:derivAtPoint}
\end{figure}
 
\myboldhead{Direction of motion vector field determined by $\velo$} 
The global velocity state $\velo$ sets every point in space moving in a particular direction.  We show how to find this direction as an example of how PGA kinematics works. 
The theorem implies that $\dot{\vec{P}}(t) = 2(\velo_s(t) \times \vec{P}_s(t))$, where $\vec{A} \times \vec{B} := \vec{A}\vec{B} - \vec{B}\vec{A}$ is the \emph{commutator} of $\vec{A}$ and $\vec{B}$, sometimes called the \emph{Lie bracket} in this context.  Specializing to $t=0$ and dropping the subscripts since the body and space coordinate systems agree there, we obtain $\dot{\vec{P}}(0) = 2(\velo \times \vec{P})$. Now we can apply one of the many identities of the geometric product, this one involving the product of a 2-vector and a 3-vector: $\velo \times \vec{P} = (\velo \vee \vec{P})\eye$. 

%\subsubsection{Null system} 
To understand the right-hand side we must first introduce the concept of the \emph{null system}, that plays an essential role in  this approach to rigid body mechanics. In \cite{moebius37}, M\"{o}bius  introduced the concept of a  \emph{null line}. In our terminology, a null line $\sigo$  of a global velocity state $\velo$ is a line such that a force acting along the line $\sigo$ is indifferent to  $\velo$, that is, achieves no virtual work in the presence of $\velo$. M\"{o}bius discovered that, in general, the set of all null lines passing through a point $\vec{P}$ lie in a plane of the point, the \emph{null plane} of $\vec{P}$; and the null lines lying in a plane pass through a point of the plane, the \emph{null point} of the plane.  In PGA, a null line satisfies $\sigo \wedge \velo = 0$, the null plane of $\vec{P}$ is $\velo \vee \vec{P}$,  while the null point $\vec{m}$ is $\velo \wedge \vec{m}$.  Together these two operations are part of the \emph{null polarity} defined on the whole algebra.  It is a polarity since it is grade-reversing and is an involution.

To return to our vector field: the right-hand side $(\velo \vee \vec{P})\eye$ is the product of two polarities. First, $\velo \vee \vec{P}$ is the null polarity associated to $\velo$. This is followed by the metric polarity (multiplication by $\eye$), sending an element to its \quot{orthogonal complement}. This gives the perpendicular direction to the plane $\velo \vee \vec{P}$.  But this is exactly the infinitesimal direction of motion of the point $\vec{P}$.  Consult \Fig{fig:derivAtPoint}.
 
\subsection{Rigid body mechanics}

In the interest of space, we focus here on a qualitative treatment of this topic.  Readers interested in full details should consult \cite{gunnthesis}, Ch. 9, \cite{gunn2011}, or \cite{gunnFull2010}.  

\myboldhead{Statics} We begin by rephrasing standard results of statics in this framework.  A \emph{simple force} is represented by a simple 2-vector $\pip$ where $\pip_e$ is the direction vector of the force and  $\pip_i$, considered as a 3-vector, is the \emph{moment} of the force and describes how the line of force is translated away from the origin.  A resultant of a set of forces $\pip_k$ is then given by the sum: $\pip = \sum_k{\pip_k}$.  The forces are in equilibrium $\iff \pip = 0$.  The resultant is a simple force $\iff \pip \wedge \pip = 0$.  Finally, the resultant is a force couple $\iff \pip = \pip_i$, that is, $\pip$ is ideal. Hence, just as  ideal elements allow PGA kinematics to handle translations and rotations uniformly, they do the same in dynamics for forces and force couples: A force couple is a force carried by an ideal line.

\myboldhead{Newtonian particles} Newtonian particles map in a similarly direct way to the bivector framework.  A particle is considered a point mass with mass $m$, located at position $\vec{R}$ and moving with velocity $\dot{\vec{R}}$. Then one defines:
%\begin{definition}
\begin{compactenum}
\item The \emph{spear} of the particle is $\vec{\Lambda} := \vec{R} \vee \dot{\vec{R}}$.
%\end{definition}
%\begin{definition}
\item The \emph{momentum state} of the particle is $\momo := m\vec{\Lambda}$.
%\end{definition}
%\begin{definition}
\item The \emph{velocity state} of the particle is $\pvelo := \vec{\Lambda}\eye$. 
%\end{definition}
%
%\begin{definition} \label{def:energy}
\item The \emph{kinetic energy} $E$ of the particle is 
%\begin{equation} \label{eqn:kinen}
$E := \frac{m}{2} \|\dot{\vec{R}}\|_\infty^{2} = -\frac{m}{2} \vec{\Lambda} \cdot \vec{\Lambda} = -\frac{1}{2}\pvelo \vee \momo$. 
%\end{equation}
\end{compactenum}
%\end{definition}
Physics proper begins by observing that in the absence of external forces, $\vec{\Lambda}$, $\momo$, $\pvelo$, and $E$ are conserved quantities.

\myboldhead{Rigid body mechanics} We restrict our further comments on mechanics to a bare minimum to arrive at the Euler equations of motion. Readers interested in the details are referred to \cite{gunn2011} or \cite{gunnFull2010}. 

Mimicking standard approaches of rigid body mechanics one can define a rigid body composed of such newtonian particles.  By summation/integration one can then define analogous quantities for the rigid body, such as momentum and kinetic energy.  The \emph{inertia tensor} can then be introduced, which is a quadratic form $\vec{A}$ on the space of velocity states $\bigwedge^2$ (bi-vectors).  $\vec{A}$ acts on vectors in the body coordinate system.  Then $\vec{A}(\velo, \velo)$ is the kinetic energy of the rigid body under the influence of the global velocity state $\velo$. We write $\vec{A}(\velo)$ for the associated polarizing operator, that produces the momentum state $\momo$ of the rigid body corresponding to this velocity.  Using the fact that the momentum in space, in the absence of external forces, is conserved, one can then derive the Euler equations of motion:
\begin{align*}\label{eqn:eqnmot}
\dot{\vec{g}} &= \vec{g}\velo_c\\
\dot{\velo}_c &= 2\inert^{-1}(\inert(\velo_c) \times \velo_c )
\end{align*}

\myboldhead{Analysis}This is a set of ordinary linear differential equations that display very attractive numerical features.  The solution space of each is a 6D quantity, for a total solution space of 12 dimensions. $\vec{g}$ is an element of the 8D even-subalgebra, while $\vec{\velo_c}$ is a bi-vector itself.  Hence the total representation space is 14 dimensions; there is very little space for \quot{wandering off} the solution.  In fact, we have obtained excellent practical results by simply normalizing $\vec{g}$ to have norm 1, a subspace (the so-called \MOTOR group) of dimension 6 within the even subalgebra. Compare this approach to the standard linear algebra approach, where the use of 4x4 (or 3x4) matrices to represent the Euclidean group introduces many extra dimensions to the representation space, forcing the use of Lagrange multiplies or other methods to constrain the solutions. The situation with the alternative geometric algebra model CGA is similar, see \cite{gunn2017a} \S 7.2.  That is, the PGA approach provides a nearly optimally small representation space for the Euler equations, minimizing numerical problems. Furthermore, from the solution one can directly read off the axis of the resultant screw motion $\vec{g}$ or the global velocity state $\velo_c$.  %Try doing that with a 4x4 matrix.
\begin{table}
\begin{center}
\normalsize
\setlength{\extrarowheight}{4pt}
\begin{tabular}{| p{.45\columnwidth} | p{.45\columnwidth} |}
\hline
\textbf{Projective geometric algebra} & \textbf{Vector and linear alg + analytic geom.} \\
\hline
\hline
Unified representation for points, lines, and planes based on a graded exterior algebra; all are \quot{equal citizens} in the algebra.
&
The basic primitives are  points and vectors and all other primitives are built up from these. For example, lines in 3D sometimes  parametric, sometimes w/ Pl\"{u}cker coordinates. \\
\hline
exterior algebra provides robust meet and join operators that deal correctly with parallel entities
&
Meet and join operators only possible when homogeneous coordinates are used, even then tend to be \emph{ad hoc} since points have distinguished role and ideal elements  rarely integrated. \\
\hline
Unified, high-level treatment of euclidean (\quot{finite}) and ideal (\quot{infinite}) elements of all dimensions.
&
Points (euclidean) and vectors (ideal) have their own rules, user must keep track of which is which; no higher-dimensional analogues for lines and planes.\\
%\hline
%Universal expressions: formulas handle varying types of arguments seamlessly.
%&
%Formulas tend to be \emph{ad hoc} with many special cases for parallel, non-parallel; separate ones for points, lines, planes, etc.\\
\hline
Unified repn. of isometries based on sandwich operators which act uniformly on points, lines, and planes.
&
Matrix representation for isometries has different forms for points, lines, and planes.\\
\hline
Same representation for operator and operand: $\vec{m}$ is the  plane as well as the reflection in the plane.
&
Matrix representation for reflection in $\vec{m}$ is different from the vector representing the plane. \\
\hline
Compact, universal expressive formulas and constructions based on geometric product (see tables \ref{tab:pga2d}, \ref{tab:pga3d}, and \ref{tab:pga3d2}) valid for wide range of argument types and dimensions.
&
Formulas and constructions are \emph{ad hoc}, complicated, many special cases, separate formulas for points/lines/planes.\\
\hline
Faithful representation of Lie algebra and Lie group of $\Euc{n}$ is geometrically intuitive and numerically optimal.
&
Representation of Lie algebra and Lie group of $\Euc{n}$ as subgroups of matrix group $GL(n, \mathbb{R})$ is not geometrically intuitive and is numerically problematical. \\
\hline
\end{tabular}
\myvspace{.1in}
\caption{A comparison of PGA with the standard approach. }
\end{center}
\end{table}
\twocolumn
}

\bibliography{GunnRef}

\begin{thebibliography}{Gun17b}

\bibitem[Bos18]{observable}
Mike Bostock.
\newblock Observable notebooks: a reactive {J}ava{S}cript environment, 2018.
\newblock https://observablehq.com.

\bibitem[Cli73]{clifford73}
William Clifford.
\newblock A preliminary sketch of biquaternions.
\newblock {\em Proc. London Math. Soc.}, 4:381--395, 1873.

\bibitem[Cli78]{clifford78}
William Clifford.
\newblock Applications of {G}rassmann's extensive algebra.
\newblock {\em American Journal of Mathematics}, 1(4):pp. 350--358, 1878.

\bibitem[Gla90]{gg90}
Andrew~S. Glassner.
\newblock Useful 3d geometry.
\newblock In Andrew~S. Glassner, editor, {\em Graphics Gems}, pages 297--300.
  Academic Press, 1990.

\bibitem[Gra44]{grassmann44}
Hermann Grassmann.
\newblock {\em Ausdehnungslehre}.
\newblock Otto Wigand, Leipzig, 1844.

\bibitem[Gun11a]{gunnthesis}
Charles Gunn.
\newblock {\em Geometry, Kinematics, and Rigid Body Mechanics in Cayley-Klein
  Geometries}.
\newblock PhD thesis, Technical University Berlin, 2011.
\newblock \url{http://opus.kobv.de/tuberlin/volltexte/2011/3322}.

\bibitem[Gun11b]{gunn2011}
Charles Gunn.
\newblock On the homogeneous model of euclidean geometry.
\newblock In Leo Dorst and Joan Lasenby, editors, {\em A Guide to Geometric
  Algebra in Practice}, chapter~15, pages 297--327. Springer, 2011.

\bibitem[Gun11c]{gunnFull2010}
Charles Gunn.
\newblock On the homogeneous model of euclidean geometry: Extended version.
\newblock {\em \url{http://arxiv.org/abs/1101.4542}}, 2011.

\bibitem[Gun17a]{gunn2017b}
Charles Gunn.
\newblock Doing euclidean plane geometry using projective geometric algebra.
\newblock {\em Advances in Applied Clifford Algebras}, 27(2):1203--1232, 2017.
\newblock \url{http://arxiv.org/abs/1501.06511 }.

\bibitem[Gun17b]{gunn2017a}
Charles Gunn.
\newblock Geometric algebras for euclidean geometry.
\newblock {\em Advances in Applied Clifford Algebras}, 27(1):185--208, 2017.
\newblock \url{http://arxiv.org/abs/1411.6502 }.

\bibitem[Hil13]{hildebrand13}
Dieter Hildebrand.
\newblock {\em Fundamentals of Geometric Algebra Computing}.
\newblock Springer, 2013.

\bibitem[HS87]{hessob87}
David Hestenes and Garret Sobczyk.
\newblock {\em Clifford Algebra to Geometric Calculus}.
\newblock Fundamental Theories of Physics. Reidel, Dordrecht, 1987.

\bibitem[Ken17a]{ganjacs}
Steven~De Keninck.
\newblock Ganja coffeeshop, 2017.
\newblock https://enkimute.github.io/ganja.js/examples.

\bibitem[Ken17b]{ganja}
Steven~De Keninck.
\newblock Ganja: Geometric algebra for javascript, 2017.
\newblock https://github.com/enkimute/ganja.js.

\bibitem[LLD11]{lasenby2011}
Anthony Lasenby, Robert Lasenby, and Chris Doran.
\newblock Rigid body dynamics and conformal geometric algebra.
\newblock In Leo Dorst and Joan Lasenby, editors, {\em Guide to Geometric
  Algebra in Practice}, chapter~1, pages 3--25. Springer, 2011.

\bibitem[M{\"{o}}b37]{moebius37}
A.~F. M{\"{o}}bius.
\newblock {\em Lehrbuch der {S}tatik}.
\newblock G{\"{o}}schen, Leibzig, 1837.

\bibitem[Stu91]{study91}
Eduard Study.
\newblock Von den {B}ewegungen und {U}mlegungen.
\newblock {\em Mathematische Annalen}, 39:441--566, 1891.

\bibitem[Stu03]{study03}
Eduard Study.
\newblock {\em Geometrie der {D}ynamen}.
\newblock Tuebner, Leibzig, 1903.

\bibitem[Wik]{wikiAD}
Wikipedia.
\newblock \url{https://en.wikipedia.org/wiki/Automatic_differentiation}.

\bibitem[Zie85]{ziegler}
Renatus Ziegler.
\newblock {\em Die Geschichte Der Geometrischen Mechanik im 19. Jahrhundert}.
\newblock Franz Steiner Verlag, Stuttgart, 1985.

\end{thebibliography}
\bibliographystyle{alpha}

\newpage

%\end{compactenum}
\end{document}